\documentclass{article}
\usepackage{algorithm}
\usepackage{algpseudocode}
\usepackage{tikz}
\usepackage{enumitem}
\usepackage{booktabs}
\usepackage{hyperref}
\usepackage{amsmath,amssymb}
\usepackage{mathtools}
\usepackage{fullpage}
\hypersetup{
    colorlinks=true,    % False for boxed links, true for colored text
    linkcolor=blue,     % Color of internal document links
    urlcolor=cyan,      % Color of external href links
    citecolor=blue     % Color of bibliographic citations
}

\usepackage{tikz}
\usetikzlibrary{arrows.meta, positioning, calc, shapes, fit}
\usepackage{graphicx}
\usepackage[dvipsnames]{xcolor}
\usepackage[]{natbib}

\title{CARL: Chat Assisted Reinforcement Learning}
\author{Sanjiv Das$^1$, Harshad Khadilkar$^2$, Sukrit Mittal$^2$, Daniel Ostrov$^1$, Jacob Sisk$^2$}
\date{$^1$Santa Clara University\\$^2$Franklin Templeton}

\begin{document}

\maketitle

\begin{abstract}
    Failure of applied Reinforcement Learning (RL) to solve a given problem predominantly stems from unaddressed ``specification gaps," where the intent of human experts in the problem's domain is inaccurately or incompletely translated by AI into mathematical objectives, constraints, and reward functions. Failure can also stem from unaddressed ``performance gaps," where human domain experts are unable to determine the validity of the RL based solution or where parameters (i.e., hyperparameters or weights on the reward terms comprising the reward) are poorly chosen. Chat Assisted Reinforcement Learning (CARL) bridges these gaps through a human-AI co-design interface architecture that enforces interactive specification prior to code generation and then enables subsequent verification and parameter alterations via iteration. More specifically, CARL leads domain experts through a structured ten-step workflow for each iteration using step-specific skill files, domain-aware suggestions via an assistant persona  (called Trillian), and compact decision summaries that maintain context across iterations without context overload. Once the domain expert confirms the Markov decision process (MDP) specified by this procedure, CARL's autonomous tool-use loop generates, tests, executes, and debugs an end-to-end RL training pipeline, followed by an interactive post-training diagnosis and revisit step that allows for verification and parameter alteration before running a new iteration if desired. We apply CARL to two disparate, path-dependent optimization domains: (1) goals-based wealth management in finance and (2) dynamic-pressure-constrained Goddard rocket control in aerospace engineering. In both of these examples, CARL demonstrates that interactive specification engineering using human domain expertise throughout the process resolves critical failure modes (such as reward hacking and state aliasing) and consistently produces viable policies where unguided, fully autonomous code generation fails. Full implementation available at \url{https://github.com/srdas/carl}.
\end{abstract}

\section{Introduction}
\label{sec:introduction}
% {\color{red} [[[Since we are no longer constraining ourselves to finance applications, this section will need to be rewritten with that in mind.]]]}
Reinforcement learning (RL) is a computational approach to optimization where an autonomous agent seeks to maximize a cumulative reward by interacting with a dynamic environment. RL is guided by a feedback loop of actions and consequences through forward simulations. The agent perceives its current {\it state}, chooses an {\it action} based on its current {\it policy}, and receives a {\it reward} that signals the quality of that choice. Over time, the agent refines its policy to favor actions that lead to the highest long-term reward, effectively teaching itself complex optimization behaviors.

Chat-assisted reinforcement learning (CARL) was conceived to answer the question: \emph{How can human experts in any domain of interest with a potentially insufficiently specified optimization question most efficiently interact with AI agents to construct a full reinforcement learning (RL) system to accurately specify and answer their question?}  
% \emph{Can an AI agent, together with a human user, construct a full reinforcement learning (RL) system from an ambiguous business objective?}  
CARL adopts a conversational user interface that asks clarifying questions, records decisions, and gradually turns the problem into a formal RL specification, while revealing its thinking in real time during the conversation.  After the specification is ready, CARL generates a complete code repository, trains the agent,  evaluates performance, and writes a report for user verification. These results can inform a new discussion through the interface, allowing for iterative improvements to the generated RL system's performance. 

The term ``most efficiently interact" from our key question above depends on the user and the application. That is, at different times users will weigh the relative importance of features like speed and quality. Therefore, CARL has a computer assistant called \emph{Trillian} that looks to provide domain knowledge-based suggestions for specifying the problem throughout the interaction with CARL. The human expert can accept, change, or reject these. This enables the human expert to be more involved (slower, but more careful) or less involved (faster, but possibly less careful) in the process, as they see fit. Users can also choose to run or not run additional iterations each time CARL generates a complete code repository and its corresponding results.

Reinforcement Learning is an increasingly important optimal control technique. Only a few optimal control problems can be solved with exact techniques like finding the exact solution to a corresponding Hamilton-Jacobi-Bellman partial differential equation. Therefore, numerical techniques are generally necessary. Indirect methods like shooting methods for solving boundary value problems for dynamical systems comprised of state and costate equations become difficult to work with when the number of the states or costates gets large or when constraints make the Pontryagin Maximum/Minimum Principle complicated. Direct methods like Dynamic Programming are limited by the curse of dimensionality, which limits the user to three or maybe four state variables and the fact that it is computed backwards in time makes it incompatible with optimal control for path-dependent phenomena. In contrast, RL is highly effective even when there are a large number of state variables and, because it works forwards in time via simulation, it can work with path-dependent phenomena. This makes RL an indispensable tool for solving optimal control problems.

However, RL has its own challenges. For example, unlike the methods above, RL is not guaranteed to generate a solution that is even a local, nonetheless global, optimum. This, by itself, is a sufficiently compelling reason to make certain that a domain expert analyzes the generated results to verify their validity.

But programming RL solutions is notoriously difficult because it shifts a domain expert's role from defining logic (where they have expertise) to shaping the model (where a specialist in RL has the necessary expertise). The RL system for each question is bespoke and has a huge number of hyperparameters (e.g., the learning rate, the discount factor, neural net architecture, random seeds, etc.) that need tuning. (See  \cite{li_reinforcement_2022} for example.) In RL, a goal (reward) is provided by the user who hopes the agent then discovers the optimal action on its own. The path the RL algorithm takes is often unpredictable, unstable, and sensitive to the slightest configuration changes. That is, the algorithm is quite brittle. When an agent receives a reward, it isn't always clear which of its many preceding actions caused that reward (because of delayed or sparse rewards). And tuning the exploration-exploitation tradeoff is also challenging. Because this work on building RL packages requires extensive development of the RL environment and related optimization code, it often takes months of work for each specific application. 

But what if this can be done automatically? If a domain expert can simply hand off a well-defined problem specification to a vibe coding tool (Claude Code, Codex, Gemini, etc.), and it produces a high-performing RL solution, RL becomes a far more potent, accessible tool. However, there is a significant problem with vibe coding: even when an expert writes out detailed specifications of the RL problem, we will see that vibe coding tools do not develop viable solutions to even simple problems. Though the software runs bug-free and produces results, checking against known solutions shows glaring inaccuracies. Even though RL has matured into a set of well‑tested algorithms, open‑source libraries, and production‑grade training platforms, vibe coding leads to code that reuses solutions in other related libraries, plausibly deviating from the given specification. This is a known problem. That is, a large portion of applied RL projects fail not because of the algorithms but because the problem is underspecified \citep{jayawardana_impact_2022, hamadanian_how_2022}.  Domain experts often provide a high‑level goal but leave out critical details such as constraints, reward shaping, or suitable state representations.  When an RL agent is instantiated with an incomplete RL specification, hidden assumptions surface only after the code has been written, leading to costly re‑engineering and iteration cycles. 

Consequently, the success of our RL modeling hinges less on the raw power of the underlying neural networks and more on the precision with which the human-AI interaction specifies the problem from the outset. We call this the {\it RL specification gap}, where the transition from a high-level goal to a functioning RL agent is where the most significant ideation and design friction occurs. That is, humans with expertise are essential for RL agents to be able to create useful, correct results. Carefully designing the interaction between human experts and AI agents is key to minimizing the RL specification gap, which is the true bottleneck for using RL effectively. CARL is specifically designed to help resolve this bottleneck.

Academics and practitioners have begun to emphasize specification engineering as a primary stage of the development life cycle \citep{li_formal_2019, ibrahim_comprehensive_2024}. This involves more than just selecting a policy gradient or Q-learning variant; it requires a rigorous translation of human intent into formal mathematical objectives. Recent work has begun examining automating RL specification refinement \citep{ambadkar_automating_2026}. This is also related to the broader paradigm of ``spec-driven development'' \citep{alur_specification-guided_2026, huang_reinforcement_2026}.  

RL misspecification leads to several issues. Without a robust feedback loop between domain experts and vibe coding tools to define safety boundaries and penalize undesirable behaviors, the agent may find shortcuts that satisfy the reward function while violating the spirit of the original task, a phenomenon known as ``reward hacking'' \citep{amodei_concrete_2016} where the agent finds a way to maximize reward without achieving the intended task. For example, a cleaning robot clears spilled food by hiding it under the refrigerator. 

Reward hacking is just one such problem. We may also insufficiently define or design the state space, leaving relevant and essential information out of the observation space. This can lead to ``state aliasing'' \citep{asri_study_2019} where the RL agent cannot distinguish between two states that are subsumed into a single state because a relevant dimension of the space has been excluded. When aliasing occurs, the agent perceives an observation $o(t)$ instead of the true state $s_i(t)$. If two different states $s_1(t)$ and $s_2(t)$ map to the same observation $o(t)$, the agent may require a different optimal action for each, but it cannot tell which state it is in. This leads to suboptimal policies and unstable value functions. 

Another issue that arises from the RL specification gap is ``constraint violation'' \citep{bai_achieving_2022, ghosh_towards_2024}. While a standard RL agent is typically programmed to maximize a cumulative reward, it may not conform to implicit rules unless they are explicitly baked into the optimization process. This creates a conflict between performance (doing the task) and safety (doing it correctly). In many cases, the environment itself is learned and this causes risks that hard constraints may be violated during the exploration phase (in training), or function perfectly in normal states, but violate constraints when it encounters a state it didn't see during training (post-training). The environments for the RL problem, also known as constrained Markov decision processes (CMDPs), have to be carefully designed. Hyperparameters also need to be set carefully. 

A recent popular example of interactive RL use is reinforcement learning with human feedback (RLHF). This is a technique that trains a model using human judgments to build a reward model, then uses RL to optimize the model toward outputs people prefer. It emerged as a practical way to look to align models with human goals, becoming especially prominent in OpenAI’s work on language models and later InstructGPT \citep{ouyang_training_2022, stiennon_learning_2022}.  It has been useful in RL because many real tasks have goals that are hard to specify precisely in a hand-designed reward function, so human feedback supplies a richer, more practical signal. Even this approach has limitations, \cite{casper_open_2023}. 
Our approach is similar to RLHF in spirit, and comprises RL specification with human feedback (RLSHF). This paper presents the CARL methodology, the software architecture that implements it, and illustrative empirical examples.

More specifically for our two main examples, we will evaluate CARL on two sequential decision problems from very different domains of expertise. In Section \ref{sec:example_gbwm} for our first main example, we consider a goals-based wealth-management problem in which an investor reallocates wealth annually between equities and bonds while deciding whether to take or skip any currently available financial goals. It is assumed that the time when each goal is available, the cost of attaining each goal, and the utility/happiness generated by attaining each financial goal are specified. In our example, each goal's utility depends on the number of earlier goals attained, making the problem path-dependent, and hence ineligible for standard solution approaches such as dynamic programming, as applied in \cite{das_dynamic_2022}. 

In Section \ref{sec:example_goddard} we consider our second main example, which is a variation of the classic Goddard rocket problem. As in the classic case, a rocket modulates its throttle to maximize its peak altitude under coupled flight dynamics. To this we add a hard dynamic-pressure constraint due to pressure limitations on the rocket's hull. The exact optimal solution can be numerically determined if there is no dynamic-pressure constraint, but RL is needed in the presence of this constraint. This control problem requires decisions about the state representation, action distribution, safety mechanism, reward scale, and training procedure. 

Together, theses examples test whether CARL can turn high-level objectives into constraint-aware RL implementations across different domains. We will document improvements in the RL solutions generated by CARL in both cases, but our goal is to provide a blueprint that others can adapt to their own domains.

The rest of the paper proceeds as follows. Section \ref{sec:background} discusses how related literature influences and connects to CARL. Section \ref{sec:tour} shows how CARL is experienced by the user. Section \ref{sec:methodology} delves into the methodology and design logic behind CARL, while Section \ref{sec:architecture} presents the system architecture behind CARL. The two main examples discussed above are developed and presented in Sections \ref{sec:example_gbwm} and \ref{sec:example_goddard}. Sections \ref{sec:discussion} and \ref{sec:future} discuss the main ideas and implications of this paper, as well as possible directions for future work.

\section{Context: Background and Work Related to CARL}
\label{sec:background}

Reinforcement learning (RL) now supports many applications where optimization comes from a machine-based learning-by-doing approach. Several domains have been aggressively employing RL. A few examples include training autonomous vehicles \citep{kiran_deep_2020}, industrial robotics \citep{kober_reinforcement_2013}, personalized healthcare \citep{murphy_optimal_2003}, optimizing game-playing AIs \citep{mnih_playing_2013, mnih_human-level_2015, silver_mastering_2016}, smart recommendation systems \citep{Zheng_drn_2018}, energy grid management \citep{vazquez-canteli_reinforcement_2019}, supply chain logistics \citep{giannoccaro_inventory_2002}, optimal card game strategies \citep{sutton_reinforcement_2018},
% popularized Blackjack as the textbook benchmark problem for introducing Monte Carlo methods and temporal-difference learning), 
operating a hydroelectric reservoir under seasonal inflow and flood risk \citep{xie_deep_2026}, real-time advertisement bidding \citep{cai_real-time_2017}, and optimizing rocket propulsion \citep{rozada_matrix_2024}.
% Video game playing has been a fertile area of RL application in which RL-trained players beat or ``solve'' games in a facile manner \citep{mnih_playing_2013, mnih_human-level_2015, silver_mastering_2016}. 
In the area of reinforcement learning with human feedback (RLHF), RL has been used with humans to tune large language models \citep{ouyang_training_2022}. Robotic planning and training is also a fast-developing area \citep{li_formal_2019}.

Just within the field of finance, reinforcement learning is being used to effectively navigate many complex, high-dimensional financial landscapes, demonstrating broad utility across portfolio optimization, credit risk reduction, and robo-advising \citep{malibari_systematic_2023}, as well as algorithmic trading \citep{moody_learning_2001}. The robotic planning and training mentioned in the previous paragraph has been applied to financial optimization \citep{ndikum_advancing_2024}. In dynamic wealth management, RL addresses goals-based optimization for asset-liability matching strategies \citep{ huang_optimal_2010}. Advanced methodologies enhance multi-period asset allocation, such as the data-agnostic ``G-Learner" for high-dimensional portfolios \citep{dixon_g-learner_2020} and Schrödinger Control Optimal Planning (SCOP) for goal-based ``satisficing" solutions \citep{halperin_scop_2023}. Beyond personal wealth, RL is equally effective in liability-driven 401(k) and pension fund management where optimal strategies must balance capital market, longevity, scheme maturity, and funding ratio risks \citep{cox_managing_2013, das_augmenting_2026, zhao_what_2021}, alongside the vital liquidity and risk controls highlighted by events like the 2022 gilt market dysfunction \citep{kodres_too_2023}.

We will explore two use cases as exemplars in this paper. One use case is optimizing dynamic investment and goal-taking choices within the finance domain. The other, from the domain of engineering, is optimally controlling thrust in a rocket with a specified amount of fuel so as to maximize the rocket's altitude given a pressure constraint on the rocket's hull. These two examples were purposely chosen from very different domains to demonstrate the general application of CARL to a wide range of RL problems. Both examples also have simplified cases where they can be solved exactly, so we can see how close RL comes to obtaining the global optimal strategy, but we will be exploring cases for both examples where exact methods fail and RL becomes the best tool to find solutions.  

The ``human-in-the-loop" aspect of CARL is essential. A number of papers have shown that effectively generating solutions with AI crucially requires the involvement of humans with domain expertise. For example, in \cite{otis_uneven_2024}, a trial with 640 small business entrepreneurs showed that low performers (those with less domain expertise) did
nearly 10\% worse in generating firm revenues and profits due to being given an AI assistant, while high performers (those with more domain expertise) may have benefited
by over 15\%. Or as noted in \cite{tambe_reskilling_2026}: ``The evidence presented in this [paper's] analysis suggests that the market rewards firms that integrate algorithmic capabilities with domain expertise, as these workers can better identify opportunities and translate
... needs into technical solutions." \cite{huang_professional_2026} shows that experienced developers do not relinquish control to agents, choosing to use their expertise to guide and supervise the agents instead, thereby always remaining in the loop. Developers with domain expertise are more positive about their experience incorporating agents into software development given that they can manage around  agents' limitations. 

A domain expert will reject AI results that do not seem right to them or are not clear to them. In \cite{lebovitz_engage_2022}, three departments of diagnostic radiologists used AI to assist in making bone age, breast cancer, and lung cancer determinations, but given the opacity of reasoning behind the AI results, the bone age and breast cancer departments stopped using AI. But a domain expert that can work along with AI and change it as necessary is far more likely to find the AI results useful.  As \cite{dietvorst_overcoming_2018} states it, people are ``more willing to use an imperfect algorithm if they are given the ability to intervene when they suspect that the algorithm has it wrong." 

The core motivation for CARL is to minimize the ``RL specification gap,'' because the transition from human intent to formal mathematical objectives is often the bottleneck that leads to project failure. This challenge is well-documented: a large portion of applied RL projects fail, not because of the algorithms, but because the problem is underspecified \citep{hamadanian_how_2022}. 
This is well stated in \cite{passi_problem_2019}: ``Formulating data science problems is an uncertain and difficult process. It requires various forms of discretionary work to translate high-level objectives or strategic goals into tractable problems, necessitating, among other things, the identification of appropriate target variables and proxies. ... [T]hese choices are rarely self-evident." \cite{passi_problem_2019} go on to point out that in data science, ``[t]he work of problem formulation---of iteratively
translating between strategic goals and tractable problems---is
anything but self-evident," noting that ``some of the most
important normative implications of data science systems find their roots in the work of problem formulation."

The need to carefully orchestrate human interaction with AI and ensure accurate specification is also highlighted in \cite{li_scientific_2024}, who note three lessons from a study of eight scientific computing projects in the life science domain: First, the burden on the human for verification increases with the complexity of the project. Second, projects need to proceed by iteration, not one-shot specification. Third, even though AI agents can undertake complex tasks, they are not effective at evaluating the outcomes of the systems they build, necessitating human control and evaluation. In a similar vein, \cite{meng_scientistone_2026} note that ``generation-verification'' systems' multi-stage pipelines are characterized by greater error propagation across stages, hence one-shot specification is disfavored compared to iterative step-by-step specification. 

From the perspective of these papers, in addition to the specification gap, there is a \emph{performance} gap that also needs to be addressed. Within each iteration of CARL, there are a number of steps designed to reduce the specification gap. After each iteration, however, CARL provides an opportunity for domain expert verification of results, as well as alterations based on this analysis of the results. CARL also uses the results to recommend any new hyperparameter values and any new weights for the reward terms if some of those terms were not relevant to the results and therefore likely underweighted. These user defined alterations along with any accepted CARL recommendations are then used in the next iteration. This iterative approach is designed to reduce the performance gap.

The research areas of human-computer interaction (HCI) and human-AI interaction are connected to this paper. CARL is a human-computer interaction (HCI)  system: its central contribution is not merely the automation of reinforcement-learning implementation, but the design of an interaction process through which users progressively communicate, inspect, and revise computational objectives. CARL emphasizes the design of interaction techniques that allow human guidance, preferences, and feedback to shape an agent's learning process. (See \cite{arzate_cruz_survey_2020} for a survey of interactive RL.) More specifically, CARL treats RL specification as a mixed-initiative, human--AI collaboration where the user contributes domain knowledge and evaluative judgments, while the system elicits formal problem definitions, decomposes them into inspectable stages, and supports iterative performance gap reduction. This design reflects established HCI guidance that AI systems should communicate their capabilities and limitations, support user control, provide opportunities for correction, and make their behavior understandable;  \cite{amershi_guidelines_2019} offers guidelines for the design of such HCI. Accordingly, CARL addresses both a usability problem and an engineering problem in that it seeks to reduce the cognitive and representational burden of translating informal goals into states, actions, transitions, rewards, and constraints, while preserving meaningful human oversight over the resulting RL specification. The research area of specification engineering is also related to this paper. For example, recent work has begun on the use of conversational interfaces to refine RL objectives \citep{ibrahim_comprehensive_2024}. This paradigm shifts the role of the developer toward shaping model behavior through precise communication, a process that CARL seeks to automate via its interactive workflow.

Typical RL pipelines are split into two stages: (1) \emph{problem formalization} and (2) \emph{system implementation}. For the second stage, libraries such as \texttt{Stable-Baselines3} and \texttt{RLlib} are used. (These implementations are based on papers such as \cite{schulman_proximal_2017} and \cite{liang_scalable_2021}. Domain-specific implementations such as \texttt{FinRL} also exist \citep{liu_finrl_2022}.)
CARL, however, focuses on the first stage.  Conversational AI has emerged as a facilitator for software engineering tasks.  \texttt{GitHub\hspace{-.3em} Copilot} and \texttt{ChatGPT} can produce code given a prompt, yet they lack an explicit stepwise workflow for complex systems \citep{bai_re-examining_2024}. These tools are strongest at local, prompt-driven tasks and remain weaker on end-to-end engineering for large, complex systems that require planning, decomposition, validation, and context maintenance \citep{khojah_beyond_2024, dong_survey_2025}. Even when provided with detailed specifications, general-purpose coding assistants can produce working code that yields results with mistakes  when compared to known closed-form solutions. CARL addresses this by decomposing the design process into multiple discrete \emph{skill} files, each representing a conversation step.  CARL aims to comprehensively manage the specification and performance gaps and provide a full prototype solution to problems that can more rapidly be taken to production.

\section{A Tour of CARL from the User's Perspective}
\label{sec:tour}

In this section we describe the experience of a domain expert as they interact with CARL. In Subsection \ref{sec:ten}, we will describe the ten-step workflow for each iteration of CARL. Subsection \ref{sec:carldemo} walks through a specific simple reinforcement learning (RL) application solved by a single iteration of CARL. Finally, in Subsection \ref{sec:aux}, we discuss a number of additional features supported within CARL.

\subsection{Ten-Step Workflow}
\label{sec:ten}

\begin{figure}[ht]
\centering
\includegraphics[width=0.96\textwidth]{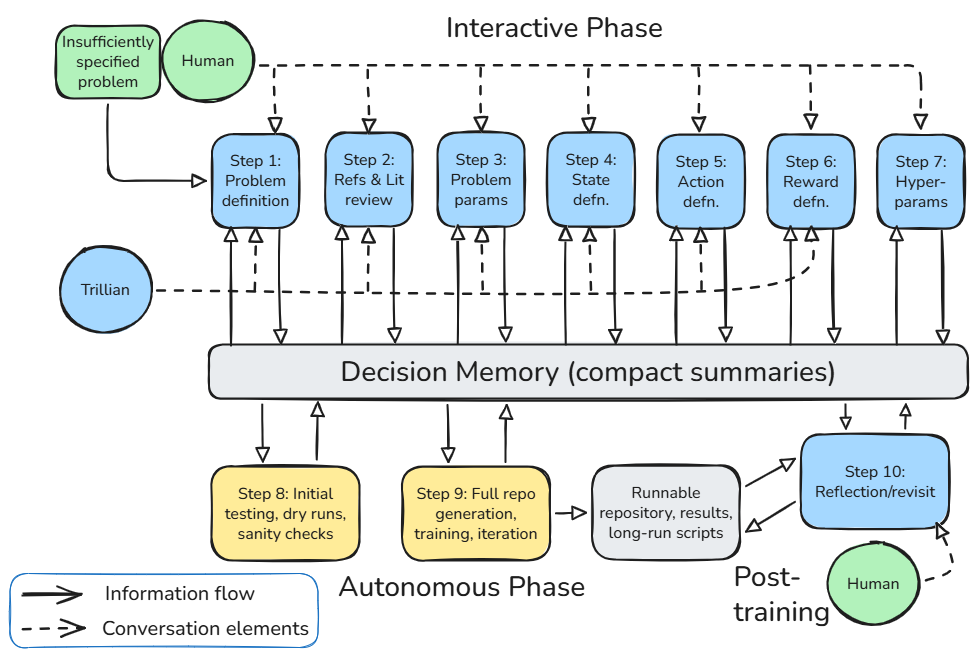}
\caption{Logical design of CARL, showing the interactive specification phase, autonomous implementation phase, Trillian reply assistance, and post-training revisit loop.}
\label{fig:carl-flowchart}
\end{figure}

Table~\ref{tab:steps} lists each of the ten steps comprising the first iteration of CARL, along with each step's associated skill and objective. Figure~\ref{fig:carl-flowchart} displays a flowchart showing how these ten steps fit into the broader CARL design. 

\begin{table}[ht]
\centering
\small
\begin{tabular}{@{}lp{0.22\textwidth}p{0.47\textwidth}r@{}}
\toprule
\textbf{Step} & \textbf{Skill} & \textbf{Objective} & \textbf{Approx. tokens}\\ \midrule
1 & Problem Definition & Clarify domain objective, dynamics, constraints & 900\\
2 & References & Gather domain knowledge, data sources, baselines & 350\\
2a & Online Search (\textbf{Optional}) & Search for related literature to assist formulation & 600\\
3 & Parameter Exploration & Define exogenous variables and randomization ranges & 700\\
4 & Information Proposal & Specify state variables and normalization & 1,150\\
5 & Action Definition & Specify action space and algorithm family & 1,250\\
6 & Reward Definition & Specify objective, reward, and shaping logic & 1,250\\
7 & ML/RL Parameters & Confirm learning settings and training budget & 2,300\\
8 & Initial Testing & Validate generated artifacts against the specification & 2,200\\
9 & Launch Training & Train, evaluate, and summarize the learned policy & 4,450\\ 
10 & Revisit (\textbf{Optional}) & Analyze results, edit MDP or hyperparameters, retrain & 5,500 \\ \bottomrule
\end{tabular}
\caption{CARL logical steps. Steps~1--7 are interactive specification steps; Steps~8 and 9 are autonomous implementation steps governed by the accumulated specification. Step 10 can be launched by the user if they are dissatisfied with the results from previous runs and wish to run another iteration. (New iterations can be done an arbitrary number of times.) CARL itself recommends whether a revisit and new iteration should occur at the end of Step 9. Token counts are approximate full-skill-file prompt sizes, rounded from a character-based estimate.}
\label{tab:steps}
\end{table}

The workflow is ordered so that each step in the iteration resolves information needed by later steps. For Steps 1--7, the Markov decision process (MDP) is not treated as a single prompt-triggered task; it is decomposed into decisions that a domain expert can inspect and revise in each step. Implementation in Steps 8 and 9 is deliberately downstream of specification. A generated repository is useful only if the state, action, reward, constraints, and evaluation protocol express the intended problem. The generated results from Steps 8 and 9 can reveal to the domain expert or to CARL further design issues that should be resolved, so Step 10 creates the option of correcting these design issues and then running an additional iteration of CARL.

More specifically, Step~1 establishes the domain problem before any RL formalism is introduced. Step~2 records user-supplied references, while Step~2a optionally broadens the design space by suggesting additional references (papers and documents) that the domain expert can choose to incorporate if desired. Steps~3--6 then construct the MDP by having the domain expert, working with Trillian and CARL, determine parameter variation, then state representation, then action definition, and, finally, reward design. Again, this order is carefully chosen. For example, because we have determined the parameter variation in Step 3, Trillian can suggest appropriately normalized state variables for the domain expert to consider in Step 4. This is important because normalized state variables are key to gradient descent (the numerical optimization method that underlies RL) working effectively. 

Step~7 closes the interactive phase by confirming the RL settings, including the hyperparameters and any training budget constraints. Since the domain expert is not assumed to be an RL expert, CARL determines these values. That is, unlike the previous steps, Trillian does not look to potentially engage in a conversation with the domain expert about these detailed aspects of RL. That said, all of these decisions are clearly stated and can easily be altered by the domain expert if they do understand how these aspects of RL work. 

Only after these decisions are agreed to by the domain expert and then recorded do Steps~8 and~9 begin. These are the steps where autonomous implementation, validation, training, and evaluation occur. Step~8 is a sanity check to ensure that the generated artifacts are consistent with the specification, after which Step~9 executes the training and evaluation protocol. At the end of Step 9, CARL reflects on how well the training went and may recommend that the user launch Step~10 to revisit the design and improve it further. The user can launch Step~10 anyway (with or without CARL's recommendation) if they wish. If Step 10 is employed, CARL will run another iteration starting from the new design Step 10 determines.

\subsection{Demonstration on a Simple Optimal Stopping Problem} \label{sec:carldemo}

The best way to truly understand how the interactive workflow for CARL works is to view the video at \url{https://youtu.be/Qx673uczp4c}. This video shows CARL solving a simple optimal stopping problem described in the next paragraph. The problem is purposely trivial for RL to solve because it is chosen to give the reader a feel for CARL's workflow. That is, it is not intended to showcase the power of CARL or of RL.

The video begins with the user giving CARL the following simple RL problem: a player can repeat, up to ten times, independently drawing a uniformly distributed integer between 1 through 50. They may decide to stop at any time and accept the current drawn number, but they must accept the tenth drawn number if they did not stop earlier. The objective is to maximize the accepted number. Dynamic programming yields the stopping strategy that produces the optimal expected value of this accepted number, which is 43.55. 

\begin{figure}[ht]
\centering
\includegraphics[width=0.96\textwidth]{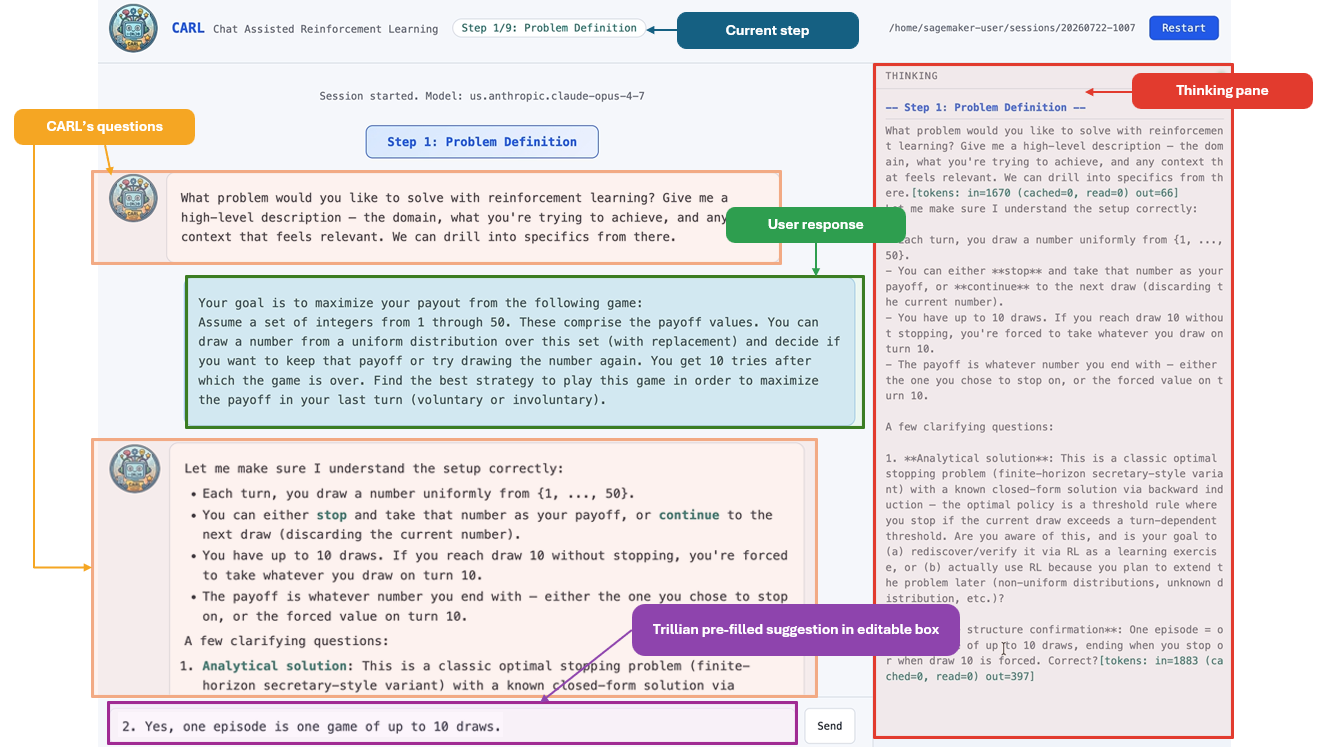}
\caption{[Time stamp \href{https://youtu.be/Qx673uczp4c?t=15}{\texttt{00:15}}] CARL web interface captured during Step 1 for our sample problem. CARL asks some clarifying questions to ensure that the rules of the game are clearly understood (better than the 2-3 sentence description initially provided). Trillian suggests answers that the user can choose to edit.}
\label{fig:carl-interface}
\end{figure}

The video shows that CARL can design the RL system to solve for the optimal strategy. Over the first seven steps of the first iteration, CARL turns the user's description of this problem into a precise optimal-stopping formulation: state observations, binary stop/continue decisions, terminal payoff, finite horizon, and an analytical threshold-policy benchmark reserved for evaluation. It decomposes the user's request into a sequence of modeling decisions: whether outside references are needed, which parameters are fixed or randomized, what information belongs in the state, what action space and algorithm fit the problem, how the reward should be defined, and which training hyperparameters and time budgets are appropriate. At each step, CARL proposes defaults, explains the tradeoffs, asks for confirmation, and then records the answer as a summary that becomes the contract for later implementation. For reference, the CARL interface is shown in Figure~\ref{fig:carl-interface} for the first of the ten steps.

In each of the first six steps, Trillian appears as the response-prefill layer between CARL and the user. After CARL asks a question, Trillian drafts a likely user reply in the input box. The user can accept that draft unchanged, which happens often for confirmations such as ``ok,'' or can edit it before sending. Two edits are especially meaningful: the user narrows the success criterion to near-optimal expected payoff rather than exact threshold recovery, and later clarifies that the analytical solution should be used only for evaluation, not training. The recording therefore shows a three-part workflow: CARL structures the problem, Trillian looks to reduce the user's response effort by prepopulating answers, and the user remains in control by accepting or modifying those answers before CARL proceeds. Steps 8 and 9 of the session show CARL carrying the specification into implementation. It builds a Double DQN environment and agent, adds baselines and tests, and then launches training. 

A more specific guide to the video, with the video's timestamps in blue, follows. Note that this serves well as a guide, but not a substitute, for watching the video:

\begin{description}[leftmargin=1.0in,style=nextline]

\item[\href{https://youtu.be/Qx673uczp4c?t=0}{\texttt{00:00}}]
{\bf Step 1} begins: CARL opens by asking the user to define the reinforcement-learning problem at a high level.

\item[\href{https://youtu.be/Qx673uczp4c?t=10}{\texttt{00:10}}]
The user specifies the game described above: uniform draws from integers 1 through 50, replacement after each draw, 10 total tries, and a stop-or-redraw choice to maximize final payoff.

\item[\href{https://youtu.be/Qx673uczp4c?t=20}{\texttt{00:20}}]
CARL reformulates the game as a finite-horizon optimal-stopping problem and asks clarifying questions about the learning goal and episode structure. 
Trillian prefills a response saying this is a learning exercise and that one episode is one game; the user accepts it.

\item[\href{https://youtu.be/Qx673uczp4c?t=65}{\texttt{01:05}}]
CARL summarizes the problem and asks what success should mean. Trillian suggests both threshold recovery and near-optimal payoff, but the user edits the response down to ``achieve near-optimal expected payoff.'' 

\item[\href{https://youtu.be/Qx673uczp4c?t=135}{\texttt{02:15}}]
Step 1 is summarized by CARL before moving to Step 2. Note that this summary is added to memory as the persistent takeaway for each step. This will be discussed further in Subsection \ref{skillz}.

\item[\href{https://youtu.be/Qx673uczp4c?t=140}{\texttt{02:20}}]
{\bf Step 2} handles references. Trillian first supplies ``no additional files,'' then the user edits the next prefill to skip the literature search explicitly.

\item[\href{https://youtu.be/Qx673uczp4c?t=210}{\texttt{03:30}}]
{\bf Step 3} begins parameter exploration. CARL separates fixed game rules from randomization and proposes randomized starting turns for balanced state coverage.

\item[\href{https://youtu.be/Qx673uczp4c?t=260}{\texttt{04:20}}]
The user accepts Trillian's prefilled choice to randomize the starting turn while keeping the horizon (i.e., the maximum number of draws) fixed at 10.

\item[\href{https://youtu.be/Qx673uczp4c?t=310}{\texttt{05:10}}]
{\bf Step 4} defines the state representation: normalized current draw number, normalized remaining turns, and an optional reciprocal remaining-turn feature. Since the problem is very small and technically requires only one state (the current draw) to be Markov, Trillian suggests keeping all three suggested state inputs and letting the training decide how much importance to give to each.
The user accepts Trillian's suggestion. 

\item[\href{https://youtu.be/Qx673uczp4c?t=410}{\texttt{06:50}}]
{\bf Step 5} defines actions. CARL proposes the natural binary action space (keep the drawn number or draw another), compares DQN with a policy-gradient alternative, and recommends DQN as the simpler fit.

\item[\href{https://youtu.be/Qx673uczp4c?t=455}{\texttt{07:35}}]
The user edits the Trillian draft to add an important constraint: do not use the analytical solution for training; use it only for evaluation. The user also asks CARL to explain the algorithm alternatives.

\item[\href{https://youtu.be/Qx673uczp4c?t=595}{\texttt{09:55}}]
{\bf Step 6} defines the reward: terminal payoff from when the draw is stopped or the tenth draw if it is never stopped, zero otherwise (if there is an runtime error), no shaping, and no discounting.
The user accepts Trillian's prefill. 

\item[\href{https://youtu.be/Qx673uczp4c?t=710}{\texttt{11:50}}]
{\bf Step 7} selects the Double DQN hyperparameters, including replay buffer size, epsilon-greedy exploration, target updates, network architecture, evaluation cadence, and diagnostics.

\item[\href{https://youtu.be/Qx673uczp4c?t=740}{\texttt{12:20}}]
The user sets concrete time budgets: No more than 10 minutes for initial testing and 20 minutes for the launch/training step.

\item[\href{https://youtu.be/Qx673uczp4c?t=760}{\texttt{12:40}}]
{\bf Step 8} begins, taking CARL into autonomous mode (with the option for the user to interject if needed). CARL reads the prior decisions, states the binding constraints, and starts creating the environment, agent, baselines, dry run, plots, and tests.

\item[\href{https://youtu.be/Qx673uczp4c?t=825}{\texttt{13:45}}]
The user adds a new instruction: output the training reward plots when done. CARL acknowledges this and validates plotting with baseline reward distributions.

\item[\href{https://youtu.be/Qx673uczp4c?t=880}{\texttt{14:40}}]
The dry run reveals a benchmark correction. During Step 7, CARL used pure LLM inference to estimate that the optimal expected value was 30.87. In Step 8, CARL actually writes a short backward induction code and precisely computes the optimal expected value, which is 43.55. CARL records the correction before proceeding.

\item[\href{https://youtu.be/Qx673uczp4c?t=1110}{\texttt{18:30}}]
Step 8 closes with passing tests, baseline plots, and a summary of code artifacts.

\item[\href{https://youtu.be/Qx673uczp4c?t=1115}{\texttt{18:35}}]
{\bf Step 9} begins. CARL turns the tested scaffolding into the full training repository, adds evaluation scripts and reward-buffer regression tests, and launches the Double DQN training run.

\item[\href{https://youtu.be/Qx673uczp4c?t=1300}{\texttt{21:40}}]
Training output shows the agent approaching the analytical optimum, with learned thresholds moving toward the backward-induction thresholds.

\item[\href{https://youtu.be/Qx673uczp4c?t=1525}{\texttt{25:25}}]
CARL reports the final launch summary: tests pass, plots are generated, checkpoints are saved, and the learned policy is near analytical performance (within one standard deviation error bound in simulation, and $99.93\%$ of the optimal expected value when the expectation is computed analytically using the trained policy). A trigger for analyzing the interaction for updates to the Trillian persona is launched, completing the CARL run.\\\textit{Remark: The main CARL output from Step 9 only appears as the summary, and the video capture misses some of the text at the top. We have added the full Step 9 summary in the video description.}

\end{description}

The final result demonstrates the stated goal: CARL's learned policy corresponds to an expected value of 43.52, which is extremely close to the optimal expected value of 43.55. We note again that the optimal expected value of 43.55 remained an evaluation benchmark rather than a training signal. Because CARL's result is so close to the analytical solution, CARL reports that there is no reason for an additional iteration, meaning there is no reason to run Step 10.

\subsection{Auxiliary Features}
\label{sec:aux}

CARL also includes several auxiliary features that support its core ten-step design logic:
\begin{itemize}
\item \textbf{Brainstorming mode}. The \texttt{/suggest} option is used in Step 1 when the user has a broad domain interest but not yet a precise RL problem. CARL searches relevant literature, synthesizes candidate formulations, and asks the user to choose or refine one. The selected formulation is then written as a reference document and routed back into the ordinary specification workflow.
\item \textbf{Resume mode}. If a session is interrupted, CARL can resume from the point at which it was interrupted, by adding a \texttt{--resume} flag during launch. It then analyzes the existing log files to identify where to pick up the thread, and executes the remaining steps.
\item \textbf{Step~10 revisit mode}. After a completed run, CARL can reopen the session for post-training diagnosis. The user can provide guidance about what to inspect or exclude, after which CARL reviews artifacts such as memory, logs, plots, and generated code. Any proposed changes are discussed interactively and prepared in a fresh revisit workspace, rather than silently modifying the completed run.
\item \textbf{Autonomous-step interjections}. Although Steps~8 and~9 run autonomously, they are not closed to the user. During code generation, testing, or training, the user can inject guidance that is added back into the message stream before CARL continues. This allows the user to redirect implementation without restarting the workflow or breaking the step boundary.
\item \textbf{Trillian persona update}. Trillian suggests domain-expert-style replies during the early specification steps, and the user's accepted or edited responses provide feedback about the usefulness of those suggestions. After a session, CARL can compare observed interactions with the current persona and update it only when there is systematic evidence that the user's persona is significantly different from Trillian's persona. The update process is intentionally conservative, using triage, minimal edits, and consistency checks to avoid persona drift. 
\item \textbf{\LaTeX{} report output}. At the end of Step 9, CARL can generate a paper-style report from the accumulated design memory and execution artifacts. It also adds an appendix after every iteration of Step 10. The report summarizes the problem specification, implementation choices, training logs, plots, evaluation results, and any literature gathered during the run. This output is treated as documentation of the completed workflow, not as part of the MDP itself. This enables further verification.
\end{itemize}

\section{Logical Design Choices Behind CARL}
\label{sec:methodology}

This section describes the logical design choices that govern CARL's workflow. (The software mechanisms used to implement these choices will be described in Section \ref{sec:architecture}.) The basic logical design principles are explained in Subsection \ref{DesignPrin}. This is followed by an explanation of how skill files and decision memory are employed within each step in Subsection \ref{skillz}. The design for how domain experts best work with CARL is described in Subsection \ref{HCAA}. Finally, in
Subsection \ref{GenCARL} we discuss how the overall design for CARL can also be applied to a variety of problems outside of reinforcement learning.

% The central design decision is to separate applied RL problem-solving into a human-guided specification phase followed by an agent-driven implementation phase. That is, CARL first helps the user translate domain intent into an explicit Markov Decision Process (MDP) and only then permits autonomous code generation, testing, and training. The following discussion therefore focuses on the sequencing of decisions, the role of skill files and summaries in preserving problem context, and the handoff from interactive formalization to autonomous execution. 

\subsection{Design Principles}
\label{DesignPrin}

CARL is organized around the observation that a successful RL solution requires three kinds of work: (1) understanding the problem and its domain, (2) formalizing the Markov decision process (MDP), and (3) implementing the resulting training system. General-purpose coding agents are increasingly capable at the third task, but CARL treats the first two tasks as logically prior. Its workflow therefore encodes the following design principles:

\begin{enumerate}[label=\arabic*.]
\item \textbf{Separate specification from implementation}. CARL does not begin by writing code. Starting with Steps 1 and 2, CARL works with the domain expert to understand the problem and its domain by eliciting the business or scientific objective, the controlled system, the episode structure, and the relevant constraints. 
\item \textbf{Enforce structure when developing MDP}. In Steps 3--7, the workflow forces separate and sequential decisions about the MDP's parameter ranges, state variables, action space, rewards, dynamics, and evaluation criteria so that implementation proceeds from an agreed formal description rather than an underspecified prompt.
\item \textbf{Constrain each conversation to one design question}. Each step is governed by a single skill (a markdown file to guide the LLM) and a checklist for that step. This prevents the agent from mixing domain clarification, MDP design, and coding into one unstructured exchange. (This is discussed further in the next subsection.)
\item \textbf{Assist the human expert with domain knowledge}. A separate persona called ``Trillian" observes the domain expert's interactions over time, learning from their behavior as well as external research sources to suggest answers to questions. The domain expert is free to accept, edit, or ignore these suggestions, but they are intended to accelerate convergence to a complete MDP specification.
\item \textbf{Carry forward decisions, not full transcripts}. At the end of each step, after the domain expert and CARL have converged on the work of the step, CARL records a compact decision summary. Later steps see these summaries as the accumulated specification, which keeps the design coherent while avoiding unnecessary context overload. (This is also discussed further in the next subsection.)
\item \textbf{Enable iterative reflection and revisiting}. Once a full flow is executed in Steps 8 and 9, training and evaluation results will be available. In Step 10, the domain expert can analyze these results to pinpoint issues that need to be addressed, and CARL can use ``revisit" logic to evaluate the artifacts, propose modifications, and generate revised code. If the domain expert, with CARL's help, decides on changes, CARL triggers a new iteration with these changes. This enables iterative improvements to the design and implementation until the performance gap is reduced to the point where the domain expert is satisfied with the results and desires no more iterations. 
\end{enumerate}

\subsection{Skill Files and Decision Memory}
\label{skillz}

Each workflow step is represented by one markdown skill file. The skill file defines the local objective of the step, the questions CARL should ask, and the checklist that must be satisfied before moving forward. Only the current skill file is active at a given point in the workflow. This is a logical control mechanism: it narrows CARL's attention to the current design decision and discourages premature movement into later tasks, especially implementation. Table~\ref{tab:skill-excerpt} gives a compact example regarding Step 6, the reward-definition skill.

\begin{table}[ht]
\centering
\small
\begin{tabular}{@{}p{0.23\textwidth}p{0.68\textwidth}@{}}
\toprule
\textbf{Skill element} & \textbf{Excerpt from Step~6: Reward Definition}\\ \midrule
Description & Design the reward with normalization and domain randomization; ask for hints or nudges for early training.\\
Design hints & Keep the reward a single readable expression; drop fixed constants; expose reward components in \texttt{info}; document discounting.\\
Reward checks & Complete reward-configuration and credit-coverage tables; ensure terminal/failure terms reach responsible agents; label unmatched oracle formulas as heuristic.\\
Randomization checks & Randomize policy-relevant parameters at reset; include randomized parameters in the observation; disable randomization during evaluation.\\ \bottomrule
\end{tabular}
\caption{Condensed excerpt from the Step~6 skill, with a local objective, design hints, and checklist items.}
\label{tab:skill-excerpt}
\end{table}

\begin{figure}[ht]
    \centering
    \includegraphics[width=0.95\linewidth]{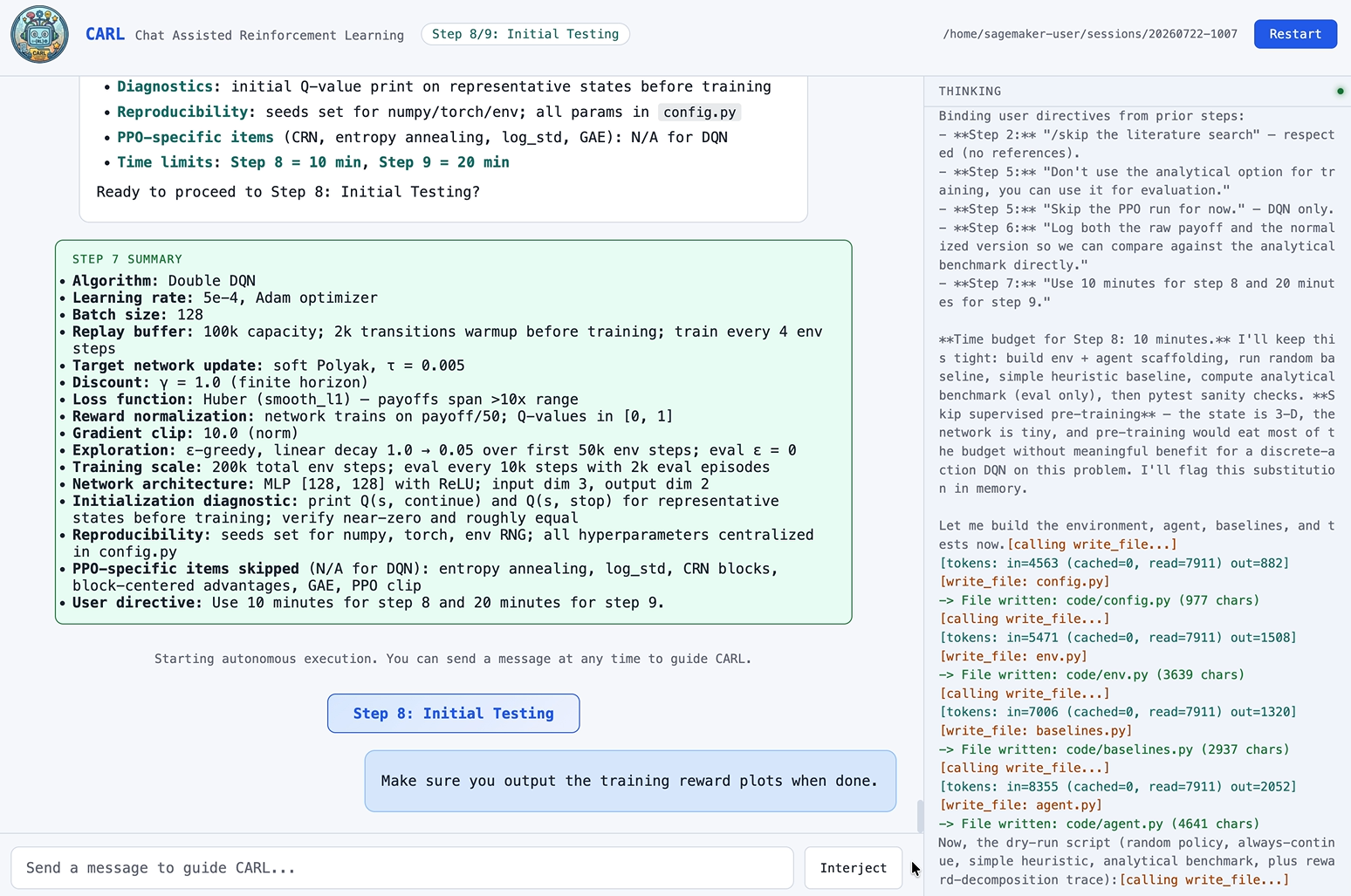}
    \caption{Sample decision summary (in green box) which is added to \texttt{memory.md}. This is the summary after Step 7 for the example problem from Subsection \ref{sec:carldemo} (time stamp \href{https://youtu.be/Qx673uczp4c?t=850}{\texttt{14:10}}). To improve context management, only the set of compact summaries from previous steps are fed to future steps. An interjection from the user is also visible at the bottom, with an instruction on the form of reporting at the end. In autonomous mode, such instructions are used for steering CARL mid-flow.}
    \label{fig:carl-memory}
\end{figure}

At the completion of each step, CARL produces a concise decision summary. These summaries form the design memory for the session. Subsequent steps receive the accumulated summaries rather than the full conversation, which makes the state of the design explicit and auditable. The summaries also define the contract between the interactive phase (Steps 1--7) and the autonomous phase (Steps 8 and 9): The autonomous phase is expected to implement the decisions already recorded, not reinterpret the original domain expert intent from scratch. As an example, Figure \ref{fig:carl-memory} shows the summary produced at the end of Step 7 for the highest drawn number example problem from Subsection \ref{sec:carldemo}.

%Table~\ref{tab:memory-excerpt} reproduces one such summary from an actual CARL run.

\subsection{Human Control and Agent Autonomy}
\label{HCAA}

\begin{figure}[ht]
    \centering
    \includegraphics[width=0.95\linewidth]{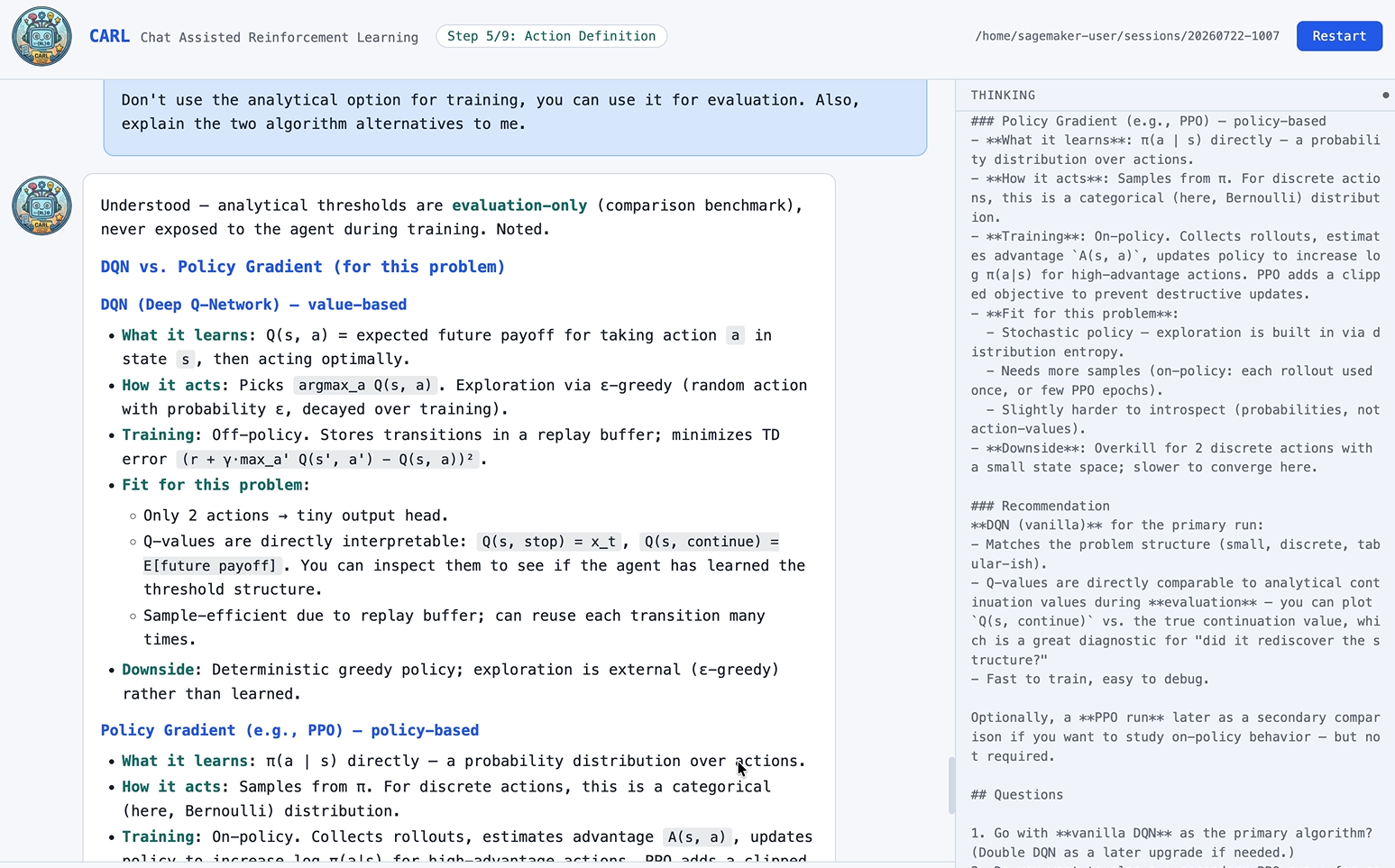}
    \caption{A sample of interactive decision-making between the user and CARL from Step 5 of the example problem in Subsection \ref{sec:carldemo} (timestamp \href{https://youtu.be/Qx673uczp4c?t=510}{\texttt{08:30}}). In this picture, CARL has previously proposed an action definition and two possible RL algorithms for training. The user reinforces the instruction not to use the known analytical solution in the training, and asks for the pros and cons of the two algorithms (DQN and policy gradient). The user also has the option (if desired) to ask for an explanation in more or less technical terms, depending on their familiarity with RL.}
    \label{fig:carl-interactive}
\end{figure}

CARL assigns authority according to the kind of decision being made. During Steps~1--7, the domain expert retains authority over domain meaning: what the objective is, which constraints matter, which variables are observable, which actions are admissible, and what constitutes reward or failure. An example of this interaction is shown in Figure \ref{fig:carl-interactive}, which is taken from Step 5 for the highest drawn number example problem in Subsection \ref{sec:carldemo}. CARL may propose formulations, defaults, or refinements, but these are presented to the domain expert for approval. Step~7 is intentionally a one-turn confirmation step, because most domain experts can approve or adjust training choices more efficiently than they can negotiate every low-level learning parameter within RL interactively. That is, the domain expert is not expected to be an RL expert as well.

After Step~7, CARL changes mode. The domain expert-approved specification becomes the governing artifact, and CARL is allowed to carry out the implementation task autonomously. This boundary reflects the asymmetry between specification errors and implementation errors. A flawed implementation of a correct MDP can often be debugged incrementally; a correct implementation of the wrong MDP optimizes the wrong problem.

\subsection{Generalizing the CARL Approach to More Than Reinforcement Learning}
\label{GenCARL}

Although CARL is presented here for reinforcement learning, its logical structure is more general. Many accelerator-style applications require a user and an AI agent to move from an ambiguous objective to a reusable artifact, such as a code repository, analysis pipeline, report, simulator, or prototype product. The same pattern can apply whenever the task benefits from step-wise conversations, bespoke skill files for each stage, explicit decision memory, and a clear boundary between interactive specification and autonomous execution.

In this broader view, CARL is an instance of a reusable design template: decompose the problem into ordered conversational steps; decide which steps require human judgment and which can be delegated; encode each step in a focused skill file; preserve decisions as structured memory; and let the agent produce a durable artifact only after the specification is sufficiently complete. RL makes this need especially visible because an incorrect MDP invalidates the resulting policy, but the same principle applies to other domains where code generation alone is not enough.

\section{Technical Design Choices Behind CARL}
\label{sec:architecture}

This section describes the software architecture that realizes the logical workflow in Section~\ref{sec:methodology}. The emphasis is on the orchestration choices that make CARL a controllable, model-agnostic system rather than a one-off prompt or a fixed agent application.

\subsection{Technology Stack}

CARL is implemented as a lightweight Python orchestration layer around a step-wise conversational workflow. The user-facing interface is a Flask web application, while the same core flow is also available through a terminal client. Both interfaces share the same session model: user messages, model responses, tool calls, step summaries, generated code, logs, plots, and reports are persisted inside a new directory for each session/project. The web interface keeps the chat surface responsive by sending user messages through HTTP endpoints and polling separate event and log streams while longer model calls or training jobs run in background threads.

The backend uses ordinary Python modules rather than a monolithic agent platform. The main web controller manages routes, shared state, interjections, and background work; the conversation engine manages step progression, model calls, tool execution, and summaries; the session layer persists memory and artifacts; and specialized modules handle literature search, resume/revisit flows, and report generation. This separation keeps the system simple enough to inspect while allowing the workflow to run for many turns without losing the state of the RL design.

\subsection{Models Used} \label{sec:carl_models}

CARL assigns models by role. The default CARL model is Claude Opus~4.7, used for the main step-wise workflow because it must maintain long context, follow skill constraints, and execute multi-round tool loops. The system varies its thinking configuration by step: early clarification uses a medium setting, most specification and revisit steps use a high setting, and the autonomous implementation steps use the highest setting. If Opus~4.7 is unavailable at startup, CARL probes and falls back to Opus~4.6.

Smaller or more specialized calls are separated from the main workflow. Trillian uses Claude Haiku~4.5, since reply suggestions should be fast, inexpensive, and narrowly scoped. Report-writing and other editorial synthesis tasks use Claude Sonnet~4.6, which is well suited to converting session memory, plots, logs, and references into structured prose without requiring the full autonomous-coding budget.

The choice of model is up to the user. The model choices presented here are the ones used by the system at the time of writing, and are not the only ones that are applicable. As explained below, model changes can be made easily in the code base by switching the model(s) accessed via API.

\subsection{Model-Agnostic Conversational Orchestration}

CARL uses Amazon Web Services' Bedrock Converse API as a message and tool-calling interface, but the orchestration logic is deliberately model-agnostic. Model identifiers are resolved separately from the workflow, and the controller supplies ordinary ingredients to the model runtime: a system prompt, a message history, optional tool schemas, and inference options. The important architectural unit is therefore not a vendor-provided agent application, but a CARL-controlled loop that can be pointed at different compatible models.

At the beginning of each step, CARL loads exactly one skill file from a \texttt{carl\_skills/} repository, gathers prior decision summaries from \texttt{memory.md}, and builds a step-specific system prompt. The Converse wrapper converts CARL's internal message representation into the API format, attaches tool schemas when needed, streams the response, and returns usage information. Steps~8 and~9 add code-writing and command-running tools; Step~2.5 adds a literature-search tool; Step~10 uses a separate set of revisit tools. This design differs from using Claude, ChatGPT, or another off-the-shelf agentic application directly: CARL owns the step boundaries, context policy, tool set, memory format, and completion criteria.

The loop also includes reliability mechanisms needed for long sessions. System prompts and tool schemas can be cached at the Converse boundary; model-specific options such as extended thinking are selected by step and can fall back if unsupported; truncated responses are continued; and tool results are returned to the model in the same conversation. The user can inject guidance during autonomous execution, but the underlying controller still decides when a step is summarized, saved, resumed, or advanced.

\subsection{Skill Definitions}

The skill files are more than short prompts attached to step names. They encode procedural knowledge about how to turn a vague RL problem into an implementable MDP. For example, the problem-definition skill explicitly delays state/action/reward formalization until the domain objective, dynamics, constraints, and episode structure are understood. The information-proposal skill emphasizes Markov sufficiency, scale-invariant features, fixed clipping, and avoiding redundant state encodings. The reward-definition skill requires reward configuration, credit coverage, checks for shaping failure, and explicit handling of discounting and domain randomization.

\begin{figure}[ht]
    \centering
    \includegraphics[width=0.95\linewidth]{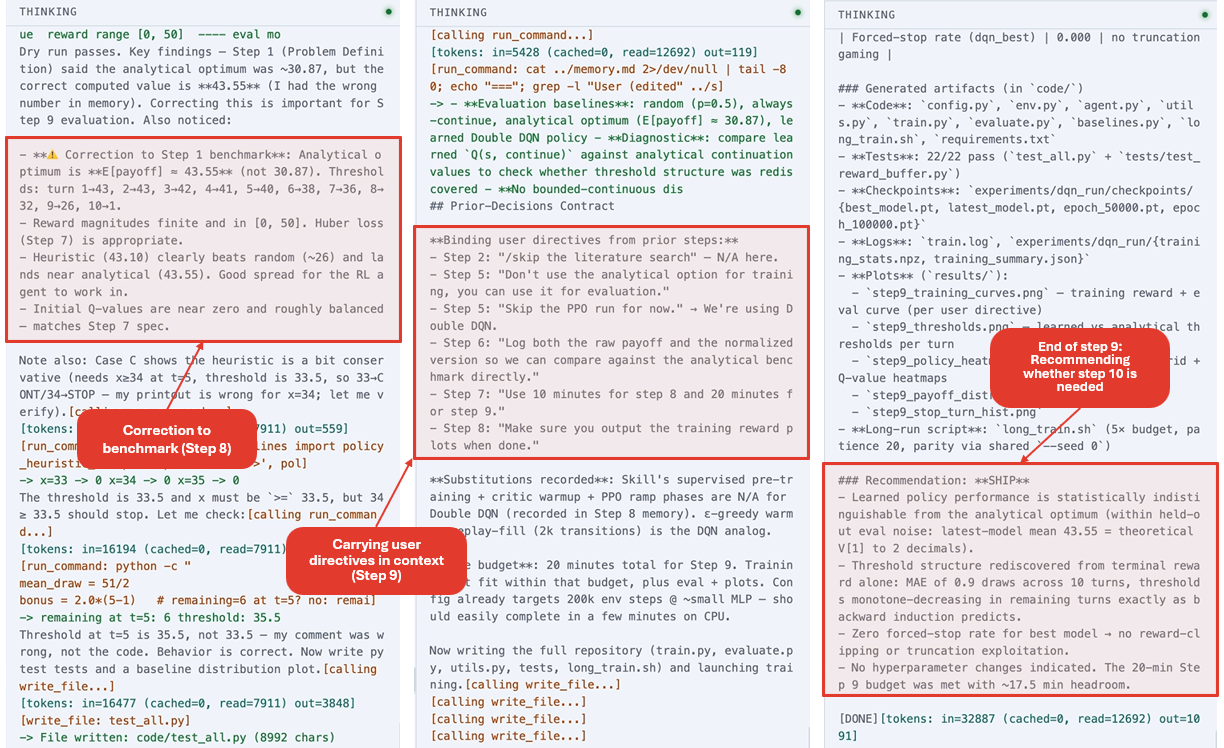}
    \caption{Screen captures of the thinking pane during autonomous execution at three different timestamps for the example problem in Subsection \ref{sec:carldemo}. On the left we can see CARL reasoning about and correcting the expected return according to the analytical dynamic programming approach, which will be used for checking how well training is progressing. In the middle is the launch of Step 9, where CARL ensures that key directives from all previous steps are respected. On the right is the end of Step 9, where CARL provides a recommendation to the user on whether the initial training is acceptable (model ready to ``ship") or whether Step 10 should be used to ``iterate" and improve the model further.}
    \label{fig:carl-thinking}
\end{figure}

Later skills become more operational but remain specification-driven. The ML/RL parameter skill asks CARL to choose hyperparameters from the problem structure rather than repeatedly querying the user. The initial-testing and launch skills instruct the agent to validate the generated environment, run dry checks, inspect reward routing, monitor training behavior, and compare against baselines. These files encode many small judgments that a general model might know individually but would not reliably apply in the right order without a staged workflow. Figure \ref{fig:carl-thinking} shows sample thinking traces as CARL progresses through the autonomous steps during the highest drawn number example in Subsection \ref{sec:carldemo}.

\subsection{Trillian Persona}

Trillian is implemented as a second model call governed by its own persona file. During the early interactive steps, CARL asks Trillian to propose a domain-expert-style user reply to CARL's latest question. The proposed reply is not binding: in the terminal it can prefill the input buffer (see Figure \ref{fig:carl-interface}), and in the web interface it is presented as a suggestion that the user may accept, edit, or ignore. CARL records whether the suggestion was accepted or edited, so these interactions become useful session evidence rather than invisible UI convenience.

The persona is intentionally scoped. Trillian is active through the specification-heavy steps and then deactivates before the autonomous implementation phase. Its role is to reduce the user's burden when answering domain questions, not to replace the user or to comment on generated code. After a run, CARL can optionally evaluate the Step~1--6 interactions against the current persona, apply a triage pass, make minimal edits when systematic evidence exists, and run a consistency check before saving an updated persona. This keeps the reply assistant adaptable without allowing uncontrolled drift.

\subsection{Resume and Revisit Modes} \label{sec:resume_revisit}

Long-running agentic systems may fail to reach completion, or complete with unsatisfactory results. To fix this, CARL includes ``resume'' and ``revisit'' steps.

CARL treats long-running sessions as recoverable stateful processes. Resume mode validates that a session directory contains both \texttt{memory.md} and \texttt{web.log}, parses completed step summaries, reads structured resume markers when available, and identifies the first unfinished step. The resumed step starts from the top with prior decisions restored, while partial code, results, and references remain available for inspection. This avoids pretending that an interrupted session is blank, while still giving the agent a clean step boundary from which to continue.

Revisit mode is a separate Step~10 workflow for completed sessions. Before artifact inspection, CARL asks the user for preflight guidance, including paths or topics to avoid. The revisit tools can list session artifacts, read text files, describe plots through a vision-capable model call, and prepare a fresh \texttt{code/revisit\_N/} workspace from a user-confirmed source directory (where $N$ takes values 1, 2, 3, ... for successive revisits). Step~10 refuses retraining commands during the interactive diagnosis phase; instead, it writes a revisit training script and associated appendix helper so the revised experiment can be run deliberately after the diagnosis is complete.

\begin{description}
    \item[\textbf{Remark.}] The skill for Step 8 (first autonomous step) provided to CARL mentions, \emph{``Propose a simple rule-based heuristic or analytical solution that computes the actions using the state variables. Use this heuristic for a short supervised pretraining phase wherever possible. The heuristic need not be optimal — just clearly better than random. The RL agent refines the policy afterward. Do NOT spend time computing exact or brute-force solutions for pretraining targets.''} This results in CARL typically writing a sensible rule-based heuristic for (i) warming up its policy output, and (ii) for tracking the quality of the trained policy, especially in order to write the current best-performing checkpoints during training. In Step 10 (the revisit step), the skill specifically asks CARL to review existing evidence from previous runs, including logs, checkpoints, evaluations, plots, and simulations. Therefore, CARL is frequently able to infer clues towards MDP refinement by observing its own behavior in contrast to the heuristic. We will see this pattern repeated when CARL is applied to the two sample problems in Sections \ref{sec:example_gbwm} and \ref{sec:example_goddard}.
\end{description}

\subsection{Literature Search and Grounding}

CARL uses Semantic Scholar\footnote{\url{https://www.semanticscholar.org/}} for literature search in two places:  the \texttt{/suggest} brainstorming path in Step~1 (discussed in Subsection \ref{sec:aux}) and the optional Step~2.5 literature-search step. Search results are filtered before use: papers must have an abstract and an open-access PDF URL, duplicate IDs are removed, and the number of pages queried is capped. Search outputs are persisted as both structured JSON and a Markdown summary inside the session's \texttt{references/} directory, which makes the retrieved literature auditable and available to later report generation.

The literature path includes several guards against hallucinations. When a model client is available, CARL fetches the first pages of each paper PDF and asks the model for a short, fixed-length summary focused on RL algorithm, state/action/reward design, and main result; the abstract is used only as a fallback. The report generator later creates BibTeX entries only from the saved structured paper data, instructs generated sections to cite only provided keys, and emits a fixed fallback sentence when no literature data exists. Thus the model can synthesize from papers, but it is not invited to invent a bibliography.

\subsection{Report Generation}

Recall that optional report generation was described at the end of Subsection \ref{sec:aux}. The reporting layer is downstream of the CARL run. It gathers the session memory, literature files, selected generated code snippets, training logs, plots, and step conversations. Plots are described through multimodal model calls before section generation, and benchmark-related context is extracted so figure descriptions can name the actual analytical relation being plotted rather than referring to a generic benchmark.

Report sections are generated as separate LaTeX files, often in parallel, and then assembled through a template-driven \texttt{main.tex}. If literature data exists, the report generator creates \texttt{references.bib} and passes the allowed citation keys into the section prompts. The final output is compiled with repeated \texttt{pdflatex}/\texttt{bibtex} passes when the toolchain is available; otherwise, the generated \texttt{.tex} files remain in the session's results directory. In architectural terms, reporting is an artifact synthesis stage: it documents what the workflow produced rather than steering the RL specification itself.

% Examples ---------------------------------------------------------
\section{Example 1: Goals Based Wealth Management (GBWM)}
\label{sec:example_gbwm}

We now illustrate the CARL workflow on a goals-based wealth management (GBWM) problem from finance. This example is intended to specifically demonstrate the failure mode motivating CARL: a reasonably well-posed problem in the finance domain that still leaves important MDP choices unresolved. Implementing the full solution requires decisions about market dynamics, goal timing, affordability, action parameterization, reward timing, and evaluation protocol. CARL's role is to make these choices explicit before autonomous implementation begins. 

For this GBWM problem, we are looking to maximize the expected value of the sum of the utilities that correspond to attained financial goals. For example, an investor may be looking to buy a motorcycle 5 years from now that will cost \$15,000 and also hope to buy a new car 10 years from now that will cost \$35,000. Because the motorcycle is less important to the investor, it must be assigned a utility, say $U_1=100$, that is less than the utility assigned to the car, say $U_2=150$. The hope is to attain both goals for a total utility of 250, but buying the motorcycle may mean there isn't enough money to buy the car, so it may be best not to buy the motorcycle to increase the chance of buying the car. 

To maximize the expected utility sum from the attained goals, which is the reward here, we must optimize two choices (actions): 1) We must optimize the choice of whether or not to attain each affordable goal available in the current year, and 2) We must optimize our investment portfolio each year, which can contain any Markovian\footnote{The term ``Markovian" means that returns depend only on current, not previous, observations.} investments, such as tax-free stock and bond positions. 

If the utility values are fixed, as is the case above, then this problem (for any number of goals and Markovian investments) can be exactly solved using dynamic programming (\cite{das_dynamic_2022}). Further, RL is excellent at closely reproducing these optimal actions known from dynamic programming, attaining an average of 97.8\% of the optimum expected utility over a wide range of goal scenarios (\cite{das_meta_2026}).

But utility values can conditionally depend on which goals were previously attained. For example, if you purchase the motorcycle, you may still want the car, but you won't want it as badly since you still have a mode of transportation. In other words, $U_2$, the utility for the car, may be 150 if the motorcycle is not purchased, but, say, 105 if it is. Because dynamic programming is computed backwards in time, it cannot address a problem with this forwards-in-time chronological dependence. But RL is computed forwards in time, making it ideal for conditional utility problems like these, and ideal for using this as an example problem for CARL.

\subsection{Our Example GBWM Problem with Conditional Utility Bonuses}
\label{sec:gbwm-demo}

An investor begins with a wealth of $W_0=100$ and chooses annually to allocate the wealth in their portfolio between two investments: an equity asset and a bond asset. The equity asset is represented by the Vanguard Total Stock Market ETF (VTI), whose evolution is modeled using geometric Brownian motion, and the bond asset is the Vanguard Total Bond Market ETF (BND), whose evolution is modeled using the Vasicek model. Each year the investor chooses a stock weight $\pi_t\in[0,1]$, so the bond weight is $1-\pi_t$. If a goal becomes available in a year, the investor must also make the binary decision of whether to take the goal, paying its cost and receiving the corresponding utility, or to skip it and preserve wealth. The problem extends over a finite horizon of $T$ years, the time when the final goal becomes available. 

The problem dynamics are formally defined as follows. Let $W_t$ denote investor
wealth, $\pi_t$ the stock weight, $S_t$ the VTI (equity/stock) price, and
$B_t$ the BND (bond) price. We use $M^{(S)}_t$ and $M^{(B)}_t$
for the Brownian motions of the stock and the bond respectively. The market
parameters are fitted from daily adjusted closing prices over 4,630 trading days,
from January~2, 2008 through May~29, 2026. 

For VTI, daily log returns are
used to determine estimates for the annual drift ($\mu_S=0.128$) and volatility ($\sigma_S=0.201$). With these fitted parameters, VTI follows the geometric Brownian motion model:
\[
  \mathrm{d}S_t = \mu_S S_t\,\mathrm{d}t + \sigma_S S_t\,\mathrm{d}M^{(S)}_t.
\]
Solving this stochastic differential equation over one year yields
\[
  S_{t+1}=S_t\exp\!\left((\mu_S-\tfrac12\sigma_S^2)+\sigma_S Z^{(S)}_{t+1}\right)
\]
where $Z^{(S)}_{t+1}\sim\mathcal{N}(0,1)$.
% $S_0=373$.

For BND, the Vasicek
parameters are obtained by regressing daily price changes on the lagged BND
price; the fitted slope gives mean reversion strength $(\kappa_B=5.20\times10^{-4})$, the intercept determines the
long-run mean $(\theta_B=73.3)$, and the residual standard deviation gives the diffusion scale $(\sigma_B=0.200)$. With these fitted parameters, BND prices follow the
Vasicek model:
\[
  \mathrm{d}B_t=\kappa_B(\theta_B-B_t)\,\mathrm{d}t+\sigma_B\,\mathrm{d}M^{(B)}_t,
\]
Solving this stochastic differential equation over the $n_h=252$ trading days that comprise a year yields:
$$
  B_{t+1}=  B_t e^{-\kappa_B n_h}+\theta_B(1-e^{-\kappa_B n_h}) +\sigma_B\sqrt{\frac{1-e^{-2\kappa_B n_h}}{2\kappa_B}}\,Z^{(B)}_{t+1}$$
where $Z^{(B)}_{t+1}\sim\mathcal{N}(0,1)$.
% where $\kappa_B=5.20\times10^{-4}$, $\theta_B=73.3$, $\sigma_B=0.200$, and
% $B_0=73.2$. 

The stock and bond Brownian motions have correlation $\rho$, represented in discrete time through
\[
  Z^{(B)}_{t+1}=\rho Z^{(S)}_{t+1}+\sqrt{1-\rho^2}Z^{(I)}_{t+1}
\]
where $\rho=0.0204$ and $Z^{(I)}_{t+1}\sim\mathcal{N}(0,1)$ is independent of $Z^{(S)}_{t+1}$. Given stock weight $\pi_t$, the gross portfolio return between times $t$ and $t+1$ is
\[
  R_{t+1}(\pi_t)=\pi_t\frac{S_{t+1}}{S_t}+(1-\pi_t)\frac{B_{t+1}}{B_t}.
\]
If goal $i$ with cost $C_i$ is available at time $t$, and we define the binary attainment decision
$$  a_i=
\left\{
\begin{array}{ll}
0  &\mbox{if goal $i$ is not taken} \\ 
1    &\mbox{if goal $i$ is taken}
\end{array}
\right.  $$
then wealth evolves as
\[
  W_{t+1}=\bigl(W_t-a_iC_i\bigr)R_{t+1}(\pi_t).
\]
The goal at the terminal time $T$ is taken whenever it is affordable since there is a gain in utility if this goal is taken and no other utilities that can be gained later.

The distinctive feature of our problem is that goal utilities are conditioned on the number of prior goals taken. If goal $i$ has base utility $U_i$ and $k_i$ earlier goals have
already been taken, the realized utility when goal $i$ is taken is
\[
  \widetilde U_i = U_i(1+0.25k_i).
\]
The objective is therefore to maximize expected cumulative realized utility,
\[
  \mathbb{E}\left[\sum_i a_i \widetilde U_i\right].
\]
% For an intuitive analogy for this utility model, imagine a warehouse business that aims to acquire properties in a single cluster. It might first buy one property, then acquire additional contiguous properties over time, eventually assembling a larger and more efficient logistics or storage operation. Earlier purchases are not strict prerequisites for later ones, but owning more contiguous properties creates economies of scale through shared infrastructure, transport, staffing, and operations. That makes subsequent purchases more valuable. The technical model above captures this idea by allowing prior goal attainment to raise the realized utility of later goals.

This creates a useful demonstration problem because early low-utility goals can still be valuable: taking them raises the payoff of later goals. A method that only sees immediate base utility may under-invest in early goal attainment, while a method that over-prioritizes early goals may deplete wealth and fail later. This also makes the problem path-dependent, ruling out standard dynamic programming solutions. It thus showcases a class of problems that RL is well-suited for. 

\subsection{Evaluation Scenarios and Algorithms}
\label{sec:gbwm-methods}

Table~\ref{tab:gbwm-examples} lists the seven evaluation scenarios (not known during training) used in this section. The first four are one-, two-, three-, and four-goal cases where the base utilities of the goals increase over time. The final three reverse the utility ordering for the multi-goal cases. This reversal is useful because it changes the economic meaning of the same goal calendar: in the increasing four-goal case, early goals mainly act as bonus multipliers for the large terminal payoff, whereas in the decreasing four-goal case the most valuable goals arrives first.

\begin{table}[ht]
\begin{center}
\small
\begin{tabular}{@{}lllc@{}}
\toprule
\textbf{Scenario} & \textbf{Goal years} & \textbf{Goal costs} & \textbf{Base goal utilities}\\
\midrule
\texttt{incr\_1goal} & $\{16\}$ & $\{300\}$ & $\{8\}$\\
\texttt{incr\_2goal} & $\{8,16\}$ & $\{150,150\}$ & $\{3,5\}$\\
\texttt{incr\_3goal} & $\{5,10,16\}$ & $\{125,125,125\}$ & $\{1,2,5\}$\\
\texttt{incr\_4goal} & $\{4,8,12,16\}$ & $\{100,100,100,100\}$ & $\{1,1,2,4\}$\\
\texttt{decr\_2goal} & $\{8,16\}$ & $\{150,150\}$ & $\{5,3\}$\\
\texttt{decr\_3goal} & $\{5,10,16\}$ & $\{125,125,125\}$ & $\{5,2,1\}$\\
\texttt{decr\_4goal} & $\{4,8,12,16\}$ & $\{100,100,100,100\}$ & $\{4,2,1,1\}$\\
\bottomrule
\end{tabular}
\end{center}
\caption{Seven benchmark evaluation scenarios used to test and compare the learned policies. The scenarios are partitioned into ``increasing configurations" where base goal values/utilities increase over time and ``decreasing configurations" where base goal values/utilities decrease over time. This range of scenarios evaluates policy robustness across contrasting economic incentives. All scenarios operate over a finite-horizon investment period ($T=16$ years) with initial wealth $W_0 = 100$. If goal $i$ with base utility $U_i$ is taken when $k_i$ prior goals have previously been taken, the realized utility earned is $\widetilde{U}_i = U_i (1 + 0.25 k_i)$. Because realized utility explicitly depends on prior goal-attainment history $k_i$, the problem is path-dependent. }
\label{tab:gbwm-examples}
\end{table}

For each evaluation scenario, we will apply six methods, formed by varying  (i) two levels for the depth of {\it specification} of the problem, and (ii) three levels for the knowledge and skills the method has access to. 

The two specification levels are full specification (\textbf{FullSpec}) and partial specification (\textbf{PartSpec}). FullSpec gives the AI algorithm (code builder) both the finance problem and the MDP-level implementation choices needed to turn it directly into a training environment. PartSpec gives a coherent domain-level brief but leaves several MDP choices open, so the builder must either infer those choices or, in CARL's case, ask the user to resolve them before implementation. Appendix~\ref{app:specs} gives the complete FullSpec and PartSpec descriptions used in the experiments. Note that PartSpec is still a well-posed problem description: it is not a vague or open-ended \textit{problem}, but a precise domain-level brief that leaves important \textit{MDP choices} open.

The three skill-access levels are:
\begin{description}
  \item[\textbf{CARL-FullSpec} and \textbf{CARL-PartSpec}] use CARL's staged workflow skills and interactive clarification loop. Importantly for the PartSpec condition, CARL can ask the user to resolve underspecified MDP choices before coding.
  \item[\textbf{CC-Custom-Skill-FullSpec} and \textbf{CC-Custom-Skill-PartSpec}] use Anthropic's Claude Code as the agentic coding engine, with the entire agentic coding harness that goes with it. (CC stands for Claude Code.) It is given a custom skill library equivalent to CARL's, except for the portions where CARL is prompted to converse with the user; the system can use those skills while building, but it cannot ask follow-up questions. It is also not forced to follow the 9-step workflow that CARL uses, although it is part of the orchestrating skill. We evaluate this regime as \textbf{CC-Custom-Skill-FullSpec}, which receives the complete problem specification, and \textbf{CC-Custom-Skill-PartSpec}, which receives the partial domain brief and must commit to the missing MDP choices non-interactively.
  \item[\textbf{CC-Only-FullSpec} and \textbf{CC-Only-PartSpec}] remove the bespoke skill structure, testing how far the same broad task can be carried by a direct agentic coding prompt to Claude Code without explicit RL workflow knowledge.
\end{description}
The models used by CARL are already specified in Section \ref{sec:architecture}. CC-Custom-Skill and CC-Only both use Opus 4.7 as the LLM, with \texttt{xhigh} thinking level. A detailed comparison of the implementations for our six methods is available in Appendix \ref{app:comparison_gbwm}. 

\subsection{CARL-PartSpec Clarification Dialogue}
\label{sec:gbwm-carlpartspec-clarifications}

Before presenting the empirical results, we demonstrate the interactive clarifications enabled by CARL through specific examples drawn from the user-CARL conversation for CARL-PartSpec. Recall that this problem starts from the sparse brief provided to the AI agent. The stepwise dialogue converted underspecified domain language into concrete MDP choices through an initial run (Steps 1-9), followed by a revisit (Step 10).
\subsubsection{Initial run}
\begin{itemize}

\item \textbf{Goal feasibility and terminal timing.} CARL asked (Step 1) whether a goal-taking action should be allowed when the investor does not have enough wealth to cover the cost of the current goal (so the goal cannot be taken no matter how much the investor wants it), and whether the terminal date includes a stock versus bond portfolio allocation decision. The user clarified that a spending goal cannot be taken unless wealth covers its amount, and that the terminal date is a final goal-settlement date rather than another investment period. CARL made a note of this and translated these conventions into RL mechanics in Step 5: infeasible take actions (where a ``take action" means choosing whether or not to attain a current financial goal) are masked, and the terminal year does not contain a portfolio allocation decision.

\item \textbf{Annual transition construction.} CARL identified the mismatch between daily fitted VTI and BND dynamics and annual investment decisions. The user clarified in Step 1 the market-modeling convention: the fitted daily processes should be converted to annual decision-step dynamics corresponding to the equations in Subsection \ref{sec:gbwm-demo}, not simulated through 252 daily substeps (for the $\sim$252 trading days per year) inside each environment transition. CARL folded this into the MDP transition by using closed-form annual equivalents for the geometric Brownian motion and Vasicek processes, including the correlated stock and bond return construction.

\item \textbf{Conditional utility arithmetic.} CARL's interpretation exposed an ambiguity in the bonus rule: should prior goals compound multiplicatively or add to the current goal's original utility? The user clarified in Step 1 the economic meaning of the bonus: each prior goal taken adds 25\% of the \emph{original base} utility to the current goal's utility, so realized utility is $U_i(1+0.25k_i)$ rather than $U_i\times 1.25^{k_i}$. CARL converted that domain rule into the reward equation in Step 6 and the Markov state component in Step 4, tracking the number of prior goals taken.

\item \textbf{Training distribution.} The user supplied domain calibration in Step 3 for what counts as realistic-but-nontrivial goal problems for training: mostly increasing utilities over time, mostly spread-out goal times, a minimum value for the cost of any single goal as a proportion of the sum of the costs of all goals, and a minimum value for the base utility of any single goal as a proportion of the sum of the base utilities of all goals. These last two minimum values need to be high enough so that each goal is detectable by the gradients used in RL training. CARL used these inputs to define the RL domain-randomization scheme for varying the time horizon ($T$), the number of goals, and, for each goal, the year when the goal may be taken, its cost, and its base utility. 

\item \textbf{Observation design.} The user (domain expert) contributed the following finance-aware feature semantics in Step 4: the wealth and BND states should be scale-comparable, future liabilities should be measured in present-valued terms, and utility should be normalized within the episode. CARL turned these requirements into normalized observation vectors using $\log(W/W_0)$, $\log(B_t/\theta_B)$, $T_{\max}$-scaled time remaining until the time horizon is reached, present-valued funding ratios, and per-goal utility fractions.

\item \textbf{Two-agent decomposition and rewards.} CARL asked in Step 5 how to separate the two economic decisions in each year: 1) whether to take a current goal and 2) how to allocate the remaining wealth between the stock and bond positions. The user clarified their desired finance timing and payoff convention: the taking decision takes into account the utility of the current goal as well as future goals, while the allocation decision only affects future goals' feasibility. CARL proposed the corresponding RL structure: a binary goal-taking agent (0 for not taking the goal; 1 for taking the goal) and a continuous allocation agent (corresponding to the fraction of the portfolio in stock, as opposed to bonds), with the goal-taking agent credited for current plus future goals' utilities and the investment agent credited only for future goals' utilities.

\item \textbf{Action boundaries and masks.} The user specified all the following the portfolio constraints and feasibility logic in Step 1: long-only allocation between bond and stock positions, neither leverage nor shorting, no unaffordable goal taking, and greedy goal taking in the last year $t=T$ since it is  guaranteed to be optimal. CARL mapped these into RL actions in Step 5 by parameterizing the policy output as a beta distribution.\footnote{\url{https://en.wikipedia.org/wiki/Beta_distribution}} This means the RL agent only has to learn $\alpha$ and $\beta$, the two parameters of the beta distribution, during training. Actions are sampled from the distribution during training (ensuring exploration), though only the distribution mean is chosen for deterministic actions during evaluation. By choosing the beta distribution, the investment allocation agent action is automatically clipped so that $\pi_t\in[0,1]$, which ensures adherence to the user-specified constraints of ``no shorting and no leverage."

\item \textbf{Reward scale and shaping.} The user kept the objective at the level of the sum of realized goal utilities rather than wealth accumulation or a shaped proxy. In Step 6, CARL therefore set the RL problem to the undiscounted finite-horizon objective with discount $\gamma=1$ (meaning there is no discounting), no shaping rewards, no clips or floors, and no early-training hints, so training targets the same conditional-utility criterion used in evaluation.
\end{itemize}

\subsubsection{Post-training policy diagnosis} 

After the first training run, the user flagged a finance-specific anomaly with CARL-PartSpec, leading to a conversation that led to  improvement in the second run. We will discuss this in further detail in the context of the results, given in the next subsection.

As we will see in the empirical results that follow, the impact of these clarifications and choices is significant: they allow CARL to set up a solid MDP with a learnable reward optimizer rather than a technically valid but weaker MDP. We also see that designing an RL algorithm is non-trivial, as there is so much to specify, but CARL provides an interactive scaffolding that helps minimize the specification gap. 

\subsection{Results}
\label{sec:gbwm-results}

All six of our AI methods are applied to each of the seven scenarios from Table \ref{tab:gbwm-examples} using the same random number sequences for $Z^{(S)}_{t}$ and $Z^{(B)}_{t}$ from Subsection \ref{sec:gbwm-demo}, which ensures the same simulation sequence of stock and bond returns occurs in each of our 42 method-scenario pairs. Each method-scenario pair uses the average results from 1,000,000 Monte Carlo episodes. Because all six methods use identical market returns, utility differences are attributable to the method chosen rather than simulation noise.

\begin{figure}[ht]
\centering
\includegraphics[width=0.8\textwidth]{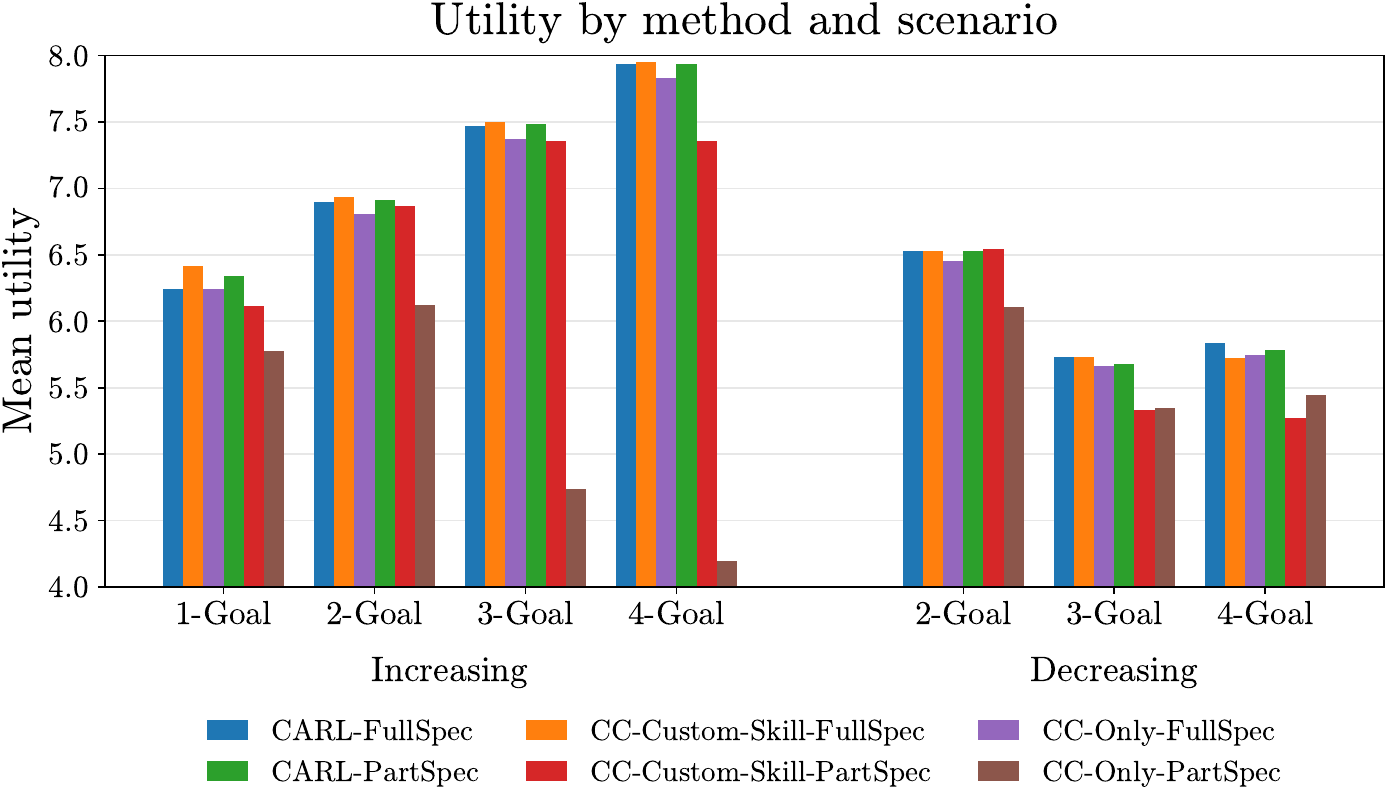}
\caption{Mean utility by method and scenario in the conditional goals-based
wealth-management example. Each bar is based on 1,000,000 random-number episodes that are identical for each method/scenario combination.}
\label{fig:gbwm-utility-bars}
\end{figure}

Figure~\ref{fig:gbwm-utility-bars} summarizes the main comparison results, which are for the mean utility sums from attained goals over 1,000,000 Monte-Carlo paths for each of the $6\times 7=42$ method/scenario combinations. The bars are grouped by scenario, with the increasing cases on the left and the decreasing cases on the right; within each scenario group, the six methods' results are shown side by side. Visually, CARL-FullSpec, CARL-PartSpec, and CC-Custom-Skill-FullSpec form a tight cluster of the top performing methods, while the partially specified non-CARL methods do comparatively worse as the number of goals increases. 

The fact that CARL performs strongly across the increasing and decreasing scenarios and that its PartSpec version tracks the FullSpec version closely are key observations. They demonstrate the central empirical point of this example's results: when the initial brief leaves important details open, CARL's interactive specification phase largely closes the gap before implementation. In contrast, the non-interactive partial-specification systems are more sensitive to missing modeling choices. The weakest case is CC-Only-PartSpec, whose mean utility falls especially sharply in the increasing three- and four-goal scenarios. The results suggest something that should not be surprising: problems from imperfect specification are compounded when fewer skills are provided. As a corollary, we observe that while frontier AI models see a large number of RL repositories as part of their training corpus, it still helps significantly to have specific guidance on building RL solutions for non-trivial problems. Our results here provide quantitative insight into this qualitative notion, highlighting the power and importance of the human-AI interaction that CARL enables.

The four-goal scenario cases are the most informative stress tests for comparing the three PartSpec methods. In the increasing four-goal case, CARL-PartSpec obtains a mean utility $7.94$, compared with $7.36$ for CC-Custom-Skill-PartSpec and $4.19$ for CC-Only-PartSpec. In the decreasing four-goal case, CARL-PartSpec obtains $5.78$, compared with $5.27$ for CC-Custom-Skill-PartSpec and $5.45$ for CC-Only-PartSpec.

\begin{figure}[ht]
\centering
\includegraphics[width=0.75\textwidth]{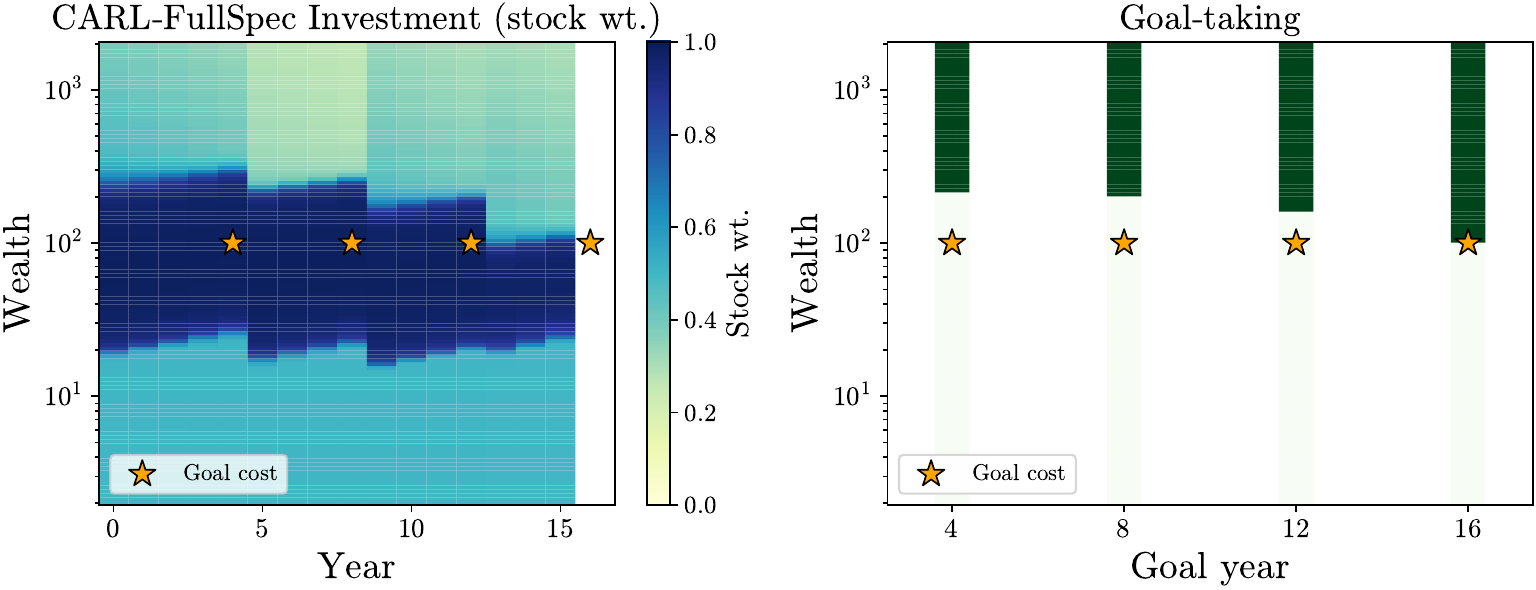}\\[4pt]
\includegraphics[width=0.75\textwidth]{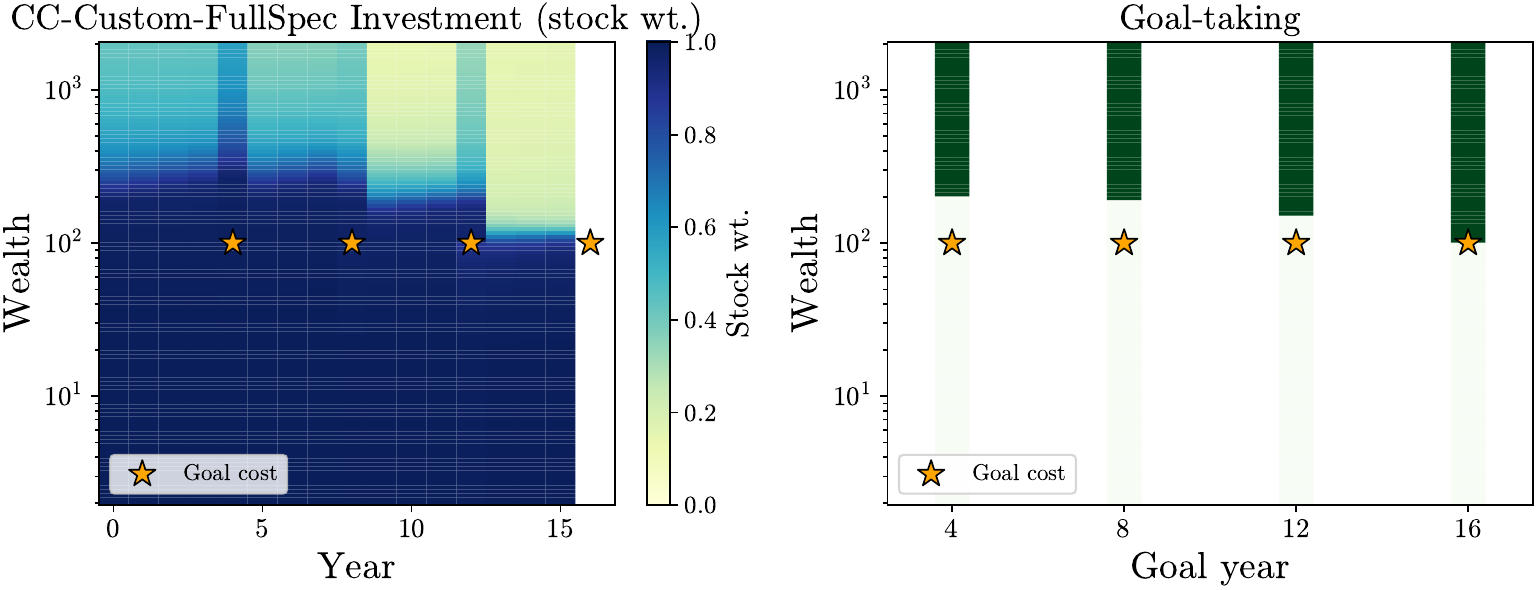}\\[4pt]
\includegraphics[width=0.75\textwidth]{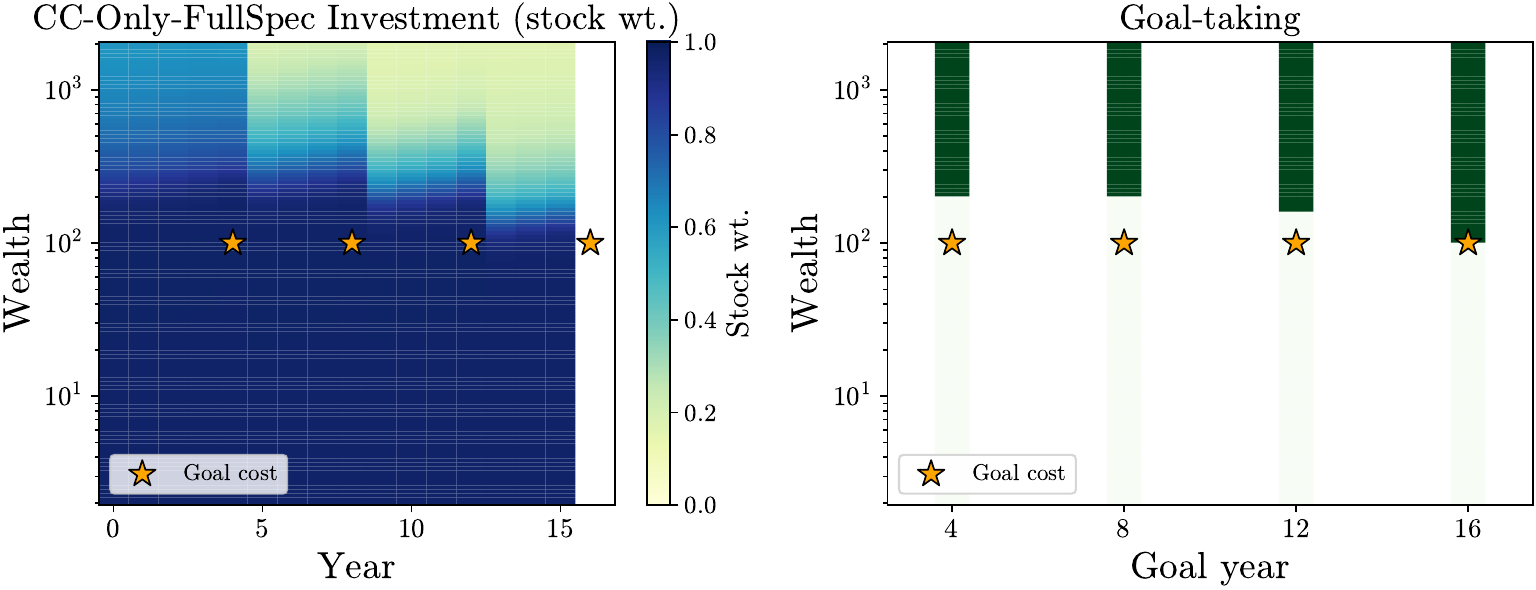}
\caption{Policies (i.e., actions) for the increasing four-goal scenario with the three fully specified methods. From top to bottom: CARL, CC-Custom-Skill, and CC-Only. The left panel in each row shows the investment policy; the right panel shows the goal-taking policy. The star shaped markers denote the goal costs --- which are each 100 --- in the four goal years.} 
\label{fig:gbwm-policy-fullspec}
\end{figure}

\begin{figure}[ht]
\centering
\includegraphics[width=0.75\textwidth]{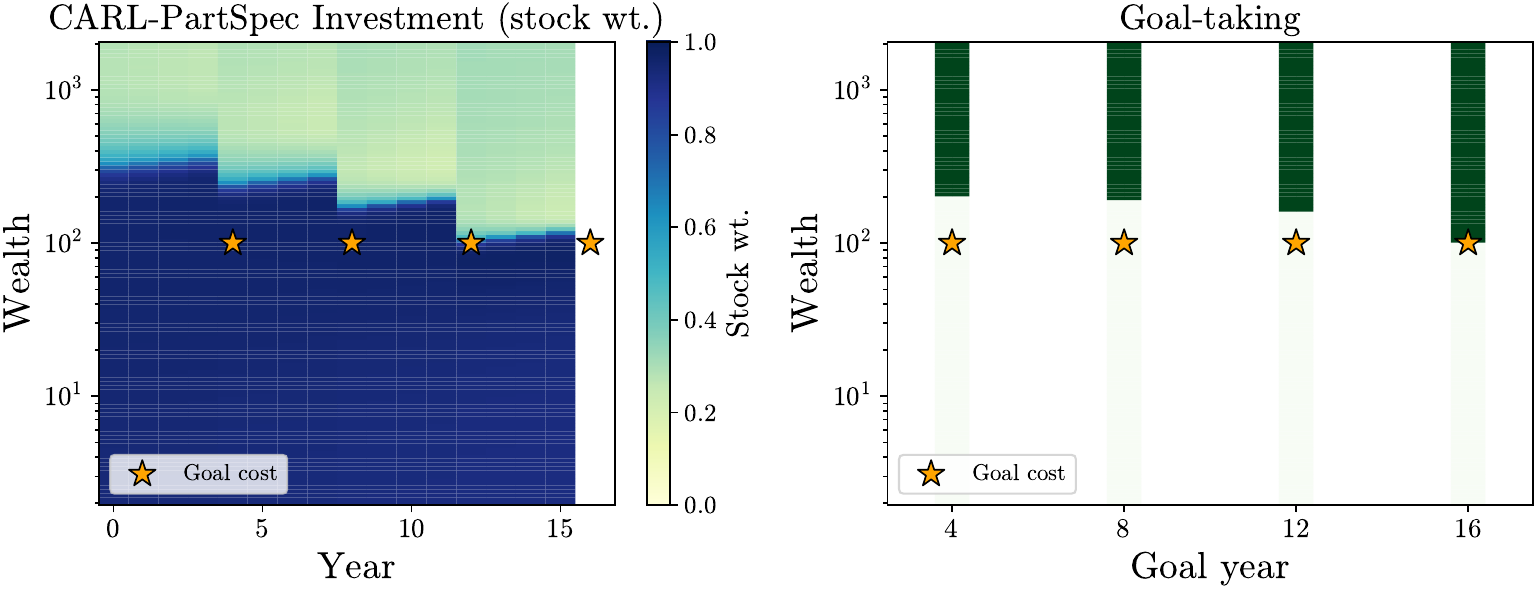}\\[4pt]
\includegraphics[width=0.75\textwidth]{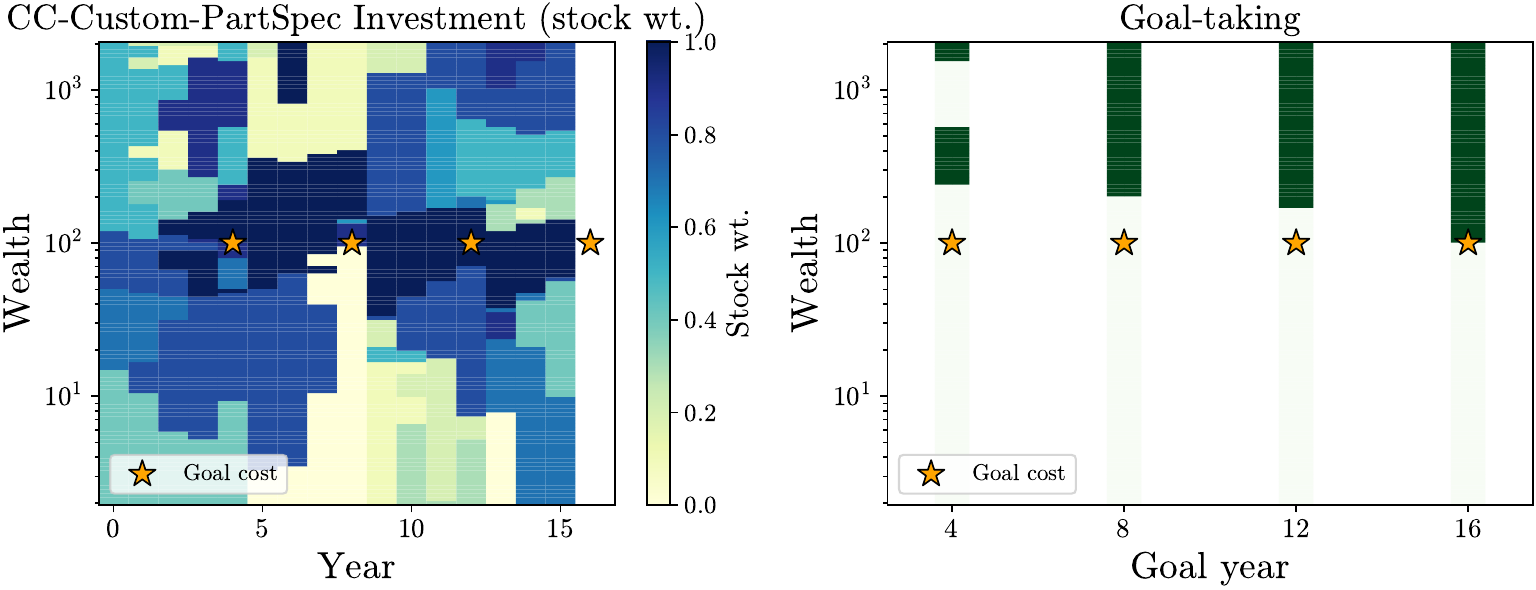}\\[4pt]
\includegraphics[width=0.75\textwidth]{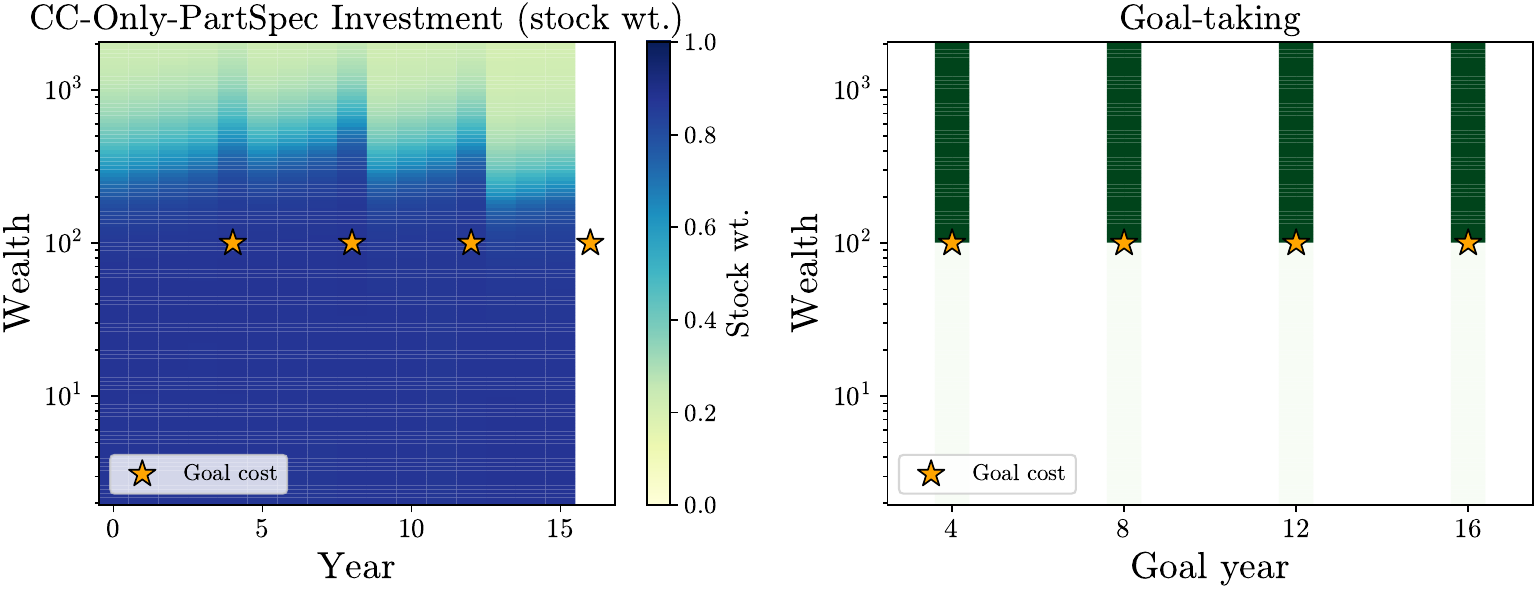}
\caption{Policies (i.e., actions) for the increasing four-goal scenario with the three partially specified methods. From top to bottom: CARL, CC-Custom-Skill, and CC-Only. The left panel in each row shows the investment policy; the right panel shows the goal-taking policy. The star shaped markers denote the goal costs --- which are each 100 --- in the four goal years.}
\label{fig:gbwm-policy-partspec}
\end{figure}

Figures~\ref{fig:gbwm-policy-fullspec} and \ref{fig:gbwm-policy-partspec} help explain the performance gaps in the four-goal scenario. We start with the investment choices, which are shown in the left panels of the figures. First we note that the differences between the six methods at high and low wealth values are immaterial to the outcomes because no matter what stock weight is chosen, goals are unlikely to be attained at low wealth and almost certain to be attained at high wealth. We also note that the goal-taking choices, shown in the right panels of the figures, are generally similar: The missing blip at year 4 for the CC\_Custom-PartSpec case is largely irrelevant since the starting wealth of $W_0=100$ is almost certain not to reach these missing wealth levels. The one exception is CC-Only-PartSpec, which selects the greedy approach of taking any goal it can afford, and also runs at a high stock weight at higher wealth values, resulting in the worst mean utility performance, as we clearly see in Figure \ref{fig:gbwm-utility-bars}.

% is relatively smooth, staying equity heavy at low wealth and becoming more bond heavy as wealth increases or as the remaining time horizon gets shorter; the goal-taking strategy (shown in the right panel) preserves the same ordering of attainment thresholds from the early low-utility goals to the terminal goal. Under FullSpec, CC-Custom-Skill and CC-Only remain fairly coherent, but both have significant differences from CARL at high wealth. {\color{red} Do they? for goal taking I see almost no difference. For stock weight, you see the heavy bond proportion *not* in CARL at high wealth. When the wealth is near the goal, we see the differences referred to earlier. This requires more explanation.} Under PartSpec, the divergence is sharper: while CARL is substantially the same as the FullSpec version, CC-Custom-Skill-PartSpec shows visible banding from its discretized 11-bin allocation action and a more conservative first-goal take region, whereas CC-Only-PartSpec applies nearly the same affordability threshold to all four goals. These qualitative differences reflect modeling choices about action representation, state representation, and episode conventions when the information given to the AI method leaves important MDP details unspecified.

In both full and partial specification settings, CARL's investment choices show a quick transition from stocks to bonds as wealth increases, in contrast to the other four methods which have either blurrier or incoherent transitions. From previous experience, quicker transitions are more typical of optimal behavior for these types of GBWM models. 

\begin{figure}[ht]
\centering
\includegraphics[width=0.75\textwidth]{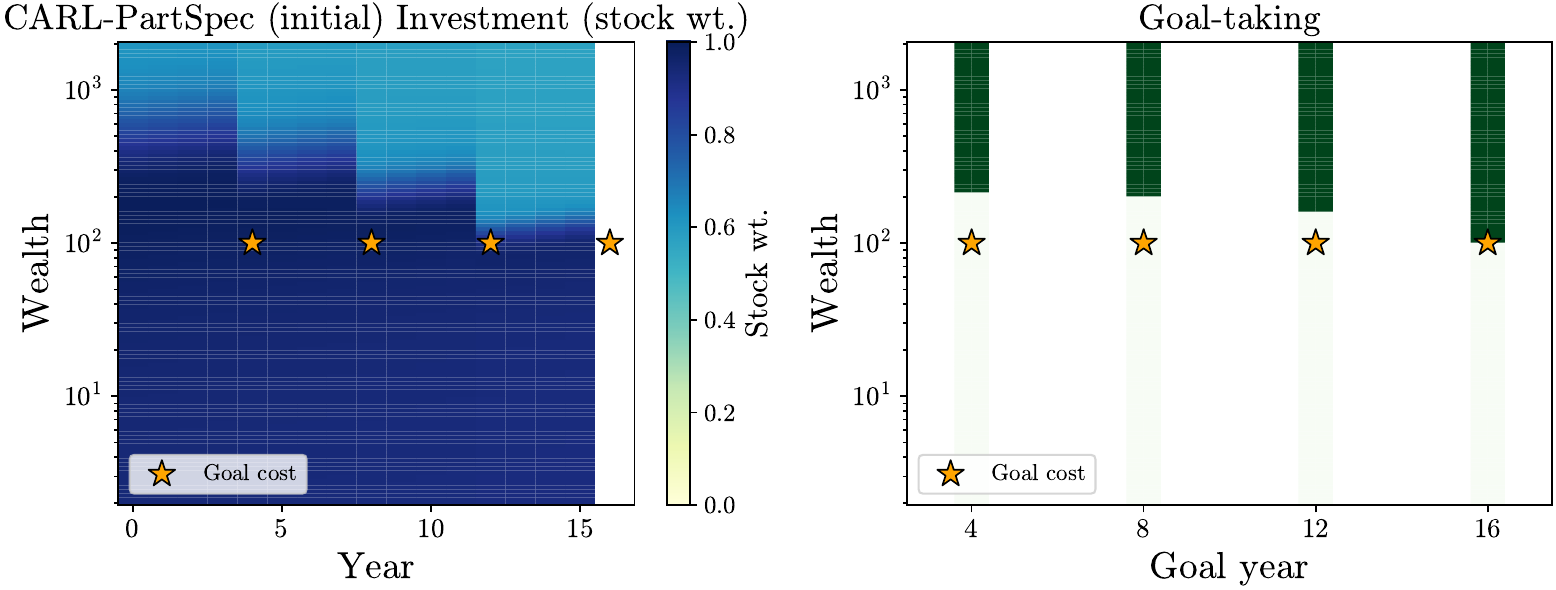}
\caption{Policies (i.e., actions) from the initial run (before the revisit) with the CARL-PartSpec method for the increasing four-goal scenario.}
\label{fig:gbwm-initial-carl-policy}
\end{figure}

Indeed, this was noticed in the CARL process. While the CARL-FullSpec case already had quick transitions, so there was no need to do more than the first 10-step run, the CARL-PartSpec case showed blurry transitions (see Figure \ref{fig:gbwm-initial-carl-policy}) that stayed somewhat heavy in stock even at high wealth after the first run. The domain expert flagged these concerns in the revisit phase (Step 10, as described in Subsection \ref{sec:resume_revisit}). CARL diagnosed this and confirmed our earlier point: staying somewhat heavy in stock at high wealth is a result of a shallow objective function since the goals will likely be taken no matter the stock fraction chosen. Therefore, the beta distribution that encodes the policy can remain near its prior mean with little ill-effect on the expected utility sum. CARL then proposed a lower beta floor, which allows more probability that the stock fraction will be closer to 0 (all bonds) or 1 (all stock). (This is discussed in more detail in Section \ref{sec:example_goddard}.) It also suggested an annealed terminal-wealth-variance shaping penalty that vanishes in the early part of training. This resulted in a stronger policy after the second run with the quicker transitions seen in Figure \ref{fig:gbwm-policy-partspec}. The second run also increased the average utility from 7.80 after the first run to 7.94. This demonstrates how the domain expert and CARL, working together, lead to better outcomes. 

Table~\ref{tab:gbwm-goal-utility} reports goal-taking rates and cumulative realized utility for the increasing and decreasing four-goal scenarios. The table makes the effect of conditional utility explicit. In the increasing case, CARL-FullSpec, CARL-PartSpec, and CC-Custom-Skill-FullSpec achieve nearly identical total utilities of $7.94$, $7.94$, and $7.95$, respectively, or about 65\% of the maximum attainable utility. CC-Custom-Skill-PartSpec reaches only $7.36$, while CC-Only-PartSpec reaches $4.19$; the latter takes the first goal in 86\% of episodes but then has lower take rates for later goals, indicating that early decisions are not followed by sustained goal attainment. The very low first-goal take rate for CC-Custom-Skill-PartSpec, $0.03$, also leaves it with less early utility and a smaller conditional bonus later in the episode. In the decreasing case, the CARL variants remain strong, reaching $5.84$ and $5.78$, while CC-Custom-Skill-FullSpec and CC-Only-FullSpec reach $5.73$ and $5.75$. The partial methods perform worse, with totals of $5.27$ for CC-Custom-Skill-PartSpec and $5.45$ for CC-Only-PartSpec. These results show how conditional utility amplifies the consequences of early goal-taking decisions and how CARL's specification process helps preserve performance when the initial problem description is incomplete.

\begin{table}[ht]
 \centering
 \footnotesize
 \setlength{\tabcolsep}{4.5pt}
 \begin{tabular}{@{}lccccccccc@{}}
 \toprule
  & \multicolumn{4}{c}{\textbf{Goal-take rate}} & \multicolumn{4}{c}{\textbf{Cumulative utility}} & \\
 \cmidrule(lr){2-5}\cmidrule(lr){6-9}
 \textbf{Method} & $t{=}4$ & $8$ & $12$ & $16$ & $t{=}4$ & $8$ & $12$ & $16$ & \textbf{\% of max}\\
 \midrule
 \multicolumn{10}{@{}l}{\textit{Increasing: base utilities $(1,1,2,4)$, $\widetilde U_{\max}=12.25$}}\\
 CARL-FullSpec            & 0.20 & 0.58 & 0.79 & 0.93 & 0.20 & 0.81 & 2.73 & 7.94 & 64.8\\
 CARL-PartSpec            & 0.21 & 0.55 & 0.78 & 0.94 & 0.21 & 0.80 & 2.68 & 7.94 & 64.8\\
 CC-Custom-Skill-FullSpec & 0.24 & 0.56 & 0.80 & 0.92 & 0.24 & 0.84 & 2.79 & 7.95 & 64.9\\
 CC-Custom-Skill-PartSpec & 0.03 & 0.53 & 0.77 & 0.93 & 0.03 & 0.57 & 2.35 & 7.36 & 60.1\\
 CC-Only-FullSpec         & 0.22 & 0.55 & 0.79 & 0.92 & 0.22 & 0.81 & 2.72 & 7.83 & 63.9\\
 CC-Only-PartSpec         & 0.86 & 0.44 & 0.35 & 0.32 & 0.86 & 1.39 & 2.34 & 4.19 & 34.2\\
 \addlinespace[2pt]
 \multicolumn{10}{@{}l}{\textit{decreasing: base utilities $(4,2,1,1)$, $\widetilde U_{\max}=9.75$}}\\
 CARL-FullSpec            & 0.86 & 0.51 & 0.42 & 0.40 & 3.44 & 4.66 & 5.22 & 5.84 & 59.9\\
 CARL-PartSpec            & 0.86 & 0.50 & 0.36 & 0.45 & 3.44 & 4.63 & 5.11 & 5.78 & 59.3\\
 CC-Custom-Skill-FullSpec & 0.79 & 0.49 & 0.42 & 0.58 & 3.16 & 4.31 & 4.87 & 5.73 & 58.7\\
 CC-Custom-Skill-PartSpec & 0.88 & 0.36 & 0.35 & 0.32 & 3.50 & 4.36 & 4.82 & 5.27 & 54.1\\
 CC-Only-FullSpec         & 0.86 & 0.50 & 0.26 & 0.51 & 3.44 & 4.64 & 5.00 & 5.75 & 59.0\\
 CC-Only-PartSpec         & 0.86 & 0.44 & 0.35 & 0.31 & 3.45 & 4.52 & 4.99 & 5.45 & 55.9\\
 \bottomrule
 \end{tabular}
 \caption{Goal-taking and utility accumulation in the four-goal scenarios, averaged over 1,000,000 episodes with identical stock and bond returns for each method. ``Goal-take rate'' is the fraction of episodes in which the goal maturing at year $t$ is taken. ``Cumulative  utility'' is mean realized utility $\widetilde U_i = U_i(1+0.25k_i)$ accumulated through year $t$, so the $t{=}T{=}16$ entry is the method's mean total utility. The final column  expresses that total as a percentage of the maximum attainable utility $\widetilde U_{\max}$ (all four goals taken).}
 \label{tab:gbwm-goal-utility}
 \end{table}

\subsection{Interpretation}
\label{sec:gbwm-interpretation}

The conditional utility GBWM example supports three conclusions. First, CARL's staged dialogue can effectively turn a domain-level description into an implementable MDP without requiring the user to write a complete mathematical specification upfront. Second, the largest relative gains appear when the initial specification is incomplete: the gap between CARL-PartSpec and the weaker partial-specification baselines is larger than the gap among the fully-specified systems. Third, the comparison illustrates why RL implementation quality and problem-specification quality are hard to separate. Poor performance
in this example is often not caused by the algorithm or hyperparameter selection, but by earlier choices that specify the MDP itself.

We believe that this example demonstrates the following principle: Rather than trusting the specification to be exhaustive up front or working hard to make it so, it is better to start with a reasonably good specification and then flesh it out further at the time of implementation, based on emerging observations. The conditional goals-based wealth management problem from this section is useful because it is small enough to inspect with policy surfaces, but rich enough for underspecification to matter. It also emphasizes that CARL is not a cosmetic front-end on top of existing agentic systems, but rather it is a critical component of the RL development pipeline.

% Examples ---------------------------------------------------------
\section{Example 2: Goddard Rocket Problem}
\label{sec:example_goddard}

Next, we study the classical Goddard rocket problem in which a single-stage rocket is launched upwards and only moves in the vertical direction. There is a maximum upward thrust, $T_{\max}$, available as long as there is fuel. (For simplicity, we will assume SI metric units throughout this section, so $T_{\max}$ is in Newtons, N.) $T_{\max}$ is assumed to remain constant given the range of reasonable altitudes reached by the rocket. A throttle command $u(t) \in [0,1]$ can be chosen making the thrust, $T(t)=u(t)T_{\max}$. The problem to be solved is to determine the throttle action, $u(t)$, that maximizes the peak altitude the rocket reaches at apogee. (That is, the peak altitude is the reward.) This optimal control problem was first posed by~\cite{goddard1919}. Its optimal solution in the unconstrained case --- meaning the rocket can withstand any structural challenges created by the pressure on its frame from the air --- is known; it has a three period \emph{bang-arc-bang} structure: initial full throttle ($u=1$) for the first period, followed in the second period by an immediate drop in $u$ so the throttle $u(t)$ can follow a known singular arc formula given later in Subsection \ref{sec:god_baselines} that depends on the drag, gravity, and other factors, followed by the final period after the fuel runs out ($u=0$) creating a free coast to the rocket's apogee~\citep{tsiotras1991, bryson1999}. The variant of this unconstrained case that we consider here adds an explicit \emph{dynamic-pressure constraint}, which is a throttling constraint due to the structural limit of what the rocket's frame can withstand. We will explain this constraint in more detail in Subsection \ref{sec:qmax}.

\subsection{Dynamics}
\label{sec:god_dynamics}

\begin{figure}
    \centering
    \includegraphics[width=0.5\linewidth]{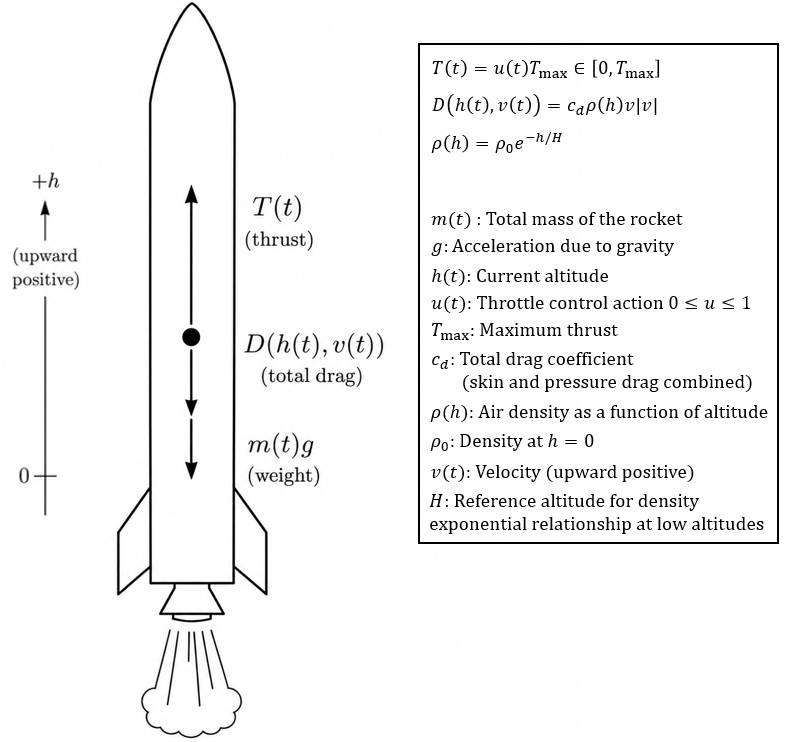}
    \caption{A free-body diagram showing the forces on the Goddard rocket. Thrust, counteracted by weight and drag, are the three forces that act on the rocket, with their imbalance leading to acceleration or deceleration.}
    \label{fig:goddard_fbd}
\end{figure}

We summarize the dynamics from \citet{goddard1919}:

The forces acting on the rocket (thrust, weight, and aerodynamic drag) are shown in Figure \ref{fig:goddard_fbd}. The formula for the aerodynamic drag force, $D$, is 
\begin{equation}
  D(h,v) = -\frac{1}{2}\,A\,c_D\, \rho(h)\, v\,|v| = -c_d\, \rho(h)\, v\,|v|
  \label{eq:drag}
\end{equation}
where 
\begin{itemize}
    \item $v$ is the velocity; its magnitude, $|v|$, is the speed. So $D$ moves in opposition to $v$ and the magnitude of $D$ depends quadratically on $|v|$.

    \item $A$ is the effective area of the rocket. 
    
    \item $c_D$ is the unitless coefficient of drag. For simplicity, we have defined the drag coefficient $c_d=\frac{1}{2}A c_D $ (which has area units, $\mathrm{m^2}$). The value of $c_d$ is constant for each scenario but can vary across scenarios. 

    \item $\rho(h)$, the atmospheric density, follows the exponential profile
    \begin{equation}
      \rho(h) = \rho_0 \, e^{-h/H}
      \label{eq:density}
    \end{equation}
    where the sea-level (i.e., $h=0$) density $\rho_0 = 1.225\ \mathrm{kg/m^3}$ and the reference altitude $H = 8500\ \mathrm{m}$.\footnote{$H$ acts as the scaling constant for the exponential decay of atmospheric density at low altitudes. Its value is empirically computed rather than derived from first principles.} 
    
\end{itemize}

These forces yield the following system of differential equations in continuous time for the dynamics of the tuple $(h,v,m)$ representing the rocket's altitude $h$, velocity $v$, and mass $m$:
\begin{align}
  \dot h &= v, \label{eq:hdot}\\
  \dot v &= \frac{T - D(h,v)}{m} - g, \label{eq:vdot}\\
  \dot m &= -\frac{T}{c_e}, \label{eq:mdot}
\end{align}
where the gravitational acceleration $g = 9.81 (\mbox{in } \mathrm{m/s^2}$), and it is assumed that the effective exhaust velocity $c_e = 0.01\, T_{\max} (\mbox{in } \mathrm{m/s})$ for our examples.\footnote{$c_e$ is roughly calibrated to fall in the exhaust velocity range given in \url{https://science.nasa.gov/learn/basics-of-space-flight/chapter3-2/}, which is also our source for  $T_\mathrm{max} =$ 200,000 N, which we use later.} Note that the mass $m$ combines the mass of the fuel, $m_{\textrm{fuel},t}$, which decreases as the fuel is depleted, and the mass of the rocket with no fuel, $m_{\mathrm{dry}}$, which, of course, stays constant. We will assume that $m_{\mathrm{dry}} = 1000\ \mathrm{kg}$ in our examples.

For each episode (i.e., simulated rocket run), we solve these three differential equations for the rocket's dynamics by using a 4th-order Runge--Kutta (RK4) numerical method with a time step of $\Delta t = 0.1$ seconds. The throttle control, $u$, is allowed to change at the beginning of each time step. Once fuel is exhausted, $T$ is forced to zero and the vehicle coasts under gravity and drag alone. When fuel runs out partway through a $\Delta t$ time step, the time step is split exactly at the fuel-out instant so the RK4 integration never overdraws the tank. 
Each episode terminates at its \emph{apogee}: the first time step at which the rocket switches from $v>0$ to $v\le 0$. The objective is to maximize the peak altitude attained over the episode.

\subsection{Dynamic-Pressure Constraint}
\label{sec:qmax}

Unlike the textbook Goddard problem, every scenario here imposes a hard limit, $Q_{\max}$, on the dynamic pressure, $q(t)$, which is the pressure on the rocket frame exerted by the air due to the rocket's movement through the atmosphere:
\begin{equation}
  q(t) = \tfrac{1}{2}\,\rho(h(t))\, |v(t)|^2 \;\le\; Q_{\max}.
  \label{eq:qmax}
\end{equation}
Note that the dynamic pressure on the rocket is proportional to the drag. To impose this $Q_{\max}$ limit, at the beginning of each $\Delta t$ step we first look to see if $T=u\cdot T_{\max}$ \emph{results} in a dynamic pressure at $t+\Delta t$ that still satisfies~\eqref{eq:qmax}. If it does, then $T=u\cdot T_{\max}$. If it does not, we set $T=T_{\lim}$, the maximum value of $T$ where the resulting dynamic pressure at $t+\Delta t$ still satisfies~\eqref{eq:qmax}. $T_{\lim}$ is found using binary search over $T \in [0, u\cdot T_{\max}]$, re-integrating the step for each trial $T$. Steps where $T=u\cdot T_{\max}$ use the nominal drag law~\eqref{eq:drag} to compute the drag $D$. Steps where $T=T_{\lim}$ (and therefore, $u$ is recomputed to be $T_{\lim}/T_{\max}$) incur a 5\% drag surcharge
($D \to 1.05\,D$) intended to model airframe vibration under constraint-limited flight. This surcharge mechanism is applied identically inside the shared evaluation environment, so every method is subject to exactly the same physical limitations.

% --------------------------------------------------------------------------
\subsection{Objective}
\label{sec:god_objective}

The control objective is to maximize the peak altitude achieved,
\begin{equation}
  \max_{u(\cdot)} \; h_{\mathrm{peak}} = \max_{t} h(t),
  \label{eq:objective}
\end{equation}
subject to~\eqref{eq:hdot}--\eqref{eq:qmax}, $u(t)\in[0,1]$, and the mass constraint $m(t)\ge m_{\mathrm{dry}}$ based on $m_{\textrm{fuel},t}\geq 0$. 

% ==========================================================================
\subsection{Evaluation Scenarios}
\label{sec:god_examples}

Four physical parameters define a scenario: the initial fuel mass $m_{\mathrm{fuel},0}$ (a.k.a., the fuel load), the maximum thrust $T_{\max}$, the drag coefficient $c_d$, and the dynamic pressure limit $Q_{\max}$ (which we will refer to, more simply, as just \emph{the pressure limit} going forward). All six methods described in the next subsection are evaluated on 18 scenarios shown in Table \ref{tab:god_examples}. These 18 scenarios were created by crossing three fuel loads with six pressure limits, while holding $T_{\max}=200{,}000\ \mathrm{N}$ and $c_d=1.5\ \mathrm{m^2}$ fixed.
% \footnote{As explained earlier, $c_d$ incorporates the effective surface area of the rocket.} 
These scenarios are not used in training: RL policies are trained under domain randomization over substantially wider parameter ranges for all four physical parameters and never see this specific grid in Table \ref{tab:god_examples} during learning.

\begin{table}[h]
\centering
\caption{18 evaluation scenarios. Each cell contains a scenario ID. The prefix denotes the fuel load (L = light, N = nominal, H = heavy); the suffix denotes the pressure limit (MT = megatight, HT = hypertight, ST = supertight, T = tight, M = medium, Lo = loose). All 18 scenarios have $T_{\max}=200{,}000\ \mathrm{N}$ and $c_d=1.5 \mathrm{m^2}$.}
\label{tab:god_examples}
\renewcommand{\arraystretch}{1.15}
\begin{tabular}{@{}l|cccccc@{}}
\toprule
& \multicolumn{6}{c}{$Q_{\max}$ (Pa)} \\
$m_{\mathrm{fuel},0}$ (kg) & 5{,}000 & 7{,}500 & 10{,}000 & 15{,}000 & 30{,}000 & 45{,}000 \\
\midrule
500  & \texttt{L\_MT} & \texttt{L\_HT} & \texttt{L\_ST} & \texttt{L\_T} & \texttt{L\_M} & \texttt{L\_Lo} \\
2000 & \texttt{N\_MT} & \texttt{N\_HT} & \texttt{N\_ST} & \texttt{N\_T} & \texttt{N\_M} & \texttt{N\_Lo} \\
4000 & \texttt{H\_MT} & \texttt{H\_HT} & \texttt{H\_ST} & \texttt{H\_T} & \texttt{H\_M} & \texttt{H\_Lo} \\
\bottomrule
\end{tabular}
\end{table}

Lower $Q_{\max}$ pressure limits forces a shallower, longer ascent (the $Q_{\max}$ constraint activates early and often); higher $m_{\mathrm{fuel},0}$ fuel loads give the rocket more total impulse but also more mass to have to accelerate. The \texttt{H\_MT} scenario in Table \ref{tab:god_examples} ($m_{\mathrm{fuel},0}=4000\ \mathrm{kg}$, $Q_{\max}=5{,}000\ \mathrm{Pa}$) is the most demanding test of constraint-aware throttling since it combines the largest fuel load with the tightest pressure limit constraint in the grid.

% ==========================================================================
\subsection{Methods}
\label{sec:goddard_methods}

We will compare the performance of six solution methods. Three of these methods are heuristic or analytical (Subsection \ref{sec:god_baselines}). The other three are learning-based methods that are trained independently of each other (Subsection \ref{sec:god_rl}). To make the comparison apples-to-apples, a common simulation environment is written to evaluate all six methods. As discussed in Subsections \ref{sec:god_dynamics} and \ref{sec:qmax}, this shared environment uses the RK4 integrator to solve the differential equation dynamics in equations \eqref{eq:hdot}--\eqref{eq:mdot} and uses binary search to find $T_{\lim}$. Wherever an action is required from the control algorithm, the environment sends the current state for the method in question (given in Table \ref{tab:god-state-features} in Appendix \ref{app:comparison_goddard}), and receives an action, $u$, in return. Since there is no stochasticity in the Goddard problem, differences in peak altitude reflect only the differences in policies, not integrator drift or mismatches in the environment implementations for different methods.

\subsubsection{Heuristic and Analytical Baseline Methods}
\label{sec:god_baselines}

For our first three methods, we consider three problem-specific baselines: (i) a simple open loop policy, (ii) a high-quality search heuristic that trades off computational time for better policies, and (iii) the known analytical solution in the case where there is no $Q_{\max}$ constraint.

\begin{description}
  \item[\textbf{Bang-Bang (heuristic)}] The simplest open-loop policy: full throttle ($u=1$) is used until fuel is exhausted, then there is an unpowered coast. While this approach ignores considering the $Q_{\max}$ constraint entirely, the environment automatically limits the thrust to conform to the $Q_{\max}$ constraint. This method is included as a simplistic lower-bound reference.

  \item[\textbf{MC-Local-Greedy (heuristic search + local improvement)}] Effectively, this method stands in for the upper-bound reference. This heuristic is an open-loop, piecewise-constant throttle schedule, chosen in the following manner: We first purposely overestimate the total time the rocket is ascending to be no more than four times the unconstrained full-throttle burn time, $t_{\mathrm{full}}=m_{\mathrm{fuel},0}c_e/T_{\max}$. We take this overestimate and divide it into 30 equal time segments. Each time segment is assigned a constant throttle value, $u \in [0,1]$, producing a 30 value throttle schedule. For each scenario, MC-Local-Greedy samples 500 candidate throttle schedules comprising uniformly random, front-loaded, and gradually decaying throttle profiles. A full-throttle schedule is also included explicitly as an additional candidate, meaning the bang-bang heuristic method is considered.  The eight throttle schedules of the 500 that produce the highest objective values (apogee altitudes) are subsequently refined using coordinate descent, where each segment is individually tested on a coarse throttle grid from 0 to 1. This polishing procedure is repeated for two passes, followed by one additional pass using a finer throttle grid whose possible values are $\{0, 0.05, 0.10,...,0.95, 1\}$. 

  This is the most expensive method by wall-clock time, but, as we will see later in Table \ref{tab:god_results}, it comes closest to the apogee altitude obtained by the Unconstrained Optimal method (described next) in scenarios where the $Q_{\max}$ constraint is not binding, while beating it when the $Q_{\max}$ constraint is more severe. That is, while the resulting best throttle schedule is an approximate model-based optimum rather than a guaranteed global optimum, we will see that there is strong evidence that this comes quite close to the global optimum.

  \item[\textbf{Unconstrained Optimal (analytic)}] This is the three-phase \emph{bang-arc-bang} exact optimal solution for the unconstrained case (i.e., no $Q_{\max}$) discussed at the beginning of this section. This is derived from Pontryagin's conditions as discussed in \cite{bonnans2008singular}. The first phase is full thrust ($u=1$). The final phase is coasting ($u=0$), which starts when the fuel runs out. The transition from the first phase to the middle (arc) phase happens at a known time, $t_{\mathrm{switch}}$. At this time $u$ drops to conform to the following closed-form singular-arc thrust law for an exponential atmosphere with quadratic drag: 
    \begin{equation}
      T=uT_{\max} = D(h,v) + \frac{m\,(c_e+v)\,c_e\,\bigl(2g + v^2/H\bigr)}
                                 {2c_e^2 + 4c_e v + v^2}.
      \label{eq:singular}
    \end{equation}
  We emphasize that $t_{\mathrm{switch}}$ and the singular arc law \eqref{eq:singular} are derived and optimal for the \emph{unconstrained} problem that has no $Q_{\max}$. Because there is a $Q_{\max}$ constraint in our scenarios, $u$ may need to be further reduced in either of the first two phases to conform to this constraint. The Unconstrained Optimal method is therefore a progressively worse solution as the $Q_{\max}$ constraint gets progressively binding (i.e., as the value of $Q_{\max}$ gets smaller). That is, for binding $Q_{\max}$ constraints, a method that optimizes around this constraint, as opposed to merely conforms to it, can potentially beat the Unconstrained Optimal method.

\textbf{Intuitive explanation of \eqref{eq:singular}:} The control theory explanation\footnote{For interested readers, the derivation of this relation is as follows: The system dynamics are defined by \eqref{eq:hdot}, \eqref{eq:vdot}, \eqref{eq:mdot}. The Hamiltonian is given by the inner product of the costate vector $[p_h, p_v, p_m]$ with the right-hand sides of these equations. Noting that the thrust $T$ appears linearly in the Hamiltonian, we collect the terms with $T$ together, and define the coefficient of $T$ as the switching function $S(t)\coloneq \left(\frac{p_v}{m}-\frac{p_m}{c_e}\right)$. The arc is ``singular" where $S(t)=0$ along with its time derivatives (since it stays 0 for a finite amount of time). The relationship \eqref{eq:singular} is then derived by solving for $\dot{S}(t)=\ddot{S}(t)=0$ simultaneously and eliminating the costate variables. See \cite{pontryagin1987mathematical} for the underlying theory.}
    for the singular arc law can be reduced to the following physical perspective: The arc defined in \eqref{eq:singular} corresponds to the singular throttle setting where the following two choices are equally worthwhile and therefore create a critical point with respect to the thrust:
    \begin{enumerate}
        \item Consuming an additional differential amount of fuel now to quickly escape from where the air is more dense and therefore drag is stronger, or
        \item Saving that fuel to consume it slightly later, once you've climbed into thinner air where the same thrust produces more net acceleration due to less drag.
      \end{enumerate}

\end{description}

\subsubsection{Learning-Based Methods}
\label{sec:god_rl}

The three learning-based methods are identical in setup to those from Section \ref{sec:example_gbwm}.

For training, all three use a beta distribution actor network output for exploration with the throttle action/control $u\in[0,1]$. This is analogous to our using the beta distribution in Section \ref{sec:example_gbwm} for exploration with the action/control of $\pi_t\in[0,1]$, the choice for the fraction of the portfolio in stock. 

Because the properties of the chosen beta distribution are particularly important for the Goddard problem, we discuss the nature of this distribution in more detail: The beta distribution has the following probability density function (pdf):
\begin{equation} \label{beta}
    f(x) = kx^{\alpha-1}(1-x)^{\beta-1} \mbox{ where } x \in [0,1].
\end{equation}
The constant $k$, which is needed to normalize the pdf so that it integrates to one, is the reciprocal of the beta function of the positive parameters $\alpha$ and $\beta$. The mean of the beta distribution is $\alpha/(\alpha+\beta)$. Using the beta distribution automatically enforces the constraint that $u\in [0,1]$. 

In particular, we note from the pdf in \eqref{beta} that if $\alpha>1$ and $\beta>1$, then $f(0) = f(1)=0$, so the distribution is \emph{unimodal} with the peak (local maximum) of the pdf being strictly between 0 and 1. This means exploration is discouraged from considering values of $u$ near 0 or 1. But if $\alpha$ and $\beta$ are between 0 and 1, then $\lim\limits_{x\to 0} f(x) = \lim\limits_{x\to 1} f(x) =\infty$, meaning exploration is particularly encouraged to consider values of $u$ near 0 or 1. Off-hand, there is no problem with the learning-based method only considering unimodal beta distributions, unless you are a domain expert. The domain expert, however, is aware of the \emph{bang-arc-bang} optimal strategy in the unconstrained case where  $u=1$ in the first phase and $u=0$ in the third phase. This means the domain expert knows that exploring near $u=0$ and $u=1$ is important, even in the constrained case.

Given an observation $s$, the actor network produces two state-dependent parameters, $\alpha(s)$ and $\beta(s)$, and samples $u\sim\mathrm{Beta}(\alpha(s),\beta(s))$ during training. Because $\alpha$ and $\beta$ must be positive, the raw actor outputs cannot be used directly since they may be negative or zero. Therefore, the implementations apply the softplus transform, $\operatorname{softplus}(x)=\log(1+\exp(x))$, which maps real number network outputs to positive numbers. For large positive network outputs, which are the inputs into the softplus transform, softplus behaves approximately like the identity function, whereas for large negative inputs the softplus output approaches zero. 

% This choice gives the policy the correct bounded action support directly, without requiring clipping to $[0,1]$. The Beta mean (used as the `deterministic' output during policy evaluation) is $\alpha/(\alpha+\beta)$, so the relative sizes of $\alpha$ and $\beta$ determine the policy's preferred throttle level. 

Below, we will see translated versions of the softplus, specifically, $\alpha=\operatorname{softplus}(\cdot)+\varepsilon$ and $\beta=\operatorname{softplus}(\cdot)+\varepsilon$, where $\varepsilon = 1$ and where $\varepsilon=0.01$. When $\varepsilon = 1$, we have the unimodal case where $\alpha$ and $\beta$ are greater than one, so $u$ is discouraged from being 0 or 1. When $\varepsilon = 0.01$, we have the ``boundary mode" case where $u$ being close to 0 or 1 is often encouraged since $\alpha$ and $\beta$ are allowed to be less than one. 

% We can also use translated versions of the softplus, such as $\alpha=\operatorname{softplus}(\cdot)+1$ and $\beta=\operatorname{softplus}(\cdot)+1$. This guarantees $\alpha,\beta>1$, yielding a Beta density with a single mode strictly inside the interval $(0,1)$. In particular, this `unimodal' parameterization makes the probability density approach zero at both $u=0$ and $u=1$, so it cannot have a maximum near the two limits. Alternatively, if we translate by $\varepsilon<<1$ to just ensure non-zero inputs to the Beta function, $\alpha=\operatorname{softplus}(\cdot)+\varepsilon$ and $\beta=\operatorname{softplus}(\cdot)+\varepsilon$, we retain positivity while allowing either parameter to be smaller than one. Depending on how $\alpha,\beta$ compare with $1$, we can get larger probability densities near $u=0$, $u=1$, or both (a U-shaped distribution). This `boundary mode' formulation is a more flexible family that includes unimodal Betas but can also represent the near bang-bang behavior that may be useful in some fuel/weight scenarios.
%
\begin{description}
  \item[\textbf{CC-Only}] uses Claude Code with Opus 4.7 as the base LLM (set to \texttt{xhigh} thinking), and the problem description from Subsections \ref{sec:god_dynamics}-\ref{sec:god_objective}. It chose to use PPO with a \emph{unimodal} beta head ($\mathrm{softplus}(\cdot) + 1$, discouraging considering $u=0$ or $u=1$), trained on a 10-dimensional normalized observation (see Table \ref{tab:god-state-features}), using Claude's own harness. 
  
  \item[\textbf{CC-Custom-Skill}] uses the same setup as the CC-Only version, with the addition of bespoke skills written by us\footnote{\url{https://github.com/srdas/carl}} for implementing RL solutions (identical to those used for CARL, except for conversational prompting). CC-Custom-Skill  also chose to use PPO with a \emph{unimodal} beta head ($\mathrm{softplus}(\cdot) + 1$, discouraging considering $u=0$ or $u=1$), trained on a different 10-dimensional observation (raw physical state plus per-episode parameters). 

  \item[\textbf{CARL}] with a custom harness, conversational capability, and direct calls to LLMs (as per Subsection \ref{sec:carl_models}). After agreeing on the formulation with the user, CARL chose PPO with a \emph{boundary-mode} beta head ($\mathrm{softplus}(\cdot) + \varepsilon$ where $\varepsilon=0.01$, often encouraging considering $u=0$ and/or $u=1$), trained on a 14-dimensional observation, which augments the raw state with two burnout shaping features. The discussion between CARL and the user, which will include creating the boundary-mode beta head and the two burnout shaping features, will be described in far fuller detail in the next subsection.
\end{description}
All three of these methods were evaluated in deterministic mode, meaning there is no exploration and the mean of the beta distribution, $\alpha/(\alpha+\beta)$, is used in place of random samples from the beta distribution. Also, all three methods were trained under domain randomization where the ranges of the per-episode  values of 
$m_{\mathrm{fuel},0}, T_{\max}, c_d$, and $Q_{\max}$ were substantially wider than the range and values used for the evaluation grid in Subsection~\ref{sec:god_examples}. A detailed comparison between the three learning method implementations' state variables is available in Appendix \ref{app:comparison_goddard}.
%Table~\ref{tab:god-state-features} directly compares the state representations, showing the union of observation features across the three policies and which method uses each.

\subsection{CARL Clarification Dialogue}
\label{sec:god-carl-clarifications}

Before presenting our empirical results, we illustrate the interactive clarifications CARL elicited while building its solution. Note that none of the other methods involve human-in-the-loop MDP design. The reference document provided at the start to all algorithms pins down the physics completely: every constant, every equation of motion, the constraint, and the evaluation grid are specified in closed form. Even so, a fully specified \emph{physics} brief leaves the \emph{MDP} substantially open (state design, action safety, reward normalization, and the hyperparameters). The stepwise dialogue turned that physics brief into a concrete MDP implementation through two phases: The initial run (Steps 1-9) and two revisits (Step 10).

\subsubsection{Initial Run}
\begin{itemize}
\item \textbf{Selecting state variables (in Step 4).} 
\emph{The observation:} CARL's state-vector proposal included both $h/H$ (altitude in scale-heights) and $\rho(h)/\rho_0$ (density ratio). It also included one state variable $q/Q_{\max}$ for the dynamic pressure information. The user gave feedback that in our simple atmosphere model, altitude and density are monotonic functions of each other, and $h$ itself is not required by the rocket controller for any decisions. Furthermore, while $q/Q_{\max}$ is sufficient information for pressure-based information, it does make the controller discover the penalty of closing the ``available headroom" in dynamic pressure. \emph{Implications for CARL:} It dropped $h/H$ which was not directly required by the controller. It kept $q/Q_{\max}$ as the primary constraint-proximity signal, but added a $1-q/Q_{\max}$ headroom feature to make learning simpler for the network. Even though the two terms are linearly dependent, their provision as direct inputs gives the network a shortcut to learn the physics of the problem.

\item \textbf{Designing the action space (in Step 5).}
\emph{The observation:} CARL initially proposed a Gaussian distribution from which the action would be sampled during training. The domain expert pointed out the importance of strictly adhering to the two physical constraints: 1) the throttle action, $u$, would need to be clipped to remain in the $[0,1]$ range (possibly causing gradient problems during training) and 2) the policy should further ensure that the dynamic pressure constraint is not violated. \emph{Implications for CARL:} It revised the action distribution from Gaussian to Beta (cleaner $[0,1]$ support) and specified a concrete two-layer throttle cap: the environment clamps physically using the tighter constraint between the natural limits $[0,1]$ and the structural limits due to dynamic pressure. For added safety, the actor caps its own sampled action at 99\% of the max-safe throttle, computed from current velocity at $0.1\,\mathrm{s}$ resolution.

\item \textbf{Designing the reward function (in Step 6).}
\emph{The observation: }CARL initially proposed a terminal-only reward $h_{\mathrm{apogee}}/H$. The user pushed back on this choice, stating that the reward would be too sparse for effective learning. Instead, the altitude increment in each time step could be immediately provided as a reward, and its sum over the episode would be identical to the final altitude (so long as no discounting was done for current vs. future rewards). The user also preauthorized a fallback in case the normalization by $H$ made the reward magnitudes too small; it was acceptable to try alternative $H$ constants. \emph{Implications for CARL: }The reward changed to $\Delta h/H$ as a step reward where $\Delta h$ is the change to the altitude in each $\Delta t$ time step. With step rewards, CARL had to specify a discount factor, $\gamma$, over time, discussed further in the next step.  The fallback for changing the value of $H$ was also utilized in Step 9 when an entropy-collapse failure made the normalized reward's per-step signal too faint for PPO's advantage estimates to resolve. 

\item \textbf{Explicit optimization directives (in Step 7).} \emph{The observation:} Rather than accept CARL's proposed $\gamma=0.999$ and moderate batch sizes, the user 1) directed $\gamma=1$ to match the telescoped-reward design exactly, reflecting the fact that there is no discounting over time since each increase in $h$ is equally important to the terminal reward, no matter when it happens and 2) explicitly required parallelized experience collection and large batch sizes for gradient stability. \emph{Implications for CARL: } CARL changed $\gamma$ to 1 and suggested 64 parallel environments, 640 episodes per PPO update, and a mini-batch size of 2048, roughly double its own initial proposal.

\item \textbf{Boundary-mass diagnosis mid-training (in Step 9).} \emph{The observation:} While Step 9's training run was in progress, the domain expert made sure CARL was aware of the concern discussed in the previous subsection, namely the importance of making sure exploration considers the $u=0$ and $u=1$ cases. Specifically, the domain expert warned CARL to ``make sure you are able to load probability mass on the limits of the beta distribution $[0,1]$ if training doesn't go well for some reason.'' \emph{Implications for CARL: } This anticipated exactly what happened next. The initial unimodal beta parameterization that CARL used led to a collapse to near-zero entropy after the first PPO update with no policy improvement, due to the fact that the probability density function for the unimodal beta distribution was always zero at the endpoints $u=0$ and $u=1$. Being specifically warned about this, CARL understood the cause of the problem and switched to the boundary mode beta head (previously discussed in Subsection \ref{sec:god_rl}) where the probability density function can be infinite at the two endpoints, and so the boundary mode beta head is what shipped in the final CARL checkpoint.
\end{itemize}

\subsubsection{Post-Training Diagnosis} \label{sec:godd_post_train}

\begin{itemize}
\item \textbf{Post-training burnout diagnosis (Step 10) and Revisit 1.} 
\emph{The observation: }After the long training run, CARL inspected the throttle plots directly and identified a specific, previously undiagnosed failure: it was running out of fuel far earlier than MC-Local-Greedy in several scenarios. The domain expert, aware of the nature of the second (arc) phase in the optimal unconstrained case, pointed out that burning all fuel early sacrifices the opportunity to apply thrust later, after the rocket has reached thinner air and can convert thrust into altitude more effectively. Therefore, the domain expert suggested that CARL should penalize early burnouts, specifically by computing a minimum burnout time $t_{\mathrm{target}}$ using the MC-Local-Greedy heuristic's own logic for deciding the ``optimization time span." Furthermore, the user emphasized that this penalty should be derivable from CARL's observations, which may need to be expanded to allow for this. \emph{Implications for CARL: }In response, CARL made several changes to the states and reward. It computed $t_{\mathrm{target}}$ using MC-Local-Greedy's logic ($4\times$ the unconstrained bang-bang burn duration). It applied a one-sided penalty (which is zero if the rocket burns out on or after $t_{\mathrm{target}}$, but is increasingly positive the earlier the rocket burns out before $t_{\mathrm{target}}$). To make this penalty predictable from observations, it added two new state features: a signed time-remaining fraction (which goes negative after crossing $t_{\mathrm{target}}$) and a ``constant-throttle-to-target'' hint (which computes the rough constant value of $u$ to burn out exactly at $t_{\mathrm{target}}$).

\item \textbf{Revisit 2: swapping the linear heuristic for a quadratic one.} \emph{The observation:} The first revisit substantially improved the results compared to the Step 9 output. However, these improved results were still somewhat behind the analytical solution in the scenarios where the $Q_{\max}$ constraint was not binding and the MC-Local-Greedy heuristic results when the $Q_{\max}$ constraint was binding. Therefore, the user asked CARL to do a small refinement: instead of using the MC-Local-Greedy heuristic's $t_{\mathrm{target}}$, which grew linearly with fuel, CARL should observe the burnout times of MC-Local-Greedy in various scenarios, and numerically fit a quadratic function for $t_{\mathrm{target}}$. \emph{Implications for CARL: } CARL updated its calibration to $t_{\mathrm{target}} = 75x - 8x^2$ with $x = m_{\mathrm{fuel},0}/1000$.
% , which is still evaluated once at episode reset and otherwise unchanged in how the one-sided penalty and the two new state features affect how fuel is consumed. 
At the fixed evaluation parameter $T_{\max}=200{,}000\,\mathrm{N}$, this raises the target minimum burnout times for the three initial fuel loads used in the 18 evaluation scenarios to $t_{\mathrm{target}}\approx\{35.5,118,172\}$ seconds, which is noticeably longer than Revisit 1's $\{20,80,160\}$ seconds. We will see the specific quantitative effects of Revists 1 and 2 in Subsection \ref{sec:god_carl_revisits}.
% , with the largest relative change at the light-fuel end ($7.1\times$ full-burn-time under the quadratic fit versus the original $4\times$).
\end{itemize}
\subsubsection{Use of heuristics for self-improvement with CARL}

As explained in Subsection \ref{sec:resume_revisit}, CARL is designed to observe the logs and plots from prior training runs (including baseline heuristics) and use insights from these runs for improving its MDP design. The same facility is provided to CC-Custom-Skill since it shares the same skill files. In this example, both CARL and CC-Custom-Skill wrote similar search \& improvement style heuristics to track learning performance. However, neither method leveraged the hint provided in Step 8 (\emph{``Use a rule-based heuristic for a short supervised training phase."}) to warm up their policies, which could have led to better performance in the initial runs. For CC-Custom-Skill, which is fully autonomous, there was no recourse for further improvement. However CARL's revisit phase worked as designed, using the human-in-the-loop to discuss the observations from plots and guide it towards a better MDP formulation.

As with the GBWM example from the previous section, the impact of these clarifications is not cosmetic: they can directly determine whether PPO training converges at all, as we saw with the beta boundary probability fix, or whether the policy is being optimized toward the right notion of progress, as we saw with the one-sided burnout penalty and its two supporting state features. The Revisit 2 change exemplifies that this iteration does not necessarily stop once training succeeds. Even a working shaping term can be revisited and its functional form swapped out in a later pass, with the surrounding penalty and state-feature machinery left untouched. A completely specified physics reference does not, by itself, specify a learnable MDP. The state representation, safety mechanism, reward scale, and training schedule all still had to be designed, diagnosed, and revised across the project's lifetime, and CARL's structured dialogue with the human user is what surfaced each of those decisions rather than silently defaulting them.

% ==========================================================================
\subsection{Results}
\label{sec:goddard_results}

We discuss both quantitative and qualitative comparisons across the 18 evaluation scenarios described earlier in Subsection \ref{sec:god_examples}.

\subsubsection{Comparison Methodology}
\label{sec:god_crn}

Unlike stochastic problems that require averaging over many random rollouts, the Goddard dynamics~\eqref{eq:hdot}-\eqref{eq:mdot} are entirely deterministic: for a fixed scenario and a fixed policy, a single rollout through the shared RK4 environment is always the same, with no sampling noise to average out. Each of the six methods is therefore rolled out exactly once per scenario, and wall-clock time is recorded alongside peak altitude. All computation is CPU-only on an AWS SageMaker \texttt{ml.t3.xlarge} instance (4 vCPU, 16 GB RAM).

\subsubsection{Comparison of CARL Revisits}
\label{sec:god_carl_revisits}

To isolate the effect of CARL's iterative development process, we compare the initial checkpoint with the checkpoints produced after Revisit~1 and Revisit~2. The peak altitudes (i.e., the apogees) for the 18 scenarios are shown in Table \ref{tab:god_carl_revisits}. \textbf{Boldface} is used to indicate for each scenario which instantiation of CARL had the highest apogee.

\begin{table}[ht]
\centering
\caption{Peak altitude (in meters) for CARL's initial run, Revisit~1, and Revisit~2. \textbf{Bold} marks the best internal CARL run for each of the 18 scenarios. On average, Revisit~1 improves over the initial run by 4.1\%, and Revisit~2 further improves over Revisit~1 by 6.9\% (and by 11.4\% compared to the initial run).}
\label{tab:god_carl_revisits}
\small
\begin{tabular}{cc|rrr}
\toprule
Fuel (kg) & $Q_{\max}$ (Pa) & CARL (initial) & CARL (Revisit~1) & CARL (Revisit~2) \\
\midrule
500 & 5{,}000 & 3366 & 3366 & \textbf{3484} \\
500 & 7{,}500 & 3205 & 3201 & \textbf{3491} \\
500 & 10{,}000 & 3044 & 3101 & \textbf{3493} \\
500 & 15{,}000 & 2843 & 3091 & \textbf{3493} \\
500 & 30{,}000 & 2883 & 3087 & \textbf{3491} \\
500 & 45{,}000 & 2883 & 3110 & \textbf{3488} \\
2000 & 5{,}000 & 14953 & 14966 & \textbf{15356} \\
2000 & 7{,}500 & 14568 & 14590 & \textbf{15439} \\
2000 & 10{,}000 & 14117 & 14121 & \textbf{15452} \\
2000 & 15{,}000 & 13943 & 14385 & \textbf{15445} \\
2000 & 30{,}000 & 14487 & 14442 & \textbf{15403} \\
2000 & 45{,}000 & 14568 & 14555 & \textbf{15378} \\
4000 & 5{,}000 & \textbf{41959} & 41879 & 41880 \\
4000 & 7{,}500 & 43168 & \textbf{43475} & 43114 \\
4000 & 10{,}000 & 41216 & 42668 & \textbf{44687} \\
4000 & 15{,}000 & 36482 & 43742 & \textbf{47140} \\
4000 & 30{,}000 & 39694 & 44960 & \textbf{47870} \\
4000 & 45{,}000 & 41791 & 45589 & \textbf{47864} \\
\bottomrule
\end{tabular}%
\end{table}

We see from the table that Revisit~2 is the strongest internal CARL run in 16 of the 18 scenarios. The two exceptions are the heavy-fuel cases at $Q_{\max}=7{,}500\ \mathrm{Pa}$, where Revisit~1 is best, and $Q_{\max}=5{,}000\ \mathrm{Pa}$, where the initial run remains marginally ahead. The high-fuel tight-constraint combination in these scenarios leaves very little room for policy optimization, with all the CARL policies tracking close to the Bang-Bang heuristic as we will see in Figure \ref{fig:megatight} in Subsection \ref{sec:god_profiles}.

\subsubsection{Comparison Across Methods}
\label{sec:god_combined}

Table~\ref{tab:god_results} compares both the apogee and, in parentheses, the runtime for each of the six methods applied to each of the 18 scenarios. The highest apogee per scenario is in boldface. A quick perusal of the boldface results in the Unconstrained Optimal method's column explains their presence: it is clear that the $Q_{\max}$ constraint in these cases is not binding because higher $Q_{\max}$ values do not change these results. Since the Unconstrained Optimal method is guaranteed to produce the correct optimal results when the $Q_{\max}$ constraint is not binding, we know that this method must produce the highest apogee for these scenarios. But we also see that as the $Q_{\max}$ constraint becomes progressively binding, the Unconstrained Optimal method becomes progressively worse, as expected, since it ignores this constraint in its strategy even though it must adhere to it in the rollout. This is particularly noticeable in the scenario where the fuel starts at 4000 kg and $Q_{\max}$ is only 5,000 Pa.

\begin{table}[ht]
\centering
\caption{Peak altitude (in meters) by method and scenario. \textbf{Bold} marks the best method per row. The episode rollout time (including thrust computation and environment stepping) is written in parentheses, in seconds. Note that times increase with the length of the episodes. The scenarios where the Unconstrained Optimal solution is beaten are the ones with tight pressure constraints, which render the ``unconstrained" assumption invalid. MC-Local-Greedy, which is a close approximation to the analytical solution, is the best in those cases. On average, CARL exceeds the performance of CC-Only by 19.9\%, and is within 3.1\% of the best-performing method per scenario, while using far less computational time than the MC-Local-Greedy method.}
\label{tab:god_results}
\renewcommand{\arraystretch}{1.15}
\small
\resizebox{\linewidth}{!}{%
\begin{tabular}{cc||ccc|ccc}
\toprule
Fuel & $Q_{\max}$ & Bang-Bang & MC-Local- & Unconstrained & CC-Only & CC-Custom- & CARL \\
 (kg) & (Pa) &  & Greedy & Optimal &  & Skill &  \\
\midrule
500 & 5{,}000 & 3372\,(0.02) & 3500\,(33.43) & \textbf{3505}\,(0.02) & 3384\,(0.11) & 3120\,(0.39) & 3484\,(0.19) \\
500 & 7{,}500 & 3201\,(0.02) & 3493\,(30.62) & \textbf{3505}\,(0.02) & 3431\,(0.11) & 3185\,(0.41) & 3491\,(0.20) \\
500 & 10{,}000 & 3024\,(0.01) & 3495\,(27.97) & \textbf{3505}\,(0.05) & 3431\,(0.13) & 3198\,(0.52) & 3493\,(0.27) \\
500 & 15{,}000 & 2735\,(0.01) & 3491\,(24.46) & \textbf{3505}\,(0.05) & 3386\,(0.10) & 3169\,(0.42) & 3493\,(0.21) \\
500 & 30{,}000 & 2265\,(0.01) & 3485\,(20.28) & \textbf{3505}\,(0.02) & 3219\,(0.07) & 2751\,(0.46) & 3491\,(0.20) \\
500 & 45{,}000 & 2076\,(0.01) & 3477\,(20.01) & \textbf{3505}\,(0.02) & 3109\,(0.07) & 293\,(0.27) & 3488\,(0.19) \\ \hline
2000 & 5{,}000 & 15005\,(0.07) & \textbf{15498}\,(118.18) & 14443\,(0.06) & 12093\,(0.36) & 14597\,(0.87) & 15356\,(0.54) \\
2000 & 7{,}500 & 14627\,(0.05) & \textbf{15728}\,(95.53) & 15717\,(0.04) & 12975\,(0.33) & 14695\,(0.78) & 15439\,(0.50) \\
2000 & 10{,}000 & 13874\,(0.05) & 15754\,(86.45) & \textbf{15805}\,(0.38) & 13435\,(0.35) & 14778\,(0.74) & 15452\,(0.51) \\
2000 & 15{,}000 & 12362\,(0.04) & 15705\,(71.19) & \textbf{15805}\,(0.42) & 13813\,(0.30) & 15235\,(0.83) & 15445\,(0.52) \\
2000 & 30{,}000 & 9461\,(0.02) & 15570\,(59.31) & \textbf{15805}\,(0.04) & 13909\,(0.24) & 15268\,(0.90) & 15403\,(0.49) \\
2000 & 45{,}000 & 8033\,(0.02) & 15568\,(55.51) & \textbf{15805}\,(0.05) & 13946\,(0.24) & 13580\,(1.08) & 15378\,(0.49) \\ \hline
4000 & 5{,}000 & 42083\,(0.11) & \textbf{44023}\,(233.60) & 27966\,(0.10) & 27294\,(0.52) & 42215\,(1.41) & 41880\,(0.92) \\
4000 & 7{,}500 & 43367\,(0.09) & \textbf{48170}\,(167.72) & 39307\,(0.09) & 30699\,(0.50) & 43737\,(1.29) & 43114\,(0.84) \\
4000 & 10{,}000 & 41337\,(0.08) & \textbf{49465}\,(142.76) & 46878\,(1.36) & 32905\,(0.40) & 42127\,(1.10) & 44687\,(0.72) \\
4000 & 15{,}000 & 35477\,(0.06) & 49707\,(119.38) & \textbf{50165}\,(1.34) & 33741\,(0.40) & 38059\,(1.02) & 47140\,(0.79) \\
4000 & 30{,}000 & 24137\,(0.04) & 49263\,(104.05) & \textbf{50165}\,(0.07) & 35099\,(0.39) & 40319\,(0.99) & 47870\,(0.79) \\
4000 & 45{,}000 & 19362\,(0.03) & 49067\,(99.44) & \textbf{50165}\,(0.07) & 37293\,(0.39) & 44681\,(1.21) & 47864\,(0.85) \\
\bottomrule
\end{tabular}%
}
\end{table}

In the cases where $Q_{\max}$ is binding, we see that the MC-Local-Greedy method produces the best results, and even when $Q_{\max}$ is not binding, it produces the second best results after the Unconstrained Optimal method. But this comes at a considerable computational cost. As seen from the wall-clock costs in the parentheses in the table, the MC-Local-Greedy method takes between 100 and 253 times as much wall-clock time as CARL does because it evaluates hundreds of candidate schedules through the RK4 integrator plus local-search polishing. This is one of the great advantages of all three learning-based methods: once they are done with training, they can run evaluations in inference mode extremely quickly. CARL produces a nice balance, producing close to optimal results (within 3.1\% of the best-performing method in each of the 18 scenarios) in a remarkably short amount of computational time, even when $Q_{\max}$ is heavily binding.

% shows that Unconstrained Optimal and MC-Local-Greedy generally sit at the top, followed by CARL, while the relative ordering among the remaining methods varies with constraint tightness. The clearest break from the unconstrained ranking is for heavier fuel loads under tight $Q_{\max}$ constraints, where MC-Local-Greedy outperforms all other methods (for reasons explained below). Unconstrained Optimal's open-loop schedule was optimized \emph{ignoring} $Q_{\max}$ and only satisfies the constraint during rollout, whereas MC-Local-Greedy searches with the constraint active from the start, which pays off when the constraint is severe enough to dominate the unconstrained-optimal switch-time choice.

% Wall-clock cost is sharply asymmetric: MC-Local-Greedy takes 20-160 seconds per scenario (it evaluates hundreds of candidate schedules through the real integrator plus local-search polishing), while the three RL policies and the Bang-Bang/Unconstrained Optimal baselines all return in well under two seconds. MC-Local-Greedy's competitive altitude therefore comes at roughly two orders of magnitude more compute than any policy-based method.

The most striking failure remains CC-Custom-Skill at $m_{\mathrm{fuel},0}=500\ \mathrm{kg}$, $Q_{\max}=45{,}000\ \mathrm{Pa}$ (light fuel, loose constraint), where it reaches only 293\,m. This is a policy generalization failure on a scenario that should ideally pose no particular difficulty; Figure~\ref{fig:loose} from the next subsection shows that poor initial throttle decisions are responsible, in particular, the throttle starts and stays below $u=0.1$ when it needs to be much higher at the start to generate needed lift. CARL, by contrast, tracks within 1\% of the Unconstrained Optimal upper bound across all six light-fuel scenarios and stays close to MC-Local-Greedy throughout the expanded grid, reflecting its burnout-aware observation features  from Subsection~\ref{sec:godd_post_train}. CC-Only's relative position depends on the regime: it outperforms CC-Custom-Skill on the light-fuel cases, but generally lags it for nominal and heavy fuel loads. Both CC-Only and CC-Custom-Skill are hurt by their using only unimodal beta distributions, in contrast to CARL's boundary mode beta distributions, as discussed previously.

An interesting feature of the simple Bang-Bang heuristic method is that its apogee increases as the $Q_{\max}$ constraint becomes tighter. Under a tight constraint, the environment's clamp on the throttle due to $T_{\lim}$ causes a 5\% drag surcharge incurred on clamped steps, but this is overcome by the fact that the constraint also increases the burn time, preventing Bang-Bang from burning out too fast, which leads to a lower altitude.

\subsubsection{Profiles Over Time}
\label{sec:god_profiles}

\begin{figure}[ht]
  \centering
  \includegraphics[width=\linewidth]{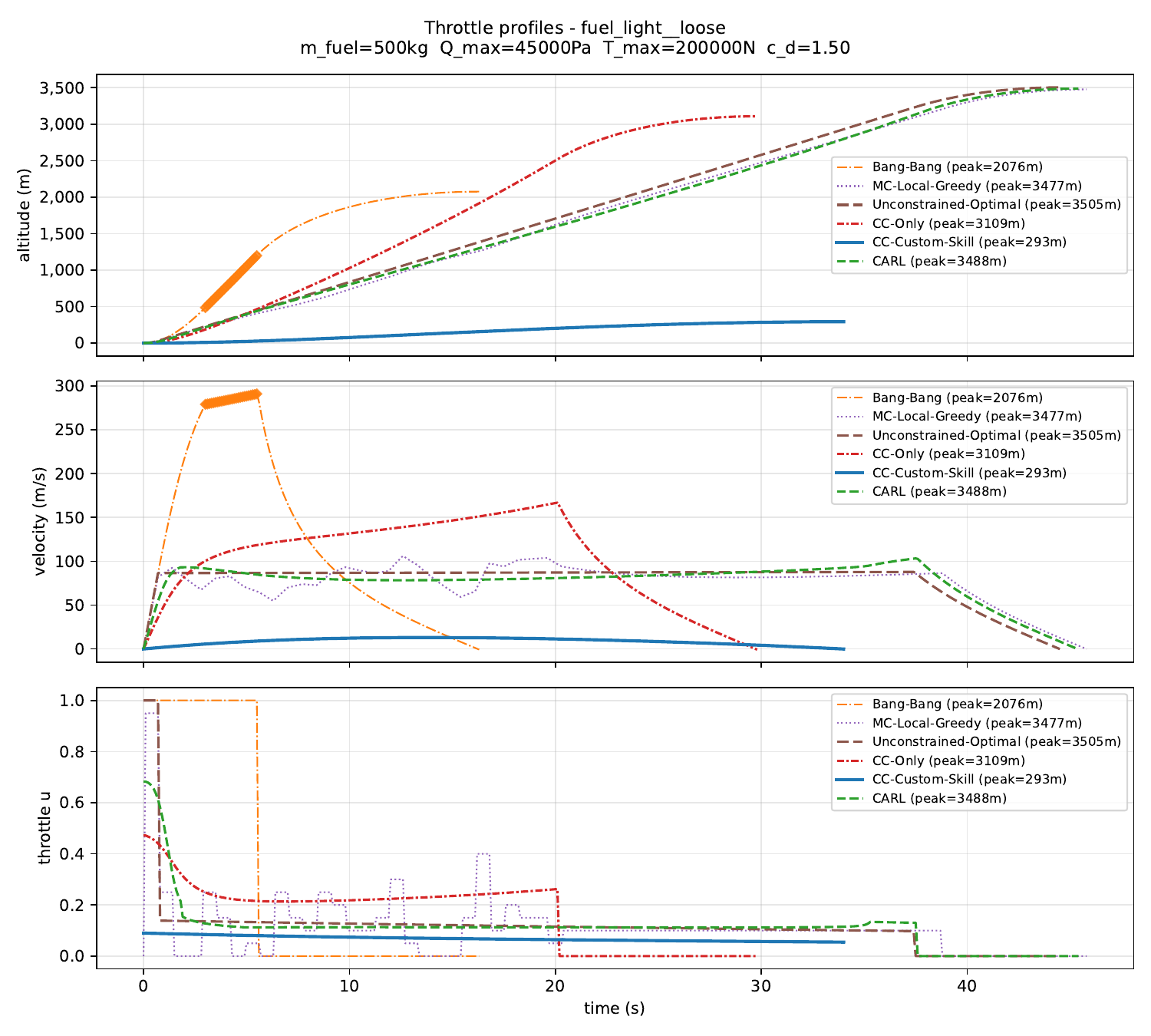}
  \caption{Profiles over time for $m_{\mathrm{fuel},0}=500\ \mathrm{kg}$,
    $Q_{\max}=45{,}000\ \mathrm{Pa}$ (lightest fuel, most loose constraint). CC-Custom-Skill's throttle collapses almost immediately, stalling the ascent at 293\,m while every other method reaches 2000-3500\,m; this is the sharpest policy failure observed anywhere in the evaluation grid.}
  \label{fig:loose}
\end{figure}

\begin{figure}[ht]
  \centering
  \includegraphics[width=\linewidth]{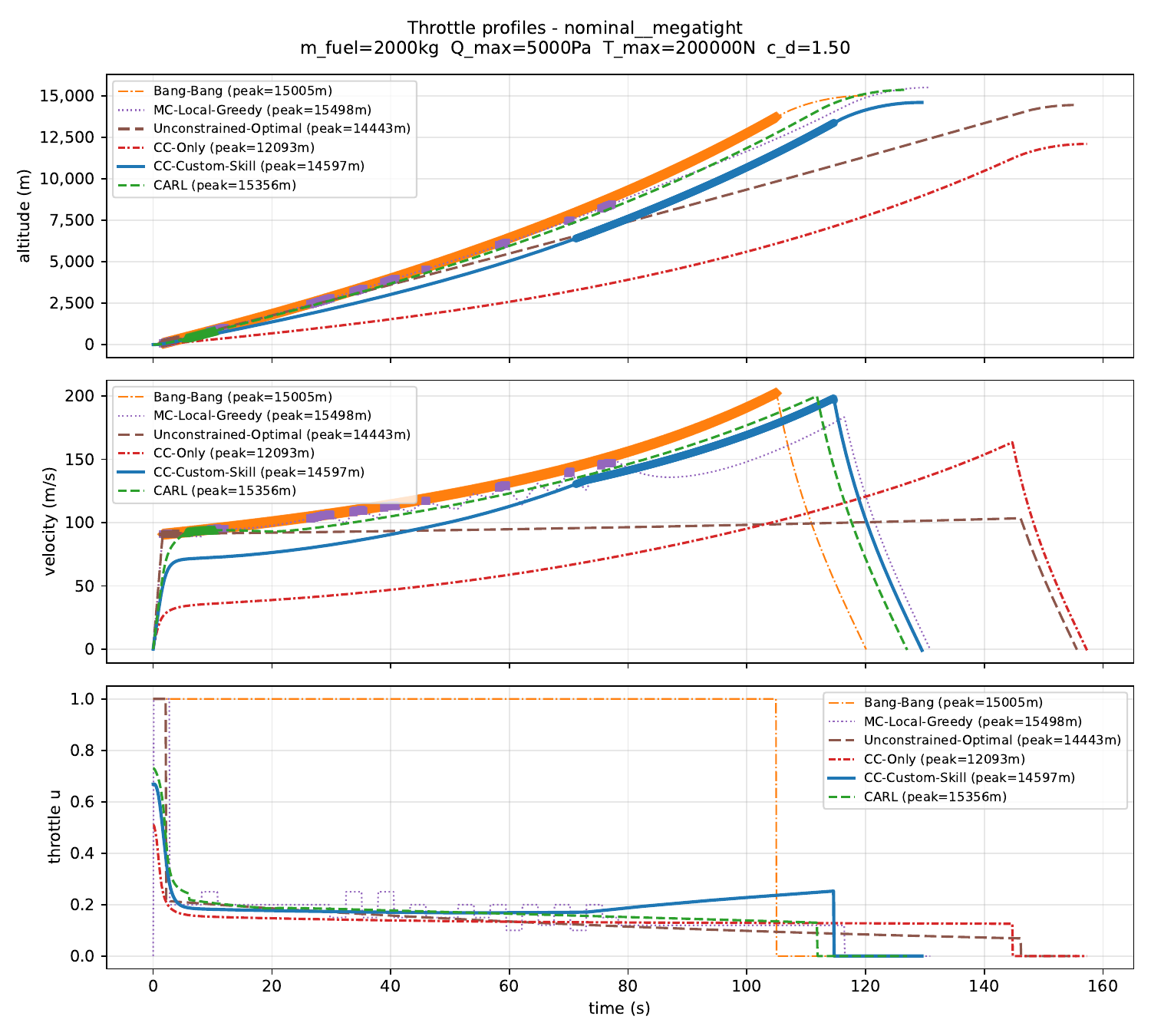}
  \caption{Profiles over time for $m_{\mathrm{fuel},0}=2000\ \mathrm{kg}$, $Q_{\max}=5{,}000\ \mathrm{Pa}$. This is one of the scenarios with the tightest $Q_{\max}$, which leads to several strategies beating the Unconstrained Optimal method, even though the constraint only affects this method between section 1--4. Interestingly, Bang-Bang is among the best performers here; CARL and MC-Local-Greedy, which exceed it, generally just manage to stay below the vibration penalty threshold.}
  \label{fig:megatight}
\end{figure}

Figures \ref{fig:loose} and \ref{fig:megatight} below correspond to two different scenarios. Both figures contain the same three-panel layout: altitude (top), velocity (middle), and throttle $u$ (bottom), all against time. In each panel, all six methods are overlaid on top of each other. Thick, solid segments on the altitude and velocity traces indicate steps where the $Q_{\max}$ constraint was active. 

Broadly speaking, the MC-Local-Greedy heuristic method tracks the Unconstrained Optimal method's solution quite closely in scenarios where the $Q_{\max}$ constraint is not binding, such as in Figure \ref{fig:loose}, although it is jittery around that solution due to its having only 30 time segments where changes can occur. While the MC-Local-Greedy method outperforms the Unconstrained Optimal method when the $Q_{\max}$ constraint is binding, it is interesting to note in scenarios such as Figure \ref{fig:megatight} that the constraint may only be affecting the Unconstrained Optimal trajectory for a very short time, for example, just during seconds 1--4 in the figure. 

Due to pushing $u=1$ as much as possible, the Bang-Bang method typically hits the $Q_{\max}$ constraint clamp for large parts of its trajectory, leading to considerable vibration-related drag. The CC-Only and CC-Custom-Skill policies have significantly different shapes and lengths compared to the Unconstrained Optimal baseline in unconstrained scenarios; CARL tracks the baseline more closely due to the changes described in Subsection \ref{sec:god-carl-clarifications}.

% \begin{figure}[ht]
%   \centering
%   \includegraphics[width=\linewidth]{throttle__fuel_heavy__tight.pdf}
%   \caption{Throttle profiles for $m_{\mathrm{fuel},0}=4000\ \mathrm{kg}$, $Q_{\max}=15{,}000\ \mathrm{Pa}$ (heavy fuel, moderately tight constraint). Bang-Bang and CC-Custom-Skill show the clamp firing (thick marker segments) during the high-dynamic-pressure phase of flight; CARL does better but still burns out significantly earlier than the two best-performing methods.}
%   \label{fig:tight}
% \end{figure}

% \begin{figure}[ht]
%   \centering
%   \includegraphics[width=\linewidth]{throttle__fuel_heavy__supertight.pdf}
%   \caption{Throttle profiles for $m_{\mathrm{fuel},0}=4000\ \mathrm{kg}$, $Q_{\max}=10{,}000\ \mathrm{Pa}$, where the constraint is hit by four of the six methods. This is one of the scenarios where MC-Local-Greedy's constraint-aware search overtakes the Unconstrained Optimal baseline (Table~\ref{tab:god_results}), reflecting the cost of ignoring the constraint.}
%   \label{fig:supertight}
% \end{figure}

\section{Conclusions}
\label{sec:discussion}

CARL demonstrates that a structured conversational workflow can efficiently and effectively turn a domain expert's intent into a runnable RL training pipeline. The two examples in the preceding sections demonstrate how this works. We show that while having a language model write RL code is now relatively easy, coming up with a working RL model from which to create the code is more nuanced and involves design choices in which the human is intimately involved to resolve the specification gap. Furthermore, this is not a one-shot effort, even with a capable LLM and an expert human; iteration is necessary to close the performance gap.

\subsection{Addressing the Specification Gap}

The central premise of CARL is that the principal barrier to applied RL adoption is not an algorithmic problem. It is \emph{specification failure}: the inability to translate a domain expert's intuition into the formal $\{$state, action, reward, dynamics$\}$ tuple that an RL algorithm requires. This premise is consistent with the findings in~\cite{hamadanian_how_2022}, which reports that a large share of industrial RL projects fail during problem formulation rather than during RL training. CARL attacks this gap by decomposing the specification task into ten discrete, sequentially dependent steps. Each of these are governed by their own skill file and checklist, so that no single step's conversation must carry the full burden of formalization. The resulting workflow enforces completeness in the specification: every component of the MDP is explicitly discussed before any code is generated.

% The \texttt{/suggest} feature takes this a step further. A user with only a vague problem description can invoke an automated literature search, receive several candidate RL formulations synthesis from published work, and select one as the starting point. This effectively lowers the entry barrier from ``I know the MDP I want'' to ``I know the domain I care about,'' while still routing the result through the same process.

\subsection{The Importance of Humans-in-the-Loop versus Automation in Reducing the Specification Gap and the Performance Gap}

CARL assigns human and machine authority according to the kind of decision being made. The domain expert controls the substantive modeling decisions in Steps~1-7, while CARL handles implementation, testing, and training after those decisions have been recorded. This separation matters because implementation errors in a correct MDP can often be debugged, but a well-trained policy for the wrong MDP optimizes the wrong problem. Step~10 then reopens the design because evidence derived from the trained policy now becomes available, serving as a verification-and-parameter-alteration step rather than treating the first completed run as final, which prevents using evidence created from the trailed policy to locate problems and potential improvements.

\subsection{Takeaways from our Two Examples}

First, the goals-based wealth management (GBWM) and Goddard Rocket examples from the previous two sections clearly show that a complete domain brief is not the same thing as a complete RL specification. In the GBWM example, the partial specification was economically meaningful but still omitted important MDP choices like feasibility rules, the terminal-goal treatment, utility arithmetic, the action boundaries, the reward timing, and state normalization. In the Goddard example, the physics, including the objective to be maximized (the height of the apogee) and the dynamic pressure constraint, were specified precisely, yet the resulting RL controller still depended on choices about the observation design, the action support at the boundaries, the reward scale, safety handling, and the training schedule. CARL is useful precisely in addressing this jump from a domain problem to an implementable MDP.

Second, the value of CARL is in its interaction with a human domain expert, not just the presence of a skill library. The comparisons used in the GBWM case separated the effectiveness of direct agentic coding, coding with custom skills, and CARL's interactive workflow. Skill libraries improve the non-interactive baselines, but the largest difference appears when the initial brief is incomplete and the missing modeling choices must be surfaced, discussed, and resolved before implementation. CARL-PartSpec remains close to CARL-FullSpec across the GBWM scenarios, whereas the partial non-CARL methods are more sensitive to the unresolved choices in the original brief.

Third, verification and parameter alteration should be viewed as part of the specification process, rather than as a postscript to training. In GBWM, inspection of the learned high-wealth allocation behavior led to changes in the policy shape and terminal-wealth regularization. In Goddard, throttle traces revealed early burnout, and the verification-and-parameter-alteration step converted that diagnosis into burnout-aware observation features and reward terms. These revisions were different from regular bug fixes; they identified defects in the policy resulting from the MDP choices.

Fourth, CARL should not be interpreted as replacing specialized optimization or model-based planning. The Goddard results are deliberately more nuanced: the MC-Local-Greedy baseline often produced higher apogee altitudes than CARL because it searches each scenario directly through the simulator, but it does so at a wall-clock cost that was at least 100 times longer than CARL. CARL's strength is that it produces reusable policies through an explicit collaborative design process. When a simulator is available and per-instance computation is acceptable, specialized search can remain a valuable alternative to RL itself (and not just to CARL). Further, as we saw, examination of the MC-Local-Greedy results was quite helpful in improving the accuracy of the final RL program produced by CARL.

\subsection{Implications of our Work for Future Deployment and Evaluation}

Our empirical results suggest that CARL is most appropriate when domain experts can judge whether computed behavior is reasonable, but cannot be expected to write a complete RL formulation at the outset, especially since RL is not likely their domain of expertise. In such settings, the central deployment question is not only whether the generated code runs, but whether the resulting policy reflects the intended decision problem. Evaluation should therefore include policy surfaces, trajectory traces, constraint-violation checks, benchmark comparisons where available, and explicit diagnoses of failure modes. The audit trail of decisions, generated artifacts, and revisit changes is also important: it lets users distinguish poor algorithmic choice from poor problem formulation. 

% Future Work ---------------------------------------------

\section{Future Work} \label{sec:future}

Several research directions are possible from the current version of CARL:

\begin{itemize}
\item \textbf{Expanding to More RL Problems}. The two main examples in this paper are deliberately different, but both still have compact state descriptions, low-dimensional action spaces, and evaluation structures that can be inspected after training. A natural next step is to apply CARL to a wider family of RL problems, such as robotic manipulation, multi-agent coordination, healthcare decision support, logistics, and recommendation systems. Such domains would stress-test whether the ten-step workflow remains effective when addressing partially-observable systems. This direction also connects to automatic reward shaping: inverse RL, preference learning, or trajectory comparisons could allow CARL to propose candidate reward functions when users can recognize good behavior but cannot easily formalize it.
\item \textbf{Expanding the Testing User Base}. The current prototype has been evaluated through detailed examples and expert-guided sessions but broader usability evidence is needed. Future studies should include users with different levels of RL knowledge, programming experience, and domain expertise, and should measure not only final policy quality but also the interaction burden imposed by the workflow. Useful measures include time to complete specification, number and type of clarification turns, frequency of user edits to Trillian suggestions, perceived control, confidence in the generated policy, and the ability of users to diagnose results during verification and parameter alteration. This would help determine where CARL lowers the barrier to applied RL and where the workflow still requires too much technical fluency.
\item \textbf{Adaptation Beyond Trillian}. Trillian currently helps by suggesting domain-expert-like responses during the specification-heavy steps, but adaptation in CARL need not be limited to a single reply persona. Future versions could adapt the whole interaction policy: deciding when to ask the user, when to propose defaults, when to surface uncertainty, when to provide alternatives, and when to move forward. One possibility is an active-learning style controller that estimates confidence in each design decision and queries the user only when the expected value of clarification is high. Another is to maintain different interaction modes for expert users, novice users, reviewers, and production owners so that CARL can adjust both the depth and style of its questions without changing the underlying specification discipline.
\item \textbf{Responsible AI}. As CARL moves toward real applications, responsible deployment requires more than code generation and reward maximization. The current examples use benchmarks, policy surfaces, trajectory traces, and failure-mode diagnosis; broader applications will need systematic evaluation protocols for problems without analytical optima. These should include out-of-sample stress tests, constraint-violation audits, reproducibility checks, uncertainty reporting, and comparisons against hand-engineered or model-based baselines where available. Explainability is part of the same responsibility: users should receive natural-language policy summaries, feature-importance or sensitivity analyses, counterfactual examples, and clear accounts of when the trained policy should not be trusted. In regulated domains such as finance and healthcare, the decision trail from user intent to MDP specification to generated artifact may be as important as the learned policy itself.
\item \textbf{CARL as a General Concept}. A final direction is to treat CARL not only as a tool for RL, but as a pattern for building structured human-AI workflows in arbitrary functional domains. In this perspective, CARL is quite generalizable: a higher-level, ``meta'' CARL could help design lower-level CARL-like entities by choosing the steps, skills, memory structure, verification checks, and autonomy boundaries appropriate for a new domain. The RL version studied here makes the idea concrete because MDP specification is visibly difficult, but the same architecture could apply to simulation design, scientific computing, financial modeling, product prototyping, or other tasks where informal intent must be converted into a durable technical artifact. The long-run research question is whether such meta-workflows can reliably generate domain-specific assistants that preserve the same core discipline: staged specification, human authority over meaning, autonomous artifact construction, and evidence-driven revision.
\end{itemize}

\bibliographystyle{chicago}
\bibliography{DBplans_Reinforcement_Learning_LDI, GBWM}

\appendix

\section{Full and Partial Specifications of the GBWM Problem} \label{app:specs}

\subsection{Full Specification}

Your goal is to learn a policy for the following specific problem:
\begin{itemize}
    \item The standard, suggested parameter values for the stock and bond evolution given above\footnote{From the viewpoint of this paper, ``above" refers to what is specified in Subsection \ref{sec:gbwm-demo}.}
    \item Two actions per time step:
        \begin{enumerate}
            \item If a goal is present in the current year, then a binary action 0/1 on whether to take the goal (add utility, subtract wealth) or skip it (no utility, but protect wealth)
            \item An investment action in [0,1] specifying the allocation between all bond (action = 0) through to all stock (action = 1), as a continuous variable
        \end{enumerate}
    \item A policy generalizable to the following cases:
        \begin{enumerate}
            \item A single-goal case with any time horizon in [10,20] years and a goal amount in [100, 500] dollars, uniformly sampled in steps of 1 year and 1 dollar respectively. 
            \item A two-goal case with any time horizon in [10,20] years and each goal amount in [50, 250] dollars, uniformly sampled in steps of 1 year and 1 dollar respectively.
            \item A three-goal case with any time horizon in [10,20] years and each goal amount in [40, 200] dollars, uniformly sampled in steps of 1 year and 1 dollar respectively.
            \item A four-goal case with any time horizon in [10,20] years and each goal amount in [25,125] dollars, uniformly sampled in steps of 1 year and 1 dollar respectively.
        \end{enumerate}
    \item In all cases, the goals can have any integer utility number (basic, before any conditional bonus) in [1,2,..8], but must sum to 8.
    \item In all scenarios, FINAL goals must be taken greedily (if enough money is available at time T, the final step).
    \item An initial wealth of 100 without loss of generalizability (all wealths are relative)
    \item A time step of 1 year for decisions, although note that the stock/bond evolution parameters may be daily frequency. You will have to carefully simulate the evolution over 1 year, including accounting for the correlation.
    \item There are no inbound/outbound cashflows during the time horizon, only the available wealth is fully invested
    \item A canonical example for each case, used for eval throughout training and for final policy plots:
        \begin{enumerate}
            \item Single-goal example of T = 16 years and goal amount = 300 dollars, utility of 8
            \item Two-goal example of T = 16 years, goal amounts of 150 dollars each, first goal at 8 years, utilities of 3 and 5 respectively
            \item Three-goal example of T = 16 years, goal amounts of 125 dollars each, intermediate goals at 5 and 10 years, utilities of 1, 2, and 5 respectively
            \item Four-goal example of T = 16 years, goal amounts of 100 dollars each, intermediate goals at 4, 8, 12 years, utilities of 1, 1, 2, and 4 respectively
        \end{enumerate}
    \item Ensure plotting of a policy heatmap (wealth and time axes for the canonical evaluation examples, with color indicating chosen weight on stock)
    \item In the evaluation plots, wealths should be plotted on log axis for clarity
    \item The total utility is only evaluated at the final time T. This means going bankrupt or achieving the goal amount in the middle of an episode does not cause termination; all episodes are rolled out to T.
\end{itemize}

\textbf{Conditional Utility Computation}

For each previous goal taken, a bonus value of 25 percent is added to the utility of each future goal. For example, if the first goal is taken and the second goal has a utility of 1, then an actual utility of 1.25 is realized for this goal if it is actually taken. If a previous two goals are taken, and a future goal has a utility of 2, then an actual utility of 3 (a 50 percent boost in total) is realized if this goal is taken.

\subsection{Partial Specification}

Your goal is to learn a policy for the following specific problem:
\begin{itemize}
    \item The standard, suggested parameter values for the stock and bond evolution given above\footnote{From the viewpoint of this paper, ``above" refers to what is specified in Subsection \ref{sec:gbwm-demo}.}
    \item Two actions per time step:
        \begin{enumerate}
            \item If a goal is present in the current year, then choose whether to take the goal or skip it
            \item An investment action specifying the allocation between all bond through to all stock
        \end{enumerate}
    \item A single policy that works for the following range of problems:
        \begin{enumerate}
            \item Between 1 and 4 goals over the time horizon, with at most one goal per year
            \item Time horizons between 10 and 20 years
            \item Goal amounts such that the problems are reasonably feasible (not requiring impossible returns)
        \end{enumerate}
    \item An initial wealth of 100 without loss of generalizability (all wealths are relative)
    \item A time step of 1 year for decisions.
    \item A canonical example for each case, used for eval throughout training and for final policy plots:
        \begin{enumerate}
            \item Single-goal example of T = 16 years and goal amount = 300 dollars, utility of 8
            \item Two-goal example of T = 16 years, goal amounts of 150 dollars each, first goal at 8 years, utilities of 3 and 5 respectively
            \item Three-goal example of T = 16 years, goal amounts of 125 dollars each, intermediate goals at 5 and 10 years, utilities of 1, 2, and 5 respectively
            \item Four-goal example of T = 16 years, goal amounts of 100 dollars each, intermediate goals at 4, 8, 12 years, utilities of 1, 1, 2, and 4 respectively
        \end{enumerate}
    \item Ensure plotting of a policy heatmap (wealth and time axes for the canonical evaluation examples, with color indicating chosen weight on stock)
    \item In the evaluation plots, wealths should be plotted on log axis for clarity
\end{itemize}

\textbf{Conditional Utility Computation}

For each previous goal taken, a bonus value of 25 percent is added to the utility of each future goal. For example, if the first goal is taken and the second goal has a utility of 1, then an actual utility of 1.25 is realized for this goal if it is actually taken. If a previous two goals are taken, and a future goal has a utility of 2, then an actual utility of 3 (a 50 percent boost in total) is realized if this goal is taken.

\section{GBWM Implementation Comparison} \label{app:comparison_gbwm}

\begin{table}[H]
\centering
\footnotesize
\renewcommand{\arraystretch}{1.16}
\begin{tabular}{@{}p{0.16\textwidth}p{0.18\textwidth}p{0.27\textwidth}p{0.31\textwidth}@{}}
\toprule
\textbf{Policy} & \textbf{RL architecture} & \textbf{State representation} & \textbf{Action choice}\\
\midrule
\textbf{CARL-FullSpec} & Two trained policies: Double DQN for goal taking and PPO for investment allocation & 24-dimensional state with log wealth, time remaining, prior goals taken, four padded goal slots, remaining-goal summaries, and due/affordable flags & For goal-taking: the DQN outputs values for skip and take, then chooses the larger one. For allocation: PPO outputs the parameters of a Beta distribution. Actions are sampled from the distribution during training, and the learned distribution mean is used for evaluation.\\
\textbf{CARL-PartSpec} & Same two-policy setup as CARL-FullSpec & 10-dimensional compressed state with wealth/time, prior goals taken, current-goal affordability, and aggregate future-goal summaries & Same action rule as CARL-FullSpec: DQN chooses skip/take by larger output value; PPO chooses a continuous stock weight by sampling during training and by mean action at evaluation.\\
\textbf{CC-Custom-Skill-FullSpec} & Single PPO actor-critic & 23-dimensional state with separate current-goal fields plus up to three future-goal slots, along with wealth/time, BND-relative price, and prior-goal terms & Two outputs: a discrete take/skip choice (Bernoulli distribution) and a continuous stock weight (Beta distribution). Training samples both choices from learned distributions; evaluation takes the goal when its learned probability exceeds 0.5 and uses the mean stock weight.\\
\textbf{CC-Custom-Skill-PartSpec} & Double DQN & 23-dimensional state with the same total size but a different layout: wealth/time, BND-relative price, prior-goal and bonus terms, plus four uniform padded goal slots & One discrete output vector with 22 values, one for each take/skip and 11-bin allocation pair. The selected action is the pair with the largest value; the allocation bin maps to a stock weight in $\{0,0.1,\ldots,1\}$. \\
\textbf{CC-Only-FullSpec} & Single PPO actor-critic with a three-layer fully connected neural network & 21-dimensional active-goal-slot state with log wealth, normalized time and horizon, goals taken, current-goal flag, and four active-goal slots & Two policy outputs: continuous stock weight (Beta distribution) and discrete take/skip (Bernoulli distribution). Training samples both choices; evaluation uses the mean stock weight and takes the goal when the take output exceeds its threshold.\\
\textbf{CC-Only-PartSpec} & Single PPO actor-critic with a two-layer fully connected neural network & 26-dimensional padded pending-goal state with six global features, including BND deviation and prior-goal bonus, plus four five-feature goal slots & Two policy outputs: discrete take/skip (Bernoulli distribution) and continuous stock weight (Beta distribution). Training samples both choices; evaluation thresholds the take output and uses the mean stock weight.\\
\bottomrule
\end{tabular}
\caption{Implementation choices made by the six AI approaches. The state dimension is the number of input variables observed by the learned policy; for CARL, which uses separate goal-taking and allocation views, the table reports the larger dimension. The action column describes the policy outputs and how they are converted into actual decisions. DQN denotes Deep Q-Network, which chooses the action with the largest learned value. Double DQN is a DQN variant that uses separate networks for action selection and value evaluation during training. PPO denotes Proximal Policy Optimization, which samples from learned action distributions during training and uses deterministic actions at evaluation. MLP denotes multilayer perceptron.}
\label{tab:gbwm-policy-implementations}
\end{table}

\subsection{State Variables for the Six Methods} \label{app:state-var}

\begin{table}[H]
\centering
\footnotesize
\renewcommand{\arraystretch}{1.1}
\setlength{\tabcolsep}{4pt}
\begin{tabular}{@{}p{0.4\textwidth}cccccc@{}}
\toprule
\textbf{State feature} & \rotatebox{60}{\textbf{CARL-FullSpec}} & \rotatebox{60}{\textbf{CARL-PartSpec}} & \rotatebox{60}{\textbf{CC-Cust.-FullSpec}} & \rotatebox{60}{\textbf{CC-Cust.-PartSpec}} & \rotatebox{60}{\textbf{CC-Only-FullSpec}} & \rotatebox{60}{\textbf{CC-Only-PartSpec}}\\
\midrule
Log wealth ($\log(W/W_0)$)                             & \checkmark & \checkmark & \checkmark & \checkmark & \checkmark & \checkmark \\
Current time within episode ($t$ normalized)           &            &            & \checkmark & \checkmark & \checkmark & \checkmark \\
Episode horizon $T$ (normalized)                       &            &            & \checkmark &            & \checkmark & \checkmark \\
Time remaining ($T-t$ normalized)                      & \checkmark & \checkmark & \checkmark & \checkmark &            & \checkmark \\
Bond / Vasicek market price state                           &            & \checkmark & \checkmark & \checkmark &            & \checkmark \\
Goals-taken count (or bonus multiplier $1+0.25k$)      & \checkmark & \checkmark & \checkmark & \checkmark & \checkmark & \checkmark \\
Number of future goals (aggregate)           & \checkmark & \checkmark &            &            &            &            \\
Goal-due-now flag                                      & \checkmark & \checkmark & \checkmark & \checkmark & \checkmark & \checkmark \\
Current-goal affordability ($W\geq C_i$?)  & \checkmark &            & \checkmark &            &            &            \\
Current-goal amount (scalar)                           &            & \checkmark & \checkmark &            &            &            \\
Current-goal utility (scalar)                          &            & \checkmark & \checkmark &            &            &            \\
Per-slot future goal: presence mask                    & \checkmark &            & \checkmark & \checkmark & \checkmark & \checkmark \\
Per-slot future goal: time-to-goal                     & \checkmark &            & \checkmark & \checkmark & \checkmark & \checkmark \\
Per-slot future goal: amount ($W$- or $W_0$-relative)  & \checkmark &            & \checkmark & \checkmark & \checkmark & \checkmark \\
Per-slot future goal: utility (normalized)             & \checkmark &            & \checkmark & \checkmark & \checkmark & \checkmark \\
Aggregate future PV / wealth slack                     & \checkmark & \checkmark &            &            &            &            \\
Aggregate remaining utility ($U_{\text{after}}/U_{\text{total}}$) &  & \checkmark &            &            &            &            \\
Time to next future goal (scalar)                      &            & \checkmark &            &            &            &            \\
Goals-taken fraction relative to episode schedule      &            &            &            & \checkmark &            &            \\
\bottomrule
\end{tabular}
\caption{State features across the six trained AI methods. States with related normalizations are grouped in the same row. (E.g., $\log(W/100)$, $\log(W/W_0)$, and clipped $\log(W/W_0)$ all count as ``log wealth''; raw count $k$, additive bonus $0.25k$, and multiplicative bonus $1+0.25k$ all count as the goals-taken feature). Some rows expand to multiple observation entries: in particular, the four ``per-slot active/future goal'' rows are each repeated once per goal slot (up to four slots per policy, so up to four scalar entries per checked cell), and the ``goals-taken count'' row is duplicated in the two policies (CC-Only-PartSpec, CC-Custom-Skill-FullSpec) that expose both the raw count and the bonus multiplier. Intuition behind the state definitions is provided in accompanying text.}
\label{tab:gbwm-state-features}
\end{table}

The state features in Table~\ref{tab:gbwm-state-features} can be read intuitively as follows:
\begin{itemize}
  \item \textbf{Log wealth} normalizes wealth by initial wealth and then takes logs, which fits the multiplicative nature of investment returns.
  \item \textbf{Current time, horizon, and time remaining} tell the policy where it is in the episode, how long the episode is, and how many years remain for investment growth or recovery from losses. The normalization constant used is ideally the maximum horizon $T_\mathrm{max}$, a constant. However, some non-interactive runs (CC-Custom-Skill or CC-Only) choose to normalize by $T$, the sampled horizon for that particular problem.
  \item \textbf{Bond / Vasicek market state} exposes whether the bond price is high or low relative to the parameter for its mean price, which matters because BND is modeled as mean-reverting to this mean parameter. An analogous stock state is not required because that process is Gaussian Brownian Motion, and hence memory-less.
  \item \textbf{Goals-taken count or bonus multiplier} records the path dependence: earlier attained goals increase the utility earned from later goals.
  \item \textbf{Number of remaining active goals} gives a compact measure of how many goals are still in the future.
  \item \textbf{Goal-due-now flag} tells the policy whether the take/skip decision is currently relevant.
  \item \textbf{Current-goal affordability} tells the policy whether the current goal can actually be purchased with available wealth, i.e. $W\geq C_i$.
  \item \textbf{Current-goal amount and utility} describe the immediate tradeoff: how much wealth the goal consumes and how much utility it provides if taken.
  \item \textbf{Per-slot future-goal presence, timing, amount, and utility} preserve the future goal calendar: the time to each goal, its cost, and its utility. Where the number of upcoming goals is smaller than the slots available to record them, $0$ values are padded to retain a fixed state size. A ``presence mask" is a binary feature indicating whether the slot is being used to indicate real goal values, or whether it is part of the padding.
  \item \textbf{Aggregate future present-value / wealth slack} summarizes whether current wealth is comfortably above or below the approximate present-value cost of remaining goals.
  \item \textbf{Aggregate remaining utility} summarizes how much reward is still available after the current point in the episode.
  \item \textbf{Time to next future goal} gives a compact urgency signal to plan for the first upcoming goal.
  \item \textbf{Goals-taken fraction relative to the episode schedule} measures progress relative to the number of goals in that episode, rather than using only the raw number already taken. This state helps disambiguate between ``1 of 2 goals taken" and ``1 of 4 goals taken." 
\end{itemize}

\section{State Variables for the Three Methods in the Goddard Problem Implementation} \label{app:comparison_goddard}

\begin{table}[H]
\centering
\footnotesize
\renewcommand{\arraystretch}{1.1}
\setlength{\tabcolsep}{4pt}
\begin{tabular}{@{}p{0.52\textwidth}ccc@{}}
\toprule
\textbf{Observation feature} & \textbf{CC-Only} & \textbf{CC-Custom-Skill} & \textbf{CARL} \\
\midrule
Altitude $h$ (normalized)                                        & \checkmark & \checkmark &            \\
Atmospheric density ratio $\rho(h)/\rho_0$                        &            & \checkmark & \checkmark \\
Velocity $v$ (normalized)                                         & \checkmark & \checkmark & \checkmark \\
Current total mass (raw, or dry-mass ratio $m_{\mathrm{dry}}/m$)  & \checkmark &            & \checkmark \\
Fuel remaining fraction                                           & \checkmark & \checkmark & \checkmark \\
Dynamic-pressure fraction $q/Q_{\max}$                            & \checkmark & \checkmark & \checkmark \\
Dynamic-pressure headroom $1-q/Q_{\max}$                          &            &            & \checkmark \\
Thrust-to-weight ratio $T_{\max}/(mg)$                            & \checkmark & \checkmark & \checkmark \\
Max thrust $T_{\max}$ (nominal-scaled)                            & \checkmark & \checkmark & \checkmark \\
Max dynamic pressure $Q_{\max}$ (nominal-scaled)                  & \checkmark & \checkmark & \checkmark \\
Drag coefficient $c_d$ (nominal-scaled or raw)                    & \checkmark & \checkmark & \checkmark \\
Initial fuel mass $m_{\mathrm{fuel},0}$ (nominal-scaled)          & \checkmark & \checkmark & \checkmark \\
Energy-based ascent-progress ratio                                &            &            & \checkmark \\
Burnout-time-remaining fraction                                   &            &            & \checkmark \\
Constant-throttle-to-burnout hint                                 &            &            & \checkmark \\
\bottomrule
\end{tabular}
\caption{State features across the three trained AI methods. States with related normalizations are grouped in the same row. (E.g. a linear nominal-scaled ratio such as $T_{\max}/200{,}000$ and a log-scaled ratio such as $\log(T_{\max}/200{,}000)$ both count as the ``max thrust'' row.) The checked cells in each column are exactly CC-Custom-Skill's 10-dimensional observation, CARL's 14-dimensional observation, and CC-Only's 10-dimensional observation (as stated in Subsection~\ref{sec:god_rl}). CARL is the only policy with a direct energy-based ascent-progress feature and the two burnout-shaping features; CC-Custom-Skill and CC-Only have no explicit notion of burnout timing and rely on fuel fraction and thrust-to-weight alone. Conversely, CARL does not observe altitude directly, relying instead on the monotonically related atmospheric density ratio, while CC-Custom-Skill and CC-Only do observe altitude directly.}
\label{tab:god-state-features}
\end{table}

The state features in Table~\ref{tab:god-state-features} can be read intuitively as follows:
\begin{itemize}
  \item \textbf{Altitude} tells the policy how far the rocket has already climbed. Normalizing by the atmospheric scale height $H$ makes the scale of the input more manageable.
  \item \textbf{Atmospheric density} decreases monotonically with altitude, but is easier to map to the physics of the problem: drag and dynamic pressure both depend directly on $\rho(h)$. A lower density also means the rocket can travel faster before reaching the same $Q_{\max}$.
  \item \textbf{Velocity} captures both progress toward apogee and risk of violating the $Q_{\max}$ constraint. Together with density, it determines dynamic pressure, while its kinetic-energy contribution also affects how much additional altitude the rocket can coast to after burnout.
  \item \textbf{Current total mass} determines how much acceleration a given thrust command produces. The same thrust is much more effective after some fuel has been burned.
  \item \textbf{Fuel remaining fraction} records the remaining propellant budget relative to the initial fuel load. This tells the policy whether it can keep burning, should conserve fuel, or is already near coast-only flight.
  \item \textbf{Dynamic-pressure fraction} $q/Q_{\max}$ tells the policy how close it is to the max-Q constraint. This is the main safety signal: values near one mean additional throttle is likely to be capped or penalized through the dynamics.
  \item \textbf{Dynamic-pressure headroom} $1-q/Q_{\max}$ exposes the remaining margin to the constraint directly. This duplicates information in $q/Q_{\max}$, but in a form that is easy for the policy to use as a throttle margin.
  \item \textbf{Thrust-to-weight ratio} $T_{\max}/(mg)$ summarizes current lift capability. It tells the policy whether full throttle can comfortably overcome gravity, and how that capability improves as mass falls.
  \item \textbf{Max thrust} identifies the episode's engine scale. Under domain randomization, the same normalized throttle can correspond to very different physical forces, so the policy needs to observe $T_{\max}$ separately.
  \item \textbf{Max dynamic pressure} identifies the episode's structural load limit. A low $Q_{\max}$ implies earlier throttle-down behavior, while a high $Q_{\max}$ lets the policy stay closer to bang-bang thrust.
  \item \textbf{Drag coefficient} tells the policy how costly high-speed atmospheric flight is in the current episode. Higher drag makes early aggressive thrust less attractive even when the $Q_{\max}$ cap is not binding.
  \item \textbf{Initial fuel mass} gives the policy the episode's starting energy budget and burn-duration scale. It also helps interpret the current fuel fraction: $50\%$ remaining means something different for a light vs. heavy rocket.
  \item \textbf{Energy-based ascent-progress ratio} combines altitude, kinetic energy, and remaining propellant energy into a compact progress measure. If all three of these quantities are converted into altitude units (meters), this state acts as a proxy for how much further altitude can be achieved. It helps distinguish trajectories that have the same altitude but very different remaining ability to climb.
  \item \textbf{Burnout-time-remaining fraction} tells the policy whether it is ahead of or behind the target fuel-depletion schedule introduced in CARL's revisit. Positive values mean the target burnout time is still ahead; negative values mean the rocket has burned past that target window.
  \item \textbf{Constant-throttle-to-burnout hint} gives the throttle that would consume the remaining fuel exactly at the target burnout time. It is an action-scale hint for modulating the fuel's burn rate.
\end{itemize}

\end{document}